\UseRawInputEncoding

\documentclass[aip,jcp,amsmath,amssymb,floatfix,reprint,citeautoscript,noeprint,superscriptaddress,twocolumn]{revtex4-2}

\usepackage{lipsum} 
\usepackage{bibentry}
\usepackage[english]{babel}
\usepackage[dvipsnames]{xcolor}
\usepackage{graphicx}
\usepackage[caption=false]{subfig} 
\usepackage{amsmath,amssymb,bm}
\usepackage[version=3]{mhchem}
\usepackage{verbatim}
\usepackage{multirow}
\usepackage{dcolumn}
\usepackage{float}
\usepackage{nicefrac}
\usepackage{siunitx}
\usepackage{booktabs}
\usepackage{chemformula}
\usepackage{wrapfig}
\usepackage{enumitem}  
\usepackage{transparent}
\usepackage[colorlinks,allcolors=black,citecolor=blue,urlcolor=blue]{hyperref}
\usepackage{tcolorbox}
\usepackage{relsize}

\DeclareSIUnit[number-unit-product = {\,}]{\amu}{amu}
\DeclareSIUnit[number-unit-product = {\,}]{\kJmol}{\kilo\joule\per\mol}
\DeclareSIUnit[number-unit-product = {\,}]{\Nsm}{\newton\second\per\meter\cubed}
\DeclareSIUnit[number-unit-product = {\,}]{\THz}{\tera\hertz}
\DeclareSIUnit[number-unit-product = {\,}]{\meV}{\milli\electronvolt}
\DeclareSIUnit[number-unit-product = {\,}]{\cal}{cal}

\newcommand{\tbf}[1]{\textbf{#1}}

\newcommand{\mbf}[1]{\boldsymbol{\mathit{#1}}}
\newcommand{\mrm}[1]{\mathrm{#1}}
\newcommand{\mcl}[1]{\mathcal{#1}}

\newcommand{\tcb}[1]{\textcolor{black}{#1}}

\newcommand{\tcn}[1]{\textcolor{black}{#1}}

\newcommand{\etal}{\emph{et al.}}

\usepackage{sansmathfonts}
\usepackage[justification=centerlast, font={sf}, labelfont={bf}]{caption}
\begin{document}

\title{A classical density functional theory for solvation across length scales}

\author{Anna T. Bui}
\affiliation{Yusuf Hamied Department of Chemistry, University of
  Cambridge, Lensfield Road, Cambridge, CB2 1EW, United Kingdom}

\author{Stephen J. Cox}
\email{sjc236@cam.ac.uk}
\affiliation{Yusuf Hamied Department of Chemistry, University of
  Cambridge, Lensfield Road, Cambridge, CB2 1EW, United Kingdom}

\date{\today}

\begin{abstract}
A central aim of multiscale modeling is to use results from the
Schr\"{o}dinger \tcb{equation} to predict phenomenology on length scales
that far exceed those of typical molecular correlations. In this work,
we present a new approach rooted in classical density functional
theory (cDFT) that allows us to accurately describe the solvation of
apolar solutes across length scales. Our approach builds on the Lum,
Chandler and Weeks (LCW) theory of hydrophobicity
[\tcb{K. Lum \etal{},} J. Phys. Chem. B \tbf{103}, 4570 (1999)] by constructing a free
energy functional that uses a slowly-varying component of the density
field as a reference. From a practical viewpoint, the theory we
present is numerically simpler and generalizes to solutes with
soft-core repulsion more easily than LCW theory. Furthermore,
by assessing the local compressibility and its critical scaling
behavior, we demonstrate that our LCW-style cDFT approach contains the
physics of critical drying, which has been emphasized as an essential
aspect of hydrophobicity by recent theories. As our approach is
parameterized on the two-body direct correlation function of the
uniform fluid and the liquid--vapor surface tension, it
straightforwardly captures the temperature dependence of solvation.
Moreover, we use our theory to describe solvation at a
first-principles level, on length scales that vastly exceed what is
accessible to molecular simulations.
\end{abstract}

\maketitle


\section*{Introduction}
Many of the most fundamental processes in nature, including protein
folding, crystallization and self-assembly, occur in solution.
Far from being an innocent bystander, the solvent often plays a
vital role in determining the static and dynamic behaviors of these
complex processes \cite{Baldwin1986,tenWolde2002,Dighe2019,Chandler2005},
owing to the delicate balance of solute--solute, solute--solvent and
solvent--solvent interactions. This provides a strong motivation to
faithfully describe solvation behavior across a broad range of fields,
from biological and chemical to physical and materials sciences.
Solutes of interest can range in length scale from microscopic
species \cite{Harris2014} and nano-particles \cite{Welch2016} to
macromolecules \cite{Patel2012} and extended
surfaces \cite{Evans2019}. Solvation is a multiscale problem.

\begin{figure}[t]
  \includegraphics[width=\linewidth]{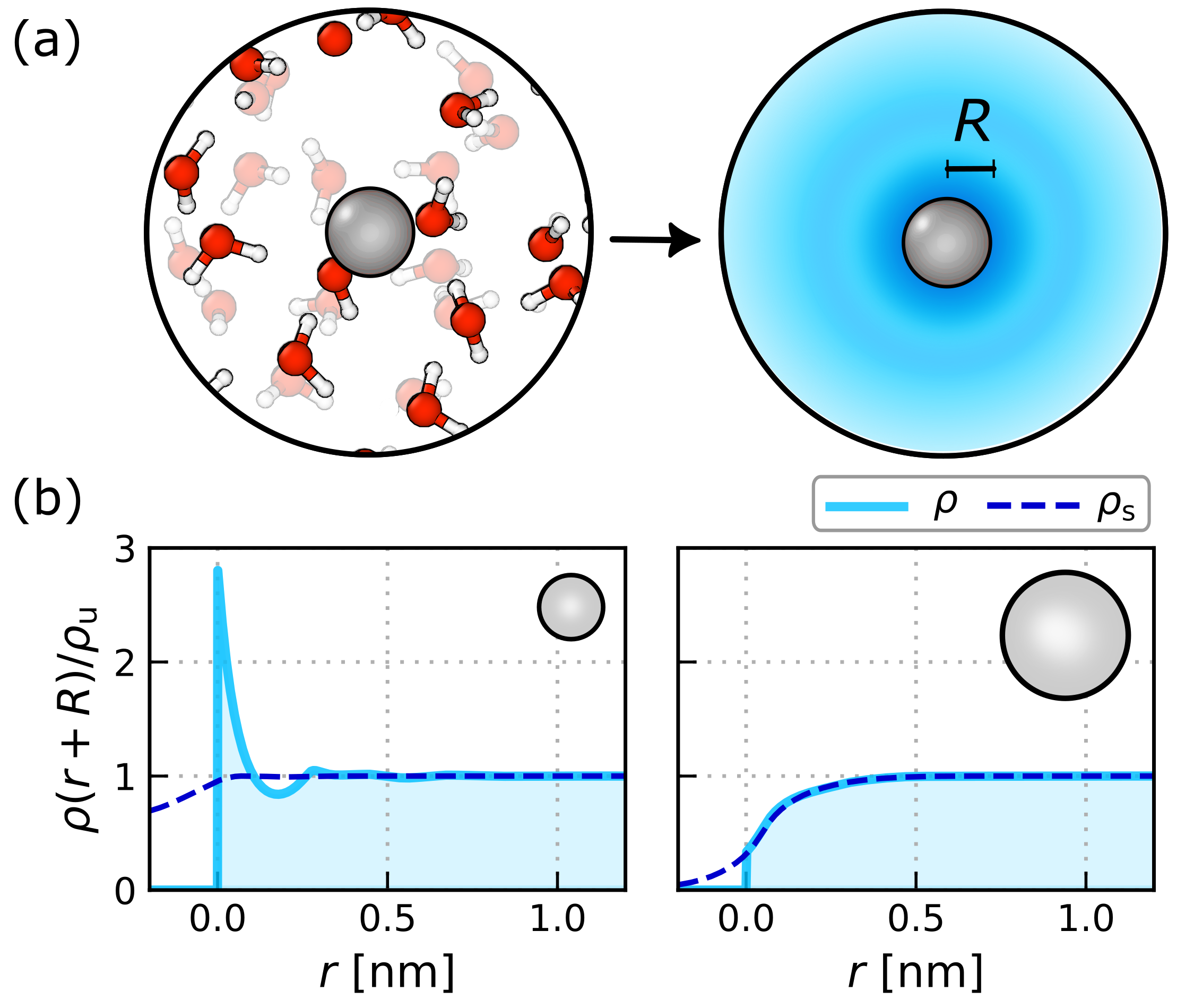}
  \caption{\textbf{``Semi--implicit'' solvation with cDFT.}
  Our aim is to accurately describe aqueous solvation without
  explicitly sampling microscopic degrees of freedom, yet still retain
  information on essential correlations. This is shown schematically
  in (a). Such an approach should be able to predict, e.g., the
  average solvent density $\rho(r)$ around a solute, as depicted in
  (b). The solid blue lines represent $\rho(r+R)/\rho_{\rm u}$ for
  hard-sphere solutes of radius $R = 0.3$\,nm (left) and $R = 3$\,nm
  (right), where $\rho_{\rm u}$ is the uniform density of the bulk
  fluid. The cDFT that we derive relies upon finding an appropriate
  slowly-varying density $\rho_{\rm s}$ (dashed blue line) that
  can act as a suitable reference system.}
\label{fig:solvation}
\end{figure}

The solvent, of course, comprises individual molecules. Molecular
simulations therefore provide a natural approach to describe
solvation, and bestow fine details at time and length scales that can
be challenging to access with experimental approaches alone. Depending
upon the approximations made in describing the intermolecular
interactions, molecular simulations provide one of the most accurate
means to estimate solvation free energies. Yet the relatively high
computational cost associated with molecular simulations makes their
routine use for solutes much larger than small organic compounds
cumbersome and inefficient \cite{Levy1998}.

Implicit solvation models alleviate the computational burden by
describing the solvent degrees of freedom as a structureless
continuum \cite{Chothia1974,Sharp1990,Sharp1991,Cramer1992}. Such
approaches are numerically efficient, making it possible to routinely
handle large macromolecules in solution.
The major drawback of implicit solvation models, however, is their
failure to account for essential solvent correlations that may hold a
prominent role in the process under investigation (see, e.g.,
Ref.~\onlinecite{tenWolde2002}). They also often make assumptions
concerning the validity of macroscopic laws applied to the microscopic
domain, which can lead to inconsistent results \cite{Wagoner2006}.
For these reasons, approaches that provide a coarse-grained
description of solvation, while retaining information about essential
molecular correlations, become very appealing. Such approaches should,
for example, capture changes in the average equilibrium density field
upon changing the solute--solvent interaction, as depicted in
Fig.~\ref{fig:solvation}, without resorting to explicitly averaging
microscopic degrees of freedom.

Motivated by both the seminal work of Lum, Chandler and Weeks
(LCW) \cite{Lum1999}, and more recent developments in molecular
density functional theory
(mDFT) \cite{Zhao2011mdft,Jeanmairet2013,Jeanmairet2013jcp,Jeanmairet2016,Borgis2020,Borgis2021},
in this article we present a classical density functional theory
(cDFT) \cite{Evans1979, EvansBook, Lutsko2010} for the solvation of
apolar solutes.  Although our approach
differs from the original LCW theory, it retains the essential feature
of appropriately accounting for both slowly- and rapidly-varying
components of the density field, as shown in
Fig.~\ref{fig:solvation}(b). From a practical viewpoint, the theory we 
present lends itself more readily to numerical evaluation than
LCW theory, including application to solutes with soft repulsive cores
and attractive tails. But as we will discuss, our approach also offers
conceptual advantages.

\section*{Relevant background theory: \lowercase{c}DFT for solvation}

The central quantity in any cDFT approach is the grand potential
functional,
\tcn{
\begin{equation}
 \label{eqn:OmegaGen}
 \varOmega_{\mathlarger{\phi_i}}[\rho] = \mcl{F}_{\mrm{intr}}[\rho] - \mu\int\!\!\mrm{d}\mbf{r}\,\rho(\mbf{r}) + \int\!\!\mrm{d}\mbf{r}\,\rho(\mbf{r})\phi_{i}(\mbf{r}),
\end{equation}
where $\rho(\mbf{r})$ is the average of a microscopic density field of
the fluid, whose chemical potential is $\mu$ and $\mcl{F}_{\rm intr}$
is the intrinsic Helmholtz free energy functional independent of the
external potential $\phi_{i}$. The grand potential functional
$\varOmega_\mathlarger{\phi_i}$ is minimized by the corresponding
equilibrium density $\rho_i$, which satisfies
\begin{equation}
    c_{i}^{(1)}(\mbf{r})= \ln\big[\Lambda^3 \rho_{i}(\mbf{r})\big] + \beta\phi_{i}(\mbf{r}) - \beta\mu, 
    \label{eqn:equilibrium_density}
\end{equation}
where $\Lambda$ is the thermal wavelength, $\beta=1/(k_{\mrm{B}}T)$, $k_{\mrm{B}}$ is the Boltzmann constant, $T$ is the temperature and $c_{i}^{(1)}$ is the one-body
direct correlation function,
\begin{equation}
   \label{eqn:1DCF-def}
    c_{i}^{(1)}(\mbf{r})
 = -\beta\frac{\delta \mcl{F}^{(\rm ex)}_{\rm intr}}{\delta\rho(\mbf{r})}\Bigg|_{\mathlarger{\rho_{i}}},
\end{equation}
with $\mcl{F}^{(\rm ex)}_{\rm intr}$ as the excess contribution to
$\mcl{F}_{\rm intr}$. Equations~\ref{eqn:OmegaGen}--\ref{eqn:1DCF-def}
are written for any general scalar external potential; in the context
of solvation, the external potential describes the solute--solvent
interactions. In such cases, we will drop the subscript $i$ for the
quantities in Eq.~\ref{eqn:equilibrium_density}.}

While cDFT is in principle an exact theory, in general, approximations for
$\mcl{F}^{\mrm{(ex)}}_{\rm intr}$ are required. For hard sphere systems,
functionals based on Rosenfeld's fundamental measure theory have
proven highly successful \cite{Rosenfeld1989,Tarazona2000,
Roth2002}. Moreover, in cases where hard spheres act as a suitable
reference fluid, attractive interactions can be reasonably treated in
a mean-field fashion \cite{EvansBook}. While approaches based on
fundamental measure theory may capture some of the essential physics
of more complex liquids such as water, it is unreasonable to expect
quantitative agreement. Instead, in such cases, the mDFT approach 
\cite{Zhao2011mdft,Jeanmairet2013,Jeanmairet2013jcp,Jeanmairet2016,Borgis2020,Borgis2021}
has shown great promise. The essential idea behind mDFT is that one can
use the two-body direct correlation function,
\begin{equation}
   c_{\rm u}^{(2)}(|\mbf{r}-\mbf{r}^\prime|)
 =  \frac{\delta c^{(1)}(\mbf{r})}{\delta\rho(\mbf{r}^\prime)}\Bigg|_{\mathlarger{\rho_{\rm u}}} = -\beta\frac{\delta^2\mcl{F}^{(\rm ex)}_{\rm intr}}{\delta\rho(\mbf{r})\delta\rho(\mbf{r}^\prime)}\Bigg|_{\mathlarger{\rho_{\rm u}}},
    \label{eqn:cu2}
\end{equation}
obtained from simulations of the bulk fluid of uniform density
$\rho_{\rm u}$
to parameterize the grand potential functional
\begin{equation}
\begin{split}
   \label{eqn:OmegaHNCandBridge}
   \varOmega_{\mathlarger{\phi}}[\rho] = &\;\varOmega_{\mathlarger 0}[\rho_{\rm u}] + \int\!\!\mrm{d}\mbf{r}\,\rho(\mbf{r})\phi(\mbf{r}) + \Delta_{\rm u}\mcl{F}_{\rm intr}^{\rm (id)}[\rho]\\
   & - \frac{k_{\rm B}T}{2}\int\!\!\mrm{d}\mbf{r}\!\!\int\!\!\mrm{d}\mbf{r}^\prime\,\delta_{\rm u}\rho(\mbf{r})\,c_{\rm u}^{(2)}(|\mbf{r}-\mbf{r}^\prime|)\,\delta_{\rm u}\rho(\mbf{r}^\prime) \\
   & + \mcl{F}_{\rm bridge}[\rho(\mbf{r})].
\end{split}
\end{equation}
In
Eq.~\ref{eqn:OmegaHNCandBridge}, $\Delta_{\rm u}\mcl{F}_{\rm
intr}^{\rm (id)}$ is the change in the ideal contribution to
$\mcl{F}_{\rm intr}$ between systems with uniform and non-uniform
density fields
\begin{equation}
  \Delta_{\rm u}\mcl{F}_{\rm
intr}^{\rm (id)}[\rho] = k_{\mrm{B}}T\!\int\!\!\mrm{d}\mbf{r}\,\left[\rho(\mbf{r})\ln\left(\frac{\rho(\mbf{r})}{\rho_{\mrm{u}}}\right) - \delta_{\rm u}\rho(\mbf{r}) \right],
\label{eqn:ideal-change}
\end{equation}
where $\delta_{\rm u}\rho(\mbf{r}) = \rho(\mbf{r})
- \rho_{\rm u}$. The ``bridge'' functional, $\mcl{F}_{\rm bridge}$,
accounts for contributions to the excess part of $\mcl{F}_{\rm intr}$
beyond quadratic order in $\delta_{\rm u}\rho(\mbf{r})$. The solvation
free energy is simply
\begin{equation}
\label{eq:solvation_free_energy}
\varOmega_{\rm solv}[\rho(\mbf{r})] = \varOmega_{\mathlarger{\phi}}[\rho(\mbf{r})] - \varOmega_{\mathlarger 0}[\rho_{\rm u}].
\end{equation}

Neglecting $\mcl{F}_{\rm bridge}$ amounts to the hypernetted-chain
approximation (HNCA) of integral equation
theories \cite{HansenMcDonaldBook}. Furthermore, upon linearizing
$\Delta_{\rm u}\mcl{F}_{\rm intr}^{\rm (id)}$, mDFT within the HNCA is
equivalent, up to a small correction factor, to Chandler's Gaussian
field theory for solvation \cite{Sergiievskyi2017,Chandler1993}. Aside
from being numerically tractable, the HNCA provides reasonable
accuracy for small solutes. However, as we have emphasized, solvation
is a multiscale problem. To see this, consider that the solute simply
excludes solvent density within a radius $R$ of its center. Whereas
within the HNCA, $\varOmega_{\rm solv}$ scales indefinitely with the
solute's \tcb{volume} \cite{Chandler1993,Hummer1996,Sergiievskyi2017,Varilly2012},
for large enough solutes we know from macroscopic theory that it
scales with the solute's surface area, i.e., $ \varOmega_{\rm
solv} \sim 4\pi\gamma R^2$ where $\gamma$ is the liquid--vapor surface
tension. Many previous studies have shown that for water under ambient
conditions, this ``hydrophobic crossover'' occurs at $R\approx
1$\,nm \cite{Lum1999, Huang2001, Huang2002, Sedlmeier2012,
Vaikuntanathan2014}.

The failure of the HNCA for large solutes arises because it cannot
describe water's proximity to its liquid--vapor coexistence at ambient
conditions \cite{Evans1983, EvansBook}.  A natural progression, then,
is to attempt to encode the physics of coexistence through
$\mcl{F}_{\rm bridge}$, while maintaining reference to the homogeneous
fluid with density $\rho_{\rm u}$.  Such an approach has been used
with some success within the mDFT framework for simple point charge
water models \cite{Borgis2020, Borgis2021}.  However, as we show in
the \tcb{\emph{Supplementary Material} (SM)}, existing bridge functionals of this kind are generally not
robust to the choice of water model.  \tcb{We also note that attempts
to use the ``hard-sphere bridge functional'' for
water \cite{Zhao2011bridge, Levesque2012, Jeanmairet2013jcp} have
proved problematic, failing to describe simultaneously the pressure
and surface tension} \cite{Borgis2020}. In this article, we will
instead follow more closely the key idea underlying LCW theory: the
fluid's density can be separated into a slowly-varying component that
can sustain interfaces and liquid--vapor coexistence, and a
rapidly-varying component that describes the local structure on
microscopic length scales \cite{Lum1999, Weeks2002}.

In the following, we will \tcb{outline} how these ideas from LCW theory can
be used to develop a cDFT approach \tcb{to describe the} solvation of apolar solutes
across both small and large length scales. We will validate our theory
against available simulation data for the solvation of both hard- and
soft-core solutes. We will show that the physics of critical drying,
which is essential for a faithful description of solvophobicity on
large length scales, is well-described. We will also demonstrate that
our approach captures a distinguishing feature of solvophobicity in
complex liquids such as water--- the ``entropic crossover''---that is
absent in simple liquids. We will use our theory to describe the
hydrophobic effect at an \emph{ab initio} level, on length scales
inaccessible to molecular simulations.

\section*{A \lowercase{c}DFT built on separation of length scales}


\subsection{Expansion about an inhomogenous density}

In the HNCA\tcb{,} the uniform fluid is assumed to act as a suitable
reference density. In principle, we can perform a similar procedure
where, instead of choosing a fluid of uniform density, we suppose that
there exists some inhomogeneous, but slowly-varying, density field
$\rho_{\rm s}(\mbf{r})$ that acts as a suitable reference. In line
with Eq.~\ref{eqn:OmegaGen}, $\rho_{\rm s}$ minimizes the grand
potential functional $\varOmega_{\mathlarger{\phi_{\rm s}}}$
prescribed by a slowly-varying external potential
$\phi_{\mrm{s}}(\mbf{r})$.  Performing the expansion around $\rho_{\rm
s}$ gives,
\begin{equation}
\begin{split}
\label{eqn:Fex_intr-1}
\mcl{F}^{(\rm ex)}_{\rm intr}[\rho] =&\; \mcl{F}^{(\rm ex)}_{\rm intr}[\rho_{\rm s}] - k_{\rm B}T\int\!\!\mrm{d}\mbf{r}\,c_{\rm s}^{(1)}(\mbf{r})\delta_{\rm s}\rho(\mbf{r}) \\
& -\frac{k_{\rm B}T}{2}\int\!\!\mrm{d}\mbf{r}\!\!\int\!\!\mrm{d}\mbf{r}^\prime\,\delta_{\rm s}\rho(\mbf{r})c_{\rm s}^{(2)}(\mbf{r},\mbf{r}^\prime)\delta_{\rm s}\rho(\mbf{r}^\prime) + \ldots,
\end{split}
\end{equation}
where $\delta_{\rm s}\rho(\mbf{r}) = \rho(\mbf{r})-\rho_{\rm
s}(\mbf{r})$, and the one- and two-body direct correlation functions
are defined in an analogous manner to
Eqs.~\ref{eqn:equilibrium_density} and~\ref{eqn:cu2}.  Upon addition
of $\mathcal{F}^{(\rm id)}_{\rm intr}$ to Eq.~\ref{eqn:Fex_intr-1} and
substitution of $c_{\mrm{s}}^{(1)}$ according to its definition
(Eq.~\ref{eqn:equilibrium_density}), we arrive at a slightly modified
version of Eq.~\ref{eqn:OmegaHNCandBridge}
\begin{equation}
\begin{split}
   \varOmega_{\mathlarger{\phi}}[\rho]=&\; \varOmega_{\mathlarger{\phi_{\mrm{s}}}}[\rho_{\rm s}] + \int\!\!\mrm{d}\mbf{r}\,\left[\phi(\mbf{r})-\phi_{\mrm{s}}(\mbf{r})\right]\rho(\mbf{r}) + \Delta_{\rm s}\mcl{F}_{\rm intr}^{\rm (id)}[\rho]\\
   & - \frac{k_{\rm B}T}{2}\int\!\!\mrm{d}\mbf{r}\!\!\int\!\!\mrm{d}\mbf{r}^\prime\,\delta_{\rm s}\rho(\mbf{r})\,c_{\rm s}^{(2)}(\mbf{r},\mbf{r}^\prime)\,\delta_{\rm s}\rho(\mbf{r}^\prime)  + \ldots,
\end{split}
\label{eqn:single-expansion}
\end{equation}
where the change in the ideal contribution $\Delta_{\rm s}\mcl{F}_{\rm
intr}^{\rm (id)}[\rho]$ is given analogously to
Eq.~\ref{eqn:ideal-change}.  If $\phi_{\rm s}$ were known (and
assuming that $c^{(2)}_{\rm s}$ can be reasonably approximated), the
reference density $\rho_{\rm s}$ would result from minimization of
$\varOmega_{\mathlarger{\phi_{\rm s}}}$ and, in turn, $\rho$ from
minimization of Eq.~\ref{eqn:single-expansion}. For example, with
$\phi_{\mrm{s}}(\mbf{r})=0$, the HNCA is recovered. In the general
context of solvation, however, the appropriate choice for
$\phi_{\mrm{s}}$ is unknown.

\subsection{Specifying an appropriate inhomogeneous reference using coexistence solutions}

Prescribing a general form for $\phi_{\rm s}$ is challenging. As we
are interested in liquid water at thermodynamic states close to
coexistence, we therefore adopt a strategy which exploits the fact
that, at coexistence, inhomogeneous density fields minimize the grand
potential when subject to appropriate boundary conditions. For
example, in a planar geometry, solutions corresponding to the free
liquid-vapor interface are obtained by minimizing the grand potential
at coexistence, subject to the conditions \[\tcb{\rho(z\to -\infty)
= \rho_{\rm v} \text{~and~} \rho(z\to \infty) = \rho_{\rm l},}\] where
$\rho_{\rm v}$ and $\rho_{\rm l}$ are the vapor and liquid densities,
respectively, at coexistence.

It is tempting to try and use such coexistence solutions directly as
the reference density $\rho_{\rm s}$. However, the above example
demonstrates the associated challenges. Imagine a liquid in contact
with a planar hard wall. Close to coexistence, the average solvent
density profile will resemble that of a \tcb{free liquid--vapor} interface
with a well-defined separation from the wall. In contrast, in the
above example, any density profile corresponding to free translation
of the interface is equally plausible. We therefore seek a procedure
that allows us to ``pick'' the appropriate coexistence solution. In
fact, with the approximations that we will make, the coexistence
solutions will ultimately only enter implicitly in our approach. To
this end, we introduce a second expansion around an ``auxiliary''
reference density, $\rho_{\rm r}$, which is an equilibrium solution at
coexistence. Expanding the slowly-varying reference density
$\rho_{\rm s}$ around $\rho_{\rm r}$, we write
\begin{equation}
 \label{eqn:cs1-expand}
  c_{\rm s}^{(1)}(\mbf{r}) = c_{\rm r}^{(1)}(\mbf{r})
  + \int\!\!\mrm{d}\mbf{r}^\prime\,c_{\rm r}^{(2)}(\mbf{r},\mbf{r}^\prime)\delta_{\rm r}\rho_{\rm s}(\mbf{r}^\prime) + \ldots,
\end{equation}
where $\delta_{\rm r}\rho_{\rm s}(\mbf{r})=\rho_{\rm
s}(\mbf{r})-\rho_{\rm r}(\mbf{r})$ and again, the one- and two-body
direct correlation functions are defined in an analogous manner to
Eqs.~\ref{eqn:equilibrium_density} and~\ref{eqn:cu2}.  Substituting
Eq.~\ref{eqn:cs1-expand} into Eq.~\ref{eqn:Fex_intr-1} and keeping
both expansions to second order, we obtain
\begin{equation}
 \label{eqn:Fex_intr-2}
 \begin{split}
 \mcl{F}^{(\rm ex)}_{\rm intr}[\rho] = &\;\mcl{F}^{(\rm ex)}_{\rm intr}[\rho_{\rm s}]
 - k_{\rm B}T\int\!\!\mrm{d}\mbf{r}\,c_{\rm r}^{(1)}(\mbf{r})\delta_{\rm s}\rho(\mbf{r}) \\
 &-\frac{k_{\rm B}T}{2}\int\!\!\mrm{d}\mbf{r}\!\int\!\!\mrm{d}\mbf{r}^\prime\,\delta_{\rm s}\rho(\mbf{r})c_{\rm s}^{(2)}(\mbf{r},\mbf{r}^\prime)\delta_{\rm s}\rho(\mbf{r}^\prime) \\
 &-k_{\rm B}T\int\!\!\mrm{d}\mbf{r}\!\int\!\!\mrm{d}\mbf{r}^\prime\,\delta_{\rm s}\rho(\mbf{r})c_{\rm r}^{(2)}(\mbf{r},\mbf{r}^\prime)\delta_{\rm r}\rho_{\rm s}(\mbf{r}^\prime).
 \end{split}
\end{equation}
The final term acts to couple differences between $\rho(\mbf{r})$ and
$\rho_{\rm s}(\mbf{r})$ with differences between $\rho_{\rm
s}(\mbf{r})$ and $\rho_{\rm r}(\mbf{r})$. As before, we will add
$\mathcal{F}^{(\rm id)}_{\rm intr}$ to Eq.~\ref{eqn:Fex_intr-2}. We
will also substitute $c_{\mrm{r}}^{(1)}$ according to
Eq.~\ref{eqn:equilibrium_density}, with $\phi_{\rm r} = \delta\mu
= \mu-\mu_{\rm coex}$, where $\mu_{\rm coex}$ is the chemical
potential at coexistence. The grand potential then reads
\begin{equation}
\begin{split}
\varOmega_{\mathlarger{\phi}}[\rho]
= &\;\mcl{F}_{\mrm{intr}}[\rho_{\mrm{s}}] - \mu\!\int\!\!\mrm{d}\mbf{r}\rho_{\mrm{s}}(\mbf{r}) -  \delta\mu\int\!\!\mrm{d}\mbf{r}\,\delta_{\mrm{s}}\rho(\mbf{r}) \\
& + \!\int\!\!\mrm{d}\mbf{r}\,\phi(\mbf{r})\rho(\mbf{r})  +  \Delta_{\rm r}\mcl{F}_{\rm
intr}^{\rm (id)}[\rho] - \Delta_{\rm r}\mcl{F}_{\rm
intr}^{\rm (id)}[\rho_{\mrm{s}}]\\
& -\frac{k_{\mrm{B}}T}{2}\!\int\!\!\mrm{d}\mbf{r}\!\!\int\!\!\mrm{d}\mbf{r}'\,\delta_{\mrm{s}}\rho(\mbf{r})\,c^{(2)}_{\mrm{s}}(\mbf{r},\mbf{r}')\,\delta_{\mrm{s}}\rho(\mbf{r}') \\
& -k_{\mrm{B}}T\!\int\!\!\mrm{d}\mbf{r}\!\!\int\!\!\mrm{d}\mbf{r}'\,\delta_{\mrm{s}}\rho(\mbf{r})\,c^{(2)}_{\mrm{r}}(\mbf{r},\mbf{r}')\,\delta_{\mrm{r}}\rho_{\mrm{s}}(\mbf{r}'), \\
\end{split}
\label{eq:grand_potential_full}
\end{equation}
where the changes in the ideal contribution are now relative to the
auxiliary reference. 
The resulting equilibrium solvent density is
\begin{equation}
\begin{split}
\rho(\mbf{r}) = 
 &\rho_{\rm r}(\mbf{r}) 
 \exp\bigg[\beta\delta\mu  +\!\!\int\!\!\mrm{d}\mbf{r}'\,c^{(2)}_{\mrm{r}}(\mbf{r},\mbf{r}')\,\delta_{\mrm{r}}\rho_{\mrm{s}}(\mbf{r}')\bigg]\\
 &   \times \exp\bigg[-\beta\phi(\mbf{r}) +
\!\!\int\!\mrm{d}\mbf{r}^\prime\,c^{(2)}_{\rm s}(\mbf{r},\mbf{r}^\prime)\,\delta_{\rm s}\rho(\mbf{r}^\prime)\bigg].
 \end{split}
\label{eqn:exact-scf}
\end{equation}
At this point, we stress the conceptual difference compared to simply
expanding around $\rho_{\rm s}$; provided that we have some procedure
for specifying $\rho_{\rm s}$ to suit our needs, we do not need to
know $\phi_{\rm s}$. Instead, we have transferred the problem to
specifying the appropriate boundary conditions for $\rho_{\rm
r}$.

In the context of solvation, how to choose boundary conditions on
$\rho_{\rm r}$ is not immediately obvious. Moreover, we appear to have
complicated matters, as we now need to deal with three density fields
($\rho$, $\rho_{\rm s}$ and $\rho_{\rm r}$). \tcb{We will now outline
a series of approximations that enable us to consider a theory
expressed explicitly in terms of $\rho$ and $\rho_{\rm s}$ only, as
well as a procedure to ``pick'' an appropriate slowly-varying
reference density.}

\tcb{The first simplifying approximation that we make,} which is reasonable
near coexistence, is
\begin{equation}
   \rho_{\rm s}(\mbf{r}) \approx \rho_{\rm r}(\mbf{r})\exp\bigg[\beta\delta\mu  +\!\!\int\!\!\mrm{d}\mbf{r}'\,c^{(2)}_{\mrm{r}}(\mbf{r},\mbf{r}')\,\delta_{\mrm{r}}\rho_{\mrm{s}}(\mbf{r}')\bigg],
\end{equation}
such that Eq.~\ref{eqn:exact-scf} can be written approximately as
\begin{equation}
\label{eqn:scf-rapid}
\rho(\mbf{r}) \approx
 \rho_{\rm s}(\mbf{r})
 \exp\left[-\beta\phi(\mbf{r}) +
\int\!\!\mrm{d}\mbf{r}^\prime\,c^{(2)}_{\rm s}(\mbf{r},\mbf{r}^\prime)\,\delta_{\rm s}\rho(\mbf{r}^\prime)\right].
\end{equation}
\tcb{To construct a procedure for specifying $\rho_{\rm s}$ that keeps it
sufficiently close to $\rho$ while maintaining a slowly-varying
nature, we introduce the following ``pseudofunctional,''}
%
\begin{equation}
\tilde{\varOmega}_{\mathlarger{\psi}}[\rho_{\mrm{s}}] 
= \mcl{F}_{\mrm{intr}}[\rho_{\mrm{s}}] - \mu\!\int\!\!\mrm{d}\mbf{r}\rho_{\mrm{s}}(\mbf{r}) + \!\int\!\!\mrm{d}\mbf{r}\,\psi(\mbf{r};[\rho,\rho_{\rm s}])\,\rho_{\mrm{s}}(\mbf{r}), 
\end{equation}
\tcb{We use the term ``pseudofunctional'' because $\psi$ is not an external
potential; it} depends parametrically on both $\rho(\mbf{r})$ and
$\rho_{\rm s}(\mbf{r})$, and penalizes differences between $\rho$ and
$\rho_{\rm s}$, while ensuring that $\rho_{\rm s}$ remains
slowly-varying. (We will specify a form for $\psi$ below.) \tcb{In
this sense, it is not a functional in the usual cDFT sense, and we
have introduced it simply as a computational tool to obtain an
appropriate slowly-varying reference.}

\tcb{As $\rho_{\rm s}$ is slowly-varying}, $\mcl{F}_{\mrm{intr}}\tcb{[\rho_{\rm s}]}$ can be
reasonably approximated by the square gradient form
\begin{equation}
\label{eqn:OmegaS-vdW}
\begin{split}
\mcl{F}_{\mrm{intr}}[\rho_{\mrm{s}}] - \mu\!\int\!\!\mrm{d}\mbf{r}\rho_{\mrm{s}}(\mbf{r})
 &\approx \int\!\!\mrm{d}\mbf{r}\,\left[\omega(\rho_{\rm s})
 + \frac{m}{2}|\nabla\rho_{\rm s}(\mbf{r})|^2\right], \\
\end{split}
\end{equation}
where $\omega(\rho_{\rm s})$ is the local grand potential
density. Assuming that $m$ is independent of $\rho_{\rm s}$,
minimizing $\tilde{\varOmega}_{\mathlarger{\psi}}$ with respect to
$\rho_{\rm s}$ gives
\begin{equation}
\label{eqn:scf-slow-exact}
 \omega^\prime(\rho_{\rm s})
 - m\nabla^2\rho_{\rm s}(\mbf{r}) + \psi(\mbf{r}) = 0,
\end{equation}
where $\omega^\prime$ indicates partial differentiation with respect
to $\rho_{\rm s}$.

Equations~\ref{eqn:scf-rapid} and~\ref{eqn:scf-slow-exact} provide a
self-consistent set of equations for $\rho$ and $\rho_{\rm s}$ that do
not feature $\rho_{\rm r}$ explicitly. To calculate the grand
potential given by Eq.~\ref{eq:grand_potential_full}, however, still
requires $\rho_{\rm r}$, owing to the ideal terms and the final double
integral. Considering the latter, as $\rho = \rho_{\rm s}$ far from
the solute, nonzero contributions to the integral are localized to
regions close to the solute. Moreover, close to coexistence, we
anticipate that $\delta_{\rm r}\rho_{\rm s}$ is small. In practice,
then, we simply ignore the final term in
Eq.~\ref{eq:grand_potential_full} when computing the free energy. By
similar reasoning, we also approximate $\Delta_{\rm r}\mcl{F}_{\rm
intr}^{\rm (id)}[\rho] - \Delta_{\rm r}\mcl{F}_{\rm intr}^{\rm
(id)}[\rho_{\mrm{s}}] \approx \Delta_{\rm s}\mcl{F}_{\rm intr}^{\rm
(id)}[\rho]$. When calculating the grand potential, we therefore adopt
the simpler approximate form
\begin{equation}
\begin{split}
\label{eqn:OmegaInPractice}
\varOmega_{\mathlarger{\phi}}[\rho]
\approx &\;\int\!\!\mrm{d}\mbf{r}\,\left[\omega(\rho_{\rm s})
 + \frac{m}{2}|\nabla\rho_{\rm s}(\mbf{r})|^2\right] -  \delta\mu\int\!\!\mrm{d}\mbf{r}\,\delta_{\mrm{s}}\rho(\mbf{r}) \\
& + \!\int\!\!\mrm{d}\mbf{r}\,\phi(\mbf{r})\rho(\mbf{r})  + \Delta_{\rm s}\mcl{F}_{\rm
intr}^{\rm (id)}[\rho] \\
& -\frac{k_{\mrm{B}}T}{2}\!\int\!\!\mrm{d}\mbf{r}\!\!\int\!\!\mrm{d}\mbf{r}'\,\delta_{\mrm{s}}\rho(\mbf{r})\,c^{(2)}_{\mrm{s}}(\mbf{r},\mbf{r}')\,\delta_{\mrm{s}}\rho(\mbf{r}').\\
\end{split}
\end{equation}
 
Equations~\ref{eqn:scf-rapid}, \ref{eqn:scf-slow-exact}
and~\ref{eqn:OmegaInPractice} provide the formal equations of our
approach for the equilibrium density and grand potential. However, for
practical calculations, a form for the potential
$\psi(\mbf{r};[\rho,\rho_{\rm s}])$ needs to be specified. \tcb{In
this work, we will not explore ways to systematically optimize
$\psi$.} \tcb{While, in principle,} the fact that $\psi$ does not enter the grand
potential explicitly offers \tcb{some flexibility in
specifying its form, it is not completely arbitrary;} as mentioned
previously, it ought to penalize differences \tcb{between}
$\rho(\mbf{r})$ and \tcb{$\rho_{\rm s}(\mbf{r})$} while ensuring that
$\rho_{\rm s}(\mbf{r})$ remains slowly-varying. \tcb{(Also see below,
where we discuss our approach in the context of LCW theory.)} With
these considerations in mind, our choice for $\psi$ is guided by the
final term in Eq.~\ref{eq:grand_potential_full}. Specifically, we
assume that $c_{\rm r}^{(2)}$ can be separated as,
\begin{equation}
\label{eqn:DCF-splitting}
  c_{\rm r}^{(2)}(\mbf{r},\mbf{r}^\prime) = c_{{\rm r},0}^{(2)}(\mbf{r},\mbf{r}^\prime) + c_{{\rm r},1}^{(2)}(\mbf{r},\mbf{r}^\prime),
\end{equation}
with $c_{{\rm r},0}^{(2)}(\mbf{r},\mbf{r}^\prime) \simeq 0$ for
$|\mbf{r}-\mbf{r}^\prime| > \ell$, where $\ell$ is a molecular length
scale, accounting for rapid variations. Conversely, $c^{(1)}_{{\rm
r},1}$ is slowly-varying over length scales comparable to $\ell$. In
keeping with the requirements for $\psi$, we therefore prescribe
\begin{equation}
\label{eqn:psi-c1}
\psi(\mbf{r}) =
  -k_{\rm B}T\int\!\!\mrm{d}\mbf{r}^\prime\,c_{{\rm r},1}^{(2)}(\mbf{r},\mbf{r}^\prime)[\rho(\mbf{r}^\prime)-\rho_{\rm s}(\mbf{r}^\prime)].
\end{equation}

\subsection{Connecting to LCW theory for a practical cDFT scheme}

With the form for $\psi$ given by Eq.~\ref{eqn:psi-c1}, together with
Eqs.~\ref{eqn:scf-rapid} and~\ref{eqn:scf-slow-exact}, the approximate
cDFT that we have derived bears a striking resemblance to LCW
theory. Indeed, $\psi$ plays the role of the ``unbalancing
potential'' in Ref.~\onlinecite{Lum1999}. The formal similarity can
be made more apparent if we introduce the coarse-graining procedure,
\begin{equation}
 \label{eqn:CG}
  \overline{\rho}(\mbf{r}) =
  \frac{\int\!\!\mrm{d}\mbf{r}^\prime\,c^{(2)}_{{\rm r},1}(\mbf{r},\mbf{r}^\prime)\rho(\mbf{r}^\prime)}
  {\int\!\!\mrm{d}\mbf{r}^\prime\,c^{(2)}_{{\rm r},1}(\mbf{r},\mbf{r}^\prime)}.
\end{equation}
Equation~\ref{eqn:scf-slow-exact} then reads
\begin{equation}
\label{eqn:scf-slow}
  \omega^\prime(\rho_{\rm s}) =
  m\nabla^2\rho_{\rm s}(\mbf{r}) +  2a(\mbf{r})[\overline{\rho}(\mbf{r})-\overline{\rho}_{\rm s}(\mbf{r})],
\end{equation}
with
\begin{equation}
 \label{eqn:energy-volume}
 2a(\mbf{r}) = k_{\rm B}T\int\!\!\mrm{d}\mbf{r}^\prime\, c^{(2)}_{{\rm r},1}(\mbf{r},\mbf{r}^\prime).
\end{equation}
\tcb{Note that, if we were interested in a Lennard--Jones fluid, this
relationship between the coarse-grained density and the correlation function is 
identical to that specified by Weeks in the context of
local molecular field theory \cite{Weeks2002}, in which $c^{(2)}_{{\rm
r}, 1}$ is treated in a random phase approximation. (Although also
note that Eqs.~\ref{eqn:scf-rapid}, \ref{eqn:scf-slow-exact}
and~\ref{eqn:OmegaInPractice} do not constitute a random phase
approximation.) In this case, there is a notion of ``unbalanced
attractive interactions,'' and in the original LCW
theory \cite{Lum1999}, it was argued that such unbalanced forces in
water could be described by a similar coarse graining, over an
appropriate length scale that describes the range of attractive
interactions.}

\tcb{Our cDFT-based approach does not emphasise the role of unbalanced
attractive forces.} In Eq.~\ref{eqn:CG}, we see that the
slowly-varying component of the two-body direct correlation function
provides a natural means to coarse-grain the density fields. However,
it is cumbersome to evaluate. Expressing $\psi$ in terms of the
coarse-grained density fields as in Eq.~\ref{eqn:scf-slow} therefore
also serves a practical purpose, as we can estimate its value with an
approximate coarse-graining, e.g.,
\begin{equation}
 \label{eqn:CG-g}
  \overline{\rho}(\mbf{r}) \approx
  \tcb{\frac{1}{(2\pi\lambda^2)^{3/2}}}\int\!\!\mrm{d}\mbf{r}^\prime\,\tcb{\exp\left(-\frac{|\mbf{r}-\mbf{r}^\prime|^2}{2\lambda^2}\right)}\rho(\mbf{r}^\prime).
\end{equation}
Along with the coarse-graining length $\lambda$, we now treat $a$ as a
parameter that needs to be determined. As detailed in the SM, we
estimate their values by using a hard sphere reference fluid for which
$c_{{\rm r},0}^{(2)}$ is known. For liquid water at 300\,K and
$\beta\delta\mu \approx 10^{-3} $ (corresponding to $\rho_{\rm
u} \approx 33.234\,\mrm{nm}^{-3}$ for SPC/E water), we obtain an
acceptable range of the coarse-graining parameters $a \approx
200$--$300$\,kJ\,cm$^3$\,mol$^{-2}$ and $\lambda \approx
0.08$--$0.11$\,nm.  \tcb{Values in} this range for $\lambda$ \tcb{are
much smaller than that used in the original LCW theory
($\lambda\approx 0.38$\,nm), although they are} comparable to the
coarse-graining length derived in Ref.~\onlinecite{Vaikuntanathan2014}
for a lattice-based version of LCW theory that emphasizes the
importance of capillary wave fluctuations.

To complete our theory for practical purposes, we adopt
\begin{equation}
    \omega(\rho_{\mrm{s}}) = \omega_{\mrm{coex}}(\rho_{\mrm{s}}) - \rho_{\mrm{s}} \delta\mu,
\end{equation}
where $\omega_{\mrm{coex}}$ takes a
quartic form,
\begin{align}
  \omega_{\rm coex}&(\rho_{\rm s}) =
  \frac{C}{2}\left(\rho_{\mrm{s}}- \rho_\mrm{l}\right)^2 \left(\rho_{\mrm{s}}- \rho_\mrm{v}\right)^2 + \nonumber \\[5pt]
  &\frac{D}{4}\left(\rho_{\mrm{s}}- \rho_\mrm{l}\right)^4 \left(\rho_{\mrm{s}}- \rho_\mrm{v}\right)^4 
  h(\rho_{\rm s}-\rho_{\rm v})h(\rho_{\rm l}-\rho_{\rm s}),
\end{align}
where $h(\rho_{\rm s})$ is the Heaviside step function. The curvature
at the minima is determined by $C$, whose value we set to be
consistent with the compressibility of the bulk fluid $\chi_{\rm
u}$. Together, $D$ and $m$ determine $\gamma$ and the shape of the
free liquid--vapor interface. Details of our parameterization procedure
are provided in the SM. In most of what follows we will treat $m$ as
a constant independent of density; for spherical solutes, this yields
the correct limiting behavior for both $R\to 0$ and $R\to\infty$. For
$R$ in the crossover regime, we will show that the extra flexibility
afforded by allowing $m$ to vary in a systematic, yet practical, way
with $\rho_{\rm s}$ yields quantitative agreement for intermediate
solute sizes.

All that is left to establish is the two-body correlation function
$c^{(2)}_{\rm s}(\mbf{r},\mbf{r}^\prime)$. We adopt the following
simple form
\begin{equation}
 c^{(2)}_\mrm{s}(\mbf{r},\mbf{r}^\prime) \approx c^{(2)}_\mrm{u}(|\mbf{r}-\mbf{r}^\prime|)\rho_{\mrm{s}}(\mbf{r}) \rho_{\mrm{s}}(\mbf{r}^\prime) \rho_{\mrm{u}}^{-2},
\end{equation}
which is exact in the limits of homogeneity and low density, and
interpolates smoothly between these two regimes. While other
reasonable approximations could be made for
$c^{(2)}_\mrm{s}(\mbf{r},\mbf{r}^\prime)$, ours is in the same spirit
as the one adopted by LCW theory for the susceptibility. This
highlights one of the key differences with
Ref.~\onlinecite{Jeanmairet2013jcp}, which also attempted to recast
ideas from LCW theory in a cDFT framework, but with correlations of a
uniform fluid. In addition, this previous work also used the
coarse-grained density $\bar{\rho}$ directly in a square gradient
functional, rather than isolating the slowly-varying component
$\rho_{\rm s}$ as we have here.

In this section, we have derived a cDFT for solvation that aims to
describe both the short-wavelength perturbations induced by effects of
excluded volume, and the physics of liquid--vapor coexistence relevant
for larger solutes. We have also described approximations for the
coarse-grained density and inhomogeneous two-body direct correlation
function that facilitate numerical evaluation. While our theory is
appropriate for arbitrarily complex forms of apolar solute--solvent
interaction, for demonstrative purposes, we will focus on spherical
solutes and planar walls. These simple geometries, however,
are sufficient to make connections with existing theories and 
highlight the advantages of our approach.

\section*{Solvation of spherical solutes, both hard and soft}

\begin{figure}[t]
  \centering  \includegraphics[width=\linewidth]{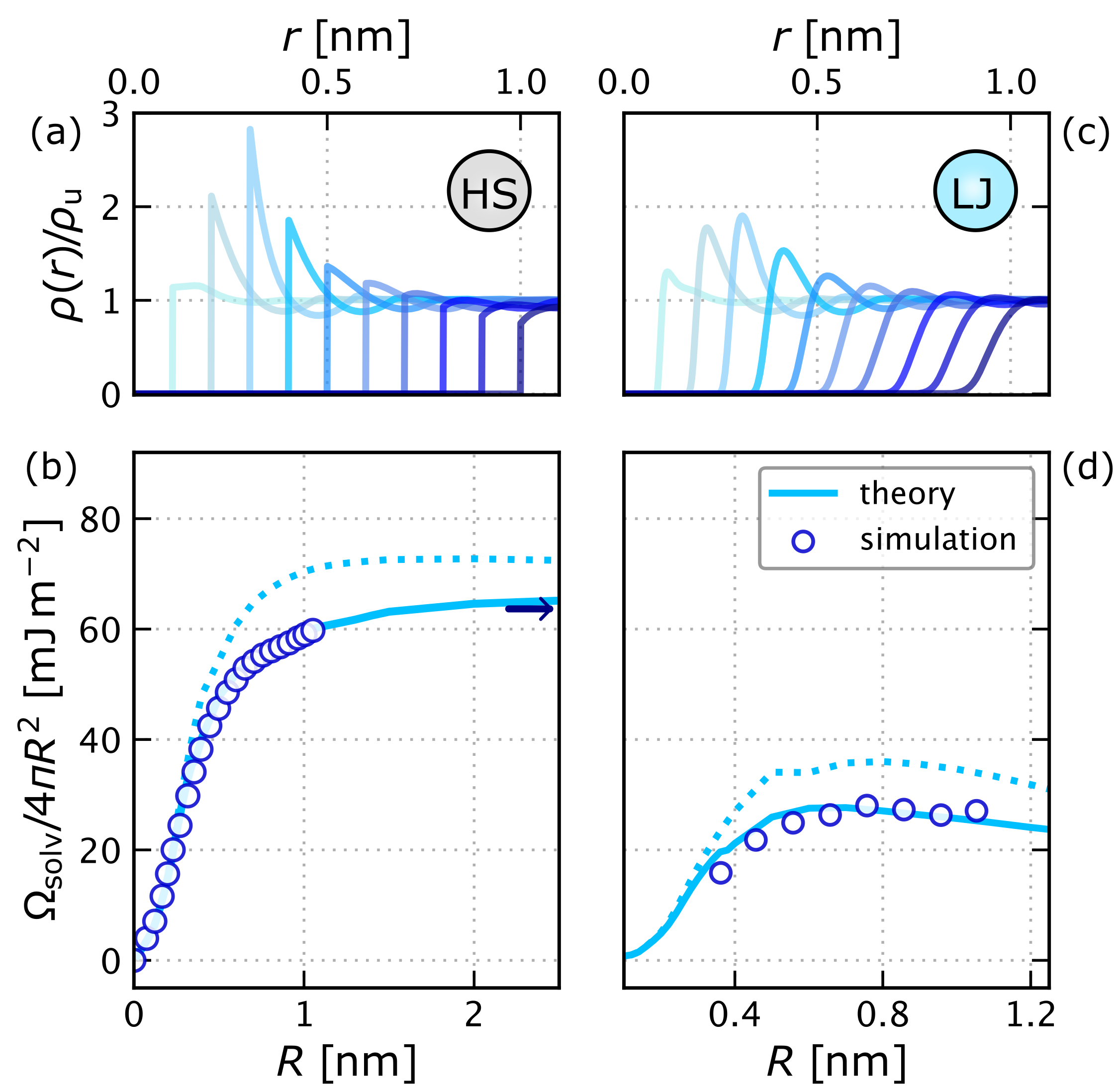}

  \caption{\textbf{Solvation of spherical apolar solutes with
  cDFT.} By solving Eqs.~\ref{eqn:scf-rapid} and~\ref{eqn:scf-slow}
  self-consistently, we have obtained $\rho(r)$ and $\varOmega_{\rm
  solv}$, shown in (a) and (b) respectively, for hard-sphere solutes
  of different $R$ centered at the origin. When using
  $m(\rho_{\rm s}^\ast)$ [(b), solid line], our theory is in
  quantitative agreement with results from molecular
  simulations \cite{Huang2001}. If we take $m$ to be independent of
  density [(b), dotted line] we find qualitative agreement, but
  $\varOmega_{\rm solv}$ is overestimated for $R\gtrsim 0.5$\,nm. In
  both cases, $\varOmega_{\rm solv}/4\pi R^2 \sim \gamma =
  63.6\,\mrm{mJ\,m^{-2}}$ (indicated by the blue
  arrow \cite{Vega2007}), as $R\to\infty$. In (c) and (d) we show
  corresponding results for a LJ solute, and again we observe good
  agreement with molecular simulations \cite{Fujita2017}. Note that we
  have used the same parameterization as (a) and (b). All results
  from our theory have been obtained for SPC/E water  
  ($T=300$\,K, $\beta\delta\mu \approx 10^{-3}$,
  $a=200$\,kJ\,cm$^3$\,mol$^{-2}$ and $\lambda=0.08$\,nm).}
\label{fig:SPCE-spherical-solute}
\end{figure}

We first consider the paradigmatic test case for any theory of
hydrophobicity: the solvation of a hard-sphere solute in water under
ambient conditions ($T=300\,\mrm{K}$, $\beta\delta\mu \approx
10^{-3}$). To this end, in Figs.~\ref{fig:SPCE-spherical-solute}(a)
and~\ref{fig:SPCE-spherical-solute}(b) we present $\rho(r)$ and
$\varOmega_{\rm solv}$ obtained from self-consistent \tcb{solutions} of
Eqs.~\ref{eqn:scf-rapid} and~\ref{eqn:scf-slow}, parameterized for a
simple point charge model of water (SPC/E \cite{Berendsen1987}), as we
increase the solute radius $R$. First treating $m$ as a constant
(dotted line in Fig.~\ref{fig:SPCE-spherical-solute}), we see that our
cDFT approach broadly captures the behavior observed in molecular
simulations \cite{Huang2001}, with $\varOmega_{\rm solv}$ increasing
with volume for small solutes, before crossing over to $\varOmega_{\rm
solv}/4\pi R^2 \sim \gamma$ for $R\gtrsim 1$\,nm. Like previous
treatments rooted in LCW theory, however, we also see that
these results overestimate $\varOmega_{\rm solv}$ in the crossover
regime.

Empirically, agreement with simulation results in the crossover regime
could be attained by reducing the value of $m$. In the context of our
theory, this implies that the optimal reference density for finite $R$
has a sharper interface than the free liquid-vapor
interface. Simply reducing $m$ in such a manner, however, would result
in an incorrect limiting behavior as $R\to\infty$. But this
observation motivates us to vary $m$, in a linear fashion, with a
characteristic feature of the slowly-varying density $\rho^{\ast}_{\rm
s}$. Specifically, $\rho^{\ast}_{\rm s}$ is the value where $\rho_{\rm
s} = \rho$ for the first time as $r$ increases. Further details
are provided in the SM. As we see in
Fig.~\ref{fig:SPCE-spherical-solute}(b) (solid line), with
$m(\rho^{\ast}_{\rm s})$ our cDFT describes the simulation data
quantitatively.


For hard spheres in water, the main advantage of our approach compared
to LCW theory is primarily conceptual: $\varOmega_{\rm solv}$ follows
directly from the minimization of $\varOmega_\mathlarger{\phi}$, as opposed to
relying on, e.g., thermodynamic integration \cite{Katsov2001b,
Huang2002, Remsing2016}. From a practical viewpoint, the cDFT approach
is also numerically simpler to implement, as discussed in detail in
Ref.~\onlinecite{Sergiievskyi2017}. But its main advantages compared
to LCW theory become apparent once we depart from the solvation of
ideal hydrophobes. In LCW theory, it is assumed that the
solute--solvent interaction can be separated into repulsive and
attractive contributions,
$\phi=\phi_{\mrm{rep}}+\phi_{\mrm{att}}$. While $\phi_{\mrm{att}}$ can
be accounted for straightforwardly in the slowly-varying part,
$\phi_{\mrm{rep}}$ must be approximated by a hard-core potential, and
the solvent density is solved subject to the constraint $\rho(r)=0$
inside the solute \cite{Chandler1993}. For cDFT approaches such as
ours, such an approximation is not needed.  Instead, we simply
minimize $\varOmega_{\mathlarger{\phi}}$ with the appropriate solute--solvent
interaction $\phi$. To illustrate this, in
Figs.~\ref{fig:SPCE-spherical-solute}(c)
and~\ref{fig:SPCE-spherical-solute}(d), we show results from our
theory for the solvation of Lennard--Jones solutes of effective radius
$R$ (see SM) with a constant well-depth of
$\epsilon=0.5\,\mrm{kJ\,mol^{-1}}$.  Using the same
parameterization for $m(\rho^\ast_{\rm s})$ as for hard-spheres, our
results for $\varOmega_{\rm solv}$ are in good agreement with
available simulation data \cite{Fujita2017}.

\section*{The physics of critical drying in an LCW-style theory}

The motivations for our work primarily stem from the desire to develop
``semi-implicit'' solvation models that retain information on
essential solvent correlations. To that end, the results we have
presented so far are promising. However, the physics of hydrophobicity
is important in its own right. While LCW theory should be considered a
seminal contribution in this area, subsequent work from Evans, Wilding
and co-workers \cite{Evans2015,Evans2015prl, Evans2016, Coe2022, Coe2022b, Coe2023} emphasizes the
central role of the critical drying transition that occurs in the
limit of a planar substrate with vanishing, or very weak, attractive
interactions with the solvent. The extent to which critical drying is
captured by LCW theory (or its subsequent lattice-based
derivatives \cite{tenWolde2002,Varilly2011,Vaikuntanathan2014}) is,
however, unclear. Although not equivalent to LCW theory, we can use
our cDFT approach to shed light on the extent to which it contains the
physics of critical drying.

A key quantity in critical drying phenomena is the local
compressibility,
\begin{equation}
  \label{eqn:ChiLoc}
  \chi(\mbf{r}) = \bigg(\frac{\partial\rho(\mbf{r})}{\partial\mu}\bigg)_T,
\end{equation}
which in practice is obtained from a finite-difference approximation
(see SM). Evans, Wilding and co-workers argue that the structure of
$\chi(\mbf{r})$ provides one of the most robust indicators of
hydrophobicity or, more generally, ``solvophobicity.''

\begin{figure}[t]
  \includegraphics[width=\linewidth]{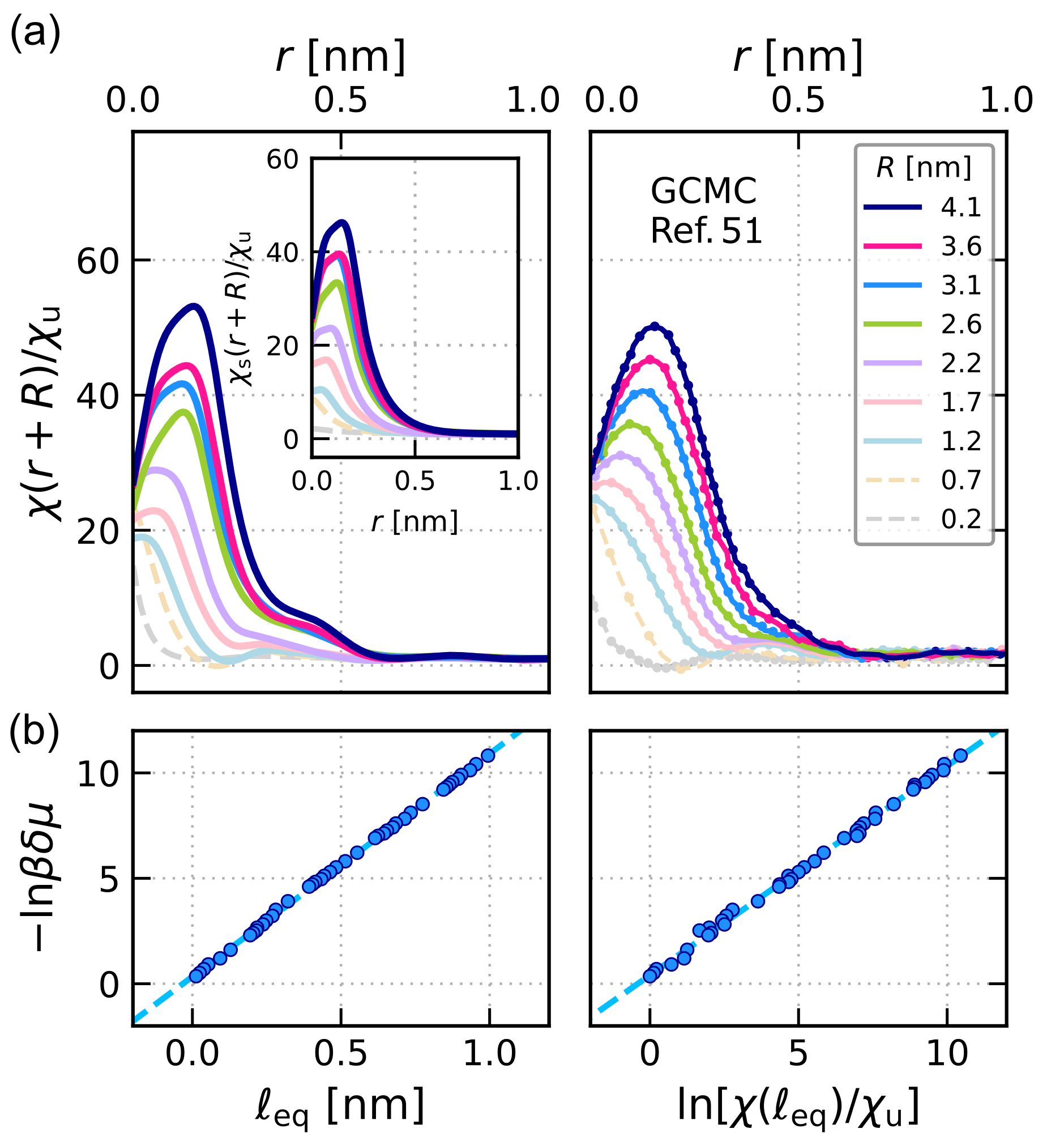}
  \caption{\textbf{The local compressibility described by an
  LCW-style cDFT.} In (a), we show $\chi(r)$ around different sized
  hard-sphere solutes, as indicated by the values of $R$ in the
  legend, centered at the origin. The results from our theory (left)
  are in good agreement with GCMC simulations from
  Ref.~\onlinecite{Coe2023} (right). The inset shows $\chi_{\rm
  s}(r)$, the contribution from the slowly-varying density, which
  becomes increasingly important as $R$ increases, though
  contributions from the rapidly varying part are still
  significant. All results are for the mW water model ($T=426$\,K,
  $\beta\delta\mu \approx 10^{-3}$, 
  $a=300$\,kJ\,cm$^3$\,mol$^{-2}$ and $\lambda=0.11$\,nm).
  In (b) we investigate the
  behavior of $\chi$ near a planar hard wall as $\delta\mu\to
  0$. Consistent with a binding potential analysis \cite{Evans2017},
  our theory yields $\ln \chi(\ell_{\rm eq}) \sim -\ln\beta\delta\mu$
  (right), where $\ell_{\rm eq}$ is the distance of the maximum in
  $\chi$ from the wall. We also observe $\ell_{\rm eq} \sim
  -\ln\beta\delta\mu$ (left). Circles show results from the theory and
  dashed lines indicate the expected scaling relation.}
  \label{fig:chi}
\end{figure}

In the specific case of solvation of hard spheres, $\chi(r)$ should
exhibit a pronounced maximum that increases in both magnitude and
position with increasing $R$. In Fig.~\ref{fig:chi}(a) we present
$\chi(r)$ from our theory, parameterized for the coarse-grained mW
water model \cite{Molinero2009} at $T = 426$\,K.\footnote{$T=426$\,K
corresponds to $T/T_{\rm c} \approx 0.46$,
with the critical point of mW at $T_{\rm c} = 917.6$\,K
(from Ref.~\onlinecite{Coe2022b}). } 
We see good agreement between the results from our theory
compared to grand canonical Monte Carlo (GCMC) simulations by
Coe \etal \cite{Coe2023} In particular, for the largest solute
investigated, $R\approx 4.1$\,nm, we see that the maximum value of
$\chi(r)$ is over 40 times larger than its bulk counterpart $\chi_{\rm
u}$. In the SM, we also present results where the strength of
attractive solute-solvent interactions is decreased, which also
compare favorably to GCMC simulations. In the context of an LCW-style
theory such as ours, we can also isolate the slowly-varying component
of the local compressibility $\chi_{\rm s}(\mbf{r})$, by replacing
$\rho(\mbf{r})$ with $\rho_{\rm s}(\mbf{r})$ in
Eq.~\ref{eqn:ChiLoc}. As one might expect, as $R$ increases, so does
the significance of $\chi_{\rm s}$. It is clear, however, that
contributions from the rapidly-varying density are still important,
even for solute sizes that far exceed $R = 1$\,nm, the colloquial
hydrophobic crossover point, as seen in the inset. 


A more stringent test of critical drying comes not from comparison to
molecular simulations, but from known scaling behaviors of $\chi$  from binding potential analyses \cite{Evans2017}. Specifically,
for a fluid in contact with a planar hard wall, it can be shown that,
close to coexistence, $\chi(\ell_{\rm eq})\sim \delta\mu^{-1}$, where
$\ell_{\rm eq}$ is the position of the maximum in $\chi$. Moreover,
$\ell_{\rm eq}\sim -\ln\tcb{\beta}\delta\mu$. As seen in Fig.~\ref{fig:chi}(b),
the LCW-style cDFT approach that we have derived obeys both of these
scaling relations. This is far from a trivial result. At face value,
the physics of critical drying and the emphasis placed on
liquid--vapor interface formation by LCW theory seem unrelated. By
recasting the essential underpinnings of LCW theory in the context of
cDFT, we begin to paint a unifying picture of these two different
views on hydrophobicity.

\section*{Multiscale solvation from first principles}

The physics of critical drying described above is an essential
component of hydrophobicity at large length scales. Moreover, it is
common to both simple and complex fluids that exhibit
solvophobicity. A distinguishing feature of solvophobicity in complex
fluids such as water, however, is the ``entropic crossover'': for
small solutes, $\varOmega_{\rm solv}$ increases with increasing $T$,
while for larger solutes it decreases. This behavior has implications
for the thermodynamics of protein folding \cite{Huang2000b}, and has
been attributed to a competition of microscopic length scales (i.e.,
solvent reorganization in the first and second solvation shells) that
is absent in simple fluids \cite{Dowdle2013}. Here, we will
demonstrate that our cDFT approach captures this entropic
crossover. Moreover, we will do it from first principles.

Factors of $k_{\rm B}T$ aside, temperature dependence in
Eqs.~\ref{eqn:scf-rapid} and~\ref{eqn:scf-slow} enters implicitly
through $c_{\rm u}^{(2)}$ and $\gamma$. As the approach that we have
developed does not assume a simple pairwise additive form for the
interatomic potential, the combination of our theory with recent
advances in machine-learned interatomic potentials (MLIPs) means we
face the exciting prospect of describing, from first principles, the
temperature dependence of solvation across the micro-, meso- and
macroscales.

\begin{figure}[t]
  \includegraphics[width=\linewidth]{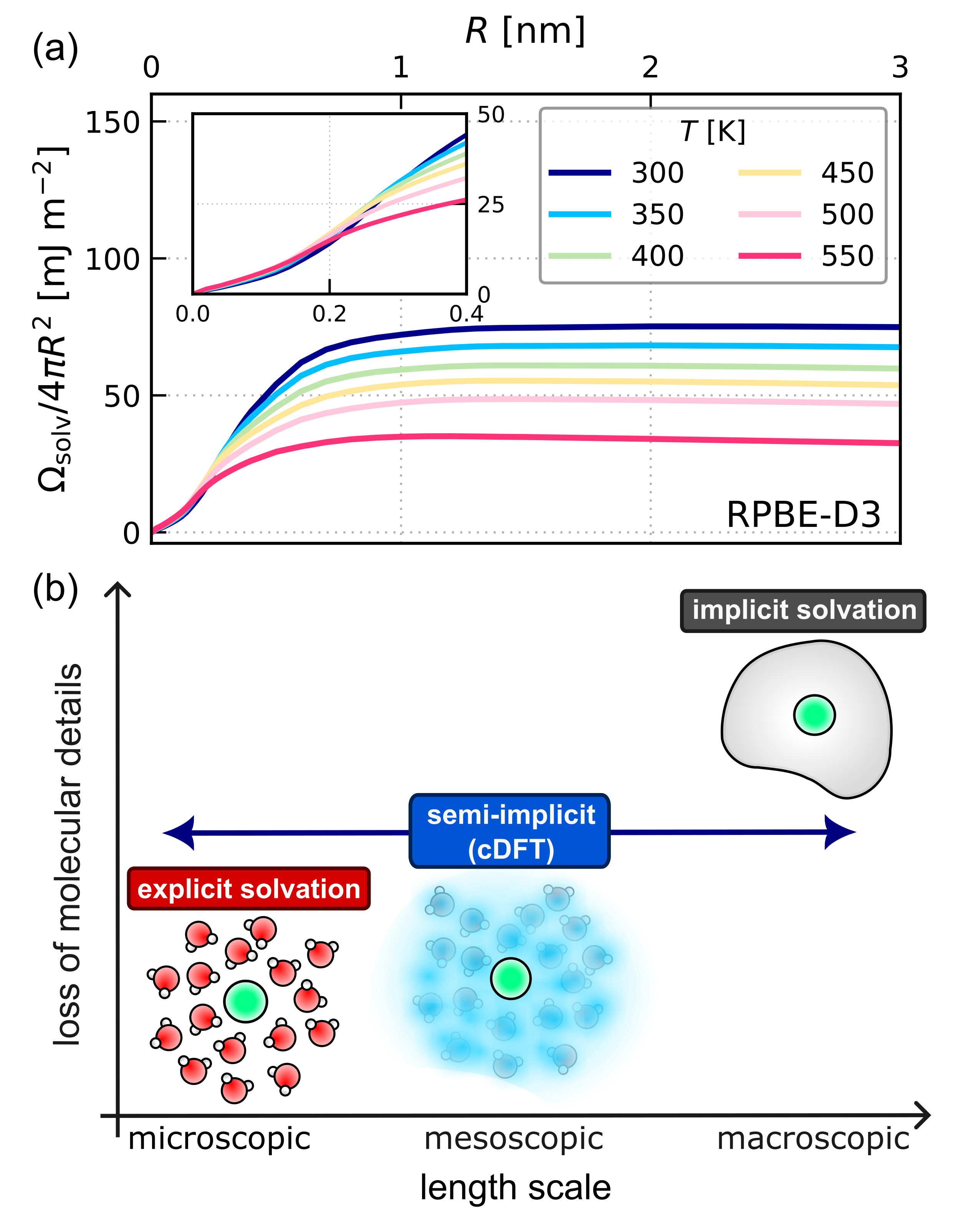}
  \caption{\textbf{Temperature-dependence of multiscale solvation
  from first principles.} Parameterizing our theory on a first
  principles representation of water (RPBE-D3), we can predict
  solvation behavior on length scales inaccessible to molecular
  simulations. This is demonstrated in (a), where we present
  $\varOmega_{\rm solv}/4\pi R^2$ for hard sphere solutes at
  different temperatures, as indicated in the legend.  The inset
  highlights the ``entropic crossover'' that is present in water but
  not simple liquids \cite{Huang2000b, Ashbaugh2009, Dowdle2013}. 
  All results are obtained for the liquid at coexistence 
  with $a=300$\,kJ\,cm$^3$\,mol$^{-2}$ and $\lambda=0.08$\,nm.
  In (b), we
  summarize the cDFT approach to multiscale modeling. Compared to
  explicit and implicit solvation approaches, the theory we present
  instead sacrifices some molecular details, but treats the remaining
  essential correlations consistently across all length
  scales. } \label{fig:abinitio}
\end{figure}

For illustrative purposes, we model water with RPBE-D3, a generalized
gradient approximation (electronic) functional \cite{Hammer1999} with
dispersion corrections \cite{Grimme2010}.  To obtain $c^{(2)}_{\rm
u}$, we have performed our own simulations of bulk water at the liquid
density along the coexistence curve, using the MLIP described in
Ref.~\onlinecite{Wohlfahrt2020}, which also provides $\gamma(T)$. In
Fig.~\ref{fig:abinitio}(a), we present results from self-consistent
solutions of Eqs.~\ref{eqn:scf-rapid} and~\ref{eqn:scf-slow} for
temperatures ranging from 300\,K to 550\,K. Note that, in the absence
of information in the crossover regime, we have simply taken $m$ to be
independent of density. For all temperatures, we observe the
hydrophobic crossover. Importantly, this occurs at progressively
smaller values of $R$ as $T$ increases; we observe the entropic
crossover. These results demonstrate that the LCW-style cDFT that we
have developed can be used to faithfully coarse-grain solvent effects
while maintaining essential molecular correlations to model
phenomenology at a first principles level across a broad range of
length scales.  This approach is summarized schematically in
Fig.~\ref{fig:abinitio}(b).

\section*{Conclusions and Outlook}

We have presented a cDFT for the solvation of apolar solutes in water,
which is accurate at both small and large length scales. Similar to
mDFT, we encode information about the liquid's small length scale
fluctuations by parameterizing our functional on the two-body direct
correlation function obtained from simulations of the bulk fluid. In
contrast to previous approaches, however, the grand potential that we
construct is not based on an expansion around the uniform bulk
fluid. Rather, from the outset, our theory acknowledges that the
perturbations induced in the solvent density field by large solutes
are too severe for the uniform fluid to act as a suitable
reference. In this work, we therefore establish a self-consistent cDFT
framework that permits the use of an inhomogeneous and slowly-varying
density field as a reference system.

The theory that we have outlined is similar in spirit to, and indeed
motivated by, the seminal work on the hydrophobic effect by Lum,
Chandler and Weeks \cite{Lum1999}. By placing the ideas of LCW theory
in the context of cDFT, not only do we gain a numerical advantage, but
we also provide conceptual insights. For example, the grand
potential that we derive (Eq.~\ref{eq:grand_potential_full}) suggests
a natural form for the ``unbalancing potential'' that specifies the
coarse-graining; the coarse-graining function is the slowly-varying
part of a two-body direct correlation function of an inhomogeneous
density field. This insight, as we explore in the SM, justifies a
coarse-graining length scale much smaller than the molecular diameter
of a water molecule that one might naively expect.

Our approach also allows us to connect the ideas of LCW theory
directly with more recent theoretical descriptions of hydrophobicity
from Evans, Wilding and co-workers \cite{Evans2016,
Coe2022}. Specifically, the local compressibility in the presence
of large hydrophobes obtained with our approach compares favorably to
results obtained by GCMC simulations \cite{Coe2023}, and we show that
its variation with chemical potential obeys known critical scaling
behaviors. We also demonstrate that our approach, similar to previous
LCW treatments \cite{Huang2000b}, captures the temperature dependence
of the hydrophobic effect. It is a curious observation that 
lattice-based theories of hydrophobicity \cite{Varilly2011,
Vaikuntanathan2014,Vaikuntanathan2016} that aim to improve LCW have
emphasized the importance of capillary wave fluctuations; these
do not enter explicitly in our theory. Nonetheless, our results for
the local compressibility, and the good agreement with solvation free
energies obtained from molecular simulations, suggest that we capture
the most salient aspects of the interfacial fluctuations necessary to
describe hydrophobicity.


A general theory of solvation should also describe the polarization
field induced by charged species such as ions. This is a challenging
problem beyond the scope of the present study. There are, however,
reasons to be optimistic. For example, the mDFT framework already
demonstrates that orientational correlations of the bulk fluid can be
used to construct density functionals that describe
polarization \cite{Jeanmairet2016}; introducing the results from our
work should amount to a straightforward modification of the present
mDFT framework. Insights from Weeks' local molecular field theory may
also prove useful in developing new functionals \cite{Rodgers2008,
Rodgers2009,rodgers2008interplay,Remsing2016,gao2020short,Cox2020} 
(see Refs.~\onlinecite{Archer2013, Remsing2016} for discussions
on the relationship
between cDFT and local molecular field theory), as could ideas from integral equation theories (see, e.g., Refs. \onlinecite{Chuev2021,Chuev2022}).
In a recent study,
Samm\"{u}ller \etal{} adopt an entirely different approach by using
deep neural networks to construct the free energy
functional \cite{sammuller2023}. Whether such an approach can be used
for polar fluids remains to be seen, but it shows great promise. In
the context of apolar solvation, our results raise the question
whether machine learning can be made easier by first separating the
functional into slowly- and rapidly-varying
contributions. Notwithstanding the obvious areas for future
development, the results of our work demonstrate a significant step
toward efficient, and fully first-principles, multiscale modeling of
solvation.

\section*{Methods}
\footnotesize 
Here we provide a brief overview of the methods used; full details are
provided in the SM. All molecular simulations have been performed
with the \texttt{LAMMPS} simulation package\cite{Thompson2022}.  To
maintain the temperature, we used the CSVR thermostat\cite{Bussi2007}.
For simulations of SPC/E water \cite{Berendsen1987}, long-ranged
electrostatic interactions were evaluated using the particle--particle
particle--mesh method\cite{hockney1988}, such that the RMS error in
the forces was a factor of $10^5$ smaller than the force between two
unit charges separated by 0.1\,nm\cite{Kolafa1992}. The rigid
geometry of the SPC/E water molecules was imposed with
the \texttt{RATTLE} algorithm\cite{Andersen1983}. For simulations of
the neural network surrogate model \cite{Wohlfahrt2020} of RPBE-D3
\cite{Hammer1999,Grimme2010}, we used the \texttt{n2p2} package 
interface\cite{Singraber2019} with \texttt{LAMMPS}.
cDFT calculations were performed with
our own bespoke code, which we have made publicly
available. Minimization was performed self-consistently using 
Picard iteration.

\section*{Supplementary Material}

\tcb{Supplementary material includes further details on parameterizing the functional, simulation details, 
further details on the numerical implementation of the functional, a comparison to molecular density functional theory, 
and details on evaluating the local compressibility.}

\section*{Data Availability}

\normalsize

Code for minimizing the functionals can be accessed at
\url{https://github.com/annatbui/LCW-cDFT}. 
Simulation input files and the direct correlation functions
obtained are openly available at the University of Cambridge Data
Repository, \url{https://doi.org/10.17863/CAM.104278}.

\section*{Acknowledgements}
We thank Robert Jack for many insightful discussions, and Bob
Evans and Nigel Wilding for comments on an initial draft of the
manuscript. 
We are grateful for computational support 
from the UK national high performance computing service, ARCHER2, for which access was obtained via the UKCP consortium and funded by
EPSRC grant ref EP/X035891/1.
A.T.B. acknowledges funding from the Oppenheimer Fund and
Peterhouse College, University of Cambridge.  S.J.C. is a Royal
Society University Research Fellow (Grant No. URF\textbackslash
R1\textbackslash 211144) at the University of Cambridge.

\bibliography{references}

\end{document}


\title{Supplementary material: A classical density functional theory for solvation across length scales}

\author{Anna T. Bui}
\affiliation{Yusuf Hamied Department of Chemistry, University of
  Cambridge, Lensfield Road, Cambridge, CB2 1EW, United Kingdom}

\author{Stephen J. Cox}
\email{sjc236@cam.ac.uk}
\affiliation{Yusuf Hamied Department of Chemistry, University of
  Cambridge, Lensfield Road, Cambridge, CB2 1EW, United Kingdom}

\date{\today}

\maketitle

\tableofcontents

\section{Square-gradient functional for the slowly-varying density field} \label{sec:slowlyvarying}

For the pseudofunctional of the slowly-varying density
$\rho_{\mrm{s}}$, we have taken the van der Waals
(square--gradient) functional for the intrinsic Helmholtz free energy
\begin{equation}
\mcl{F}_{\rm  intr}[\rho_{\rm s}] =
  \int\!\mrm{d}\mbf{r}\,\left[f_0 + f_2|\nabla\rho_{\rm s}(\mbf{r})|^2\right].
\end{equation}
In general, both $f_0$ and $f_2$ are functions of
$\rho_{\mrm{s}}(\mbf{r})$, with $f_0$ being the local Helmholtz free
energy density.  Taking the approximation $f_2 \approx m/2$,
where $m$ is a positive constant independent of density, we introduce
the local grand potential density
\begin{align}
    \omega\big(\rho_{\mrm{s}}(\mbf{r});\mu\big) &=
    f_0\big(\rho_{\rm s}(\mbf{r})\big) - \mu\rho_{\rm s}, \\[5pt]
    &= \omega_{\mrm{coex}}(\rho_{\mrm{s}}) - \rho_{\mrm{s}} \delta\mu, \label{eqn:square_gradient}
\end{align}
where $\delta\mu=\mu-\mu_{\mrm{coex}}$ describes the difference in
chemical potential from coexistence. In principle, $\omega_{\rm coex}$
can be obtained from the equation of state of the bulk fluid; to our
knowledge, this is not known for the different intermolecular
potentials for water that we explore in this work. We therefore adopt
a simple double-well form that broadly describes the physics of
liquid-vapor coexistence:
\begin{equation}
\label{eqn:local_grand_potential}
  \omega_{\rm coex}(\rho_{\rm s}) =
  \frac{C}{2}\left(\rho_{\mrm{s}}- \rho_\mrm{l}\right)^2 \left(\rho_{\mrm{s}}- \rho_\mrm{v}\right)^2 +  \frac{D}{4}\left(\rho_{\mrm{s}}- \rho_\mrm{l}\right)^4 \left(\rho_{\mrm{s}}- \rho_\mrm{v}\right)^4 
  h(\rho_{\rm s}-\rho_{\rm v})h(\rho_{\rm l}-\rho_{\rm s}),
\end{equation}
where $\rho_\mrm{l}$ and $\rho_\mrm{v}$ are the liquid and vapor
densities at coexistence, locating the minima of
$\omega_{\mrm{coex}}(\rho_{\mrm{s}})$, and $h(\rho_{\rm s})$ is the
Heaviside step function. We will now discuss the parameterization of $C$,
$D$ and $m$.

\subsection{Parameterization}

The form of $\omega_{\mrm{coex}}(\rho_{\mrm{s}})$ specified by
Eq.~\ref{eqn:local_grand_potential} is approximate. This is evident
from its symmetric form; the curvatures at $\rho_{\rm v}$ and
$\rho_{\rm l}$, as determined by $C$, are identical. Or, in other
words, the compressibilities of the vapor and liquid phases are
identical. For our purposes, we are more concerned with the
compressibility in the liquid phase, and we choose $C$ accordingly. To
be precise, denoting the bulk compressibility as $\chi_{\rm u}$, we
have
\begin{equation}
   \chi_{\rm u} = \left(\frac{\delta \rho_{\rm u}}{\delta \mu} \right)_T = \frac{1}{\omega''(\rho_\mrm{l})} = \frac{1}{C(\rho_{\rm l}-\rho_{\rm v})}.
\end{equation}
Following the standard compressibility
relation \cite{HansenMcDonaldBook} $\chi_{\mrm{u}}
= \beta \rho_{\mrm{u}} S(0)$, where $S(0)$ is the $k\rightarrow0$
limit of the liquid structure factor $S(k)$, we can write
\begin{equation}
\label{eq:C_parameter}
    C = \frac{k_{\mrm{B}} T}{\rho_{\rm u} S(0) (\rho_\mrm{l}-\rho_\mrm{v})^2}.
\end{equation}
In practice, to calculate $C$ using Eq.~\ref{eq:C_parameter}, for each
water model at a particular $T$, we perform a simulation of the bulk
liquid at density $\rho_\mrm{u}=\rho_{\mrm{l}}$ and extract $S(0)$ as
described in Section \ref{sec:simulation_details}. This ensures
consistency in $\chi_{\rm u}$ between our rapidly- and slowly-varying
functionals. We note that using an asymmetric form of square gradient
theory would allow one to distinguish the compressibilities of the
liquid and vapor phases, as well as differences in density gradients
on either side of the free liquid-vapor
interface \cite{Parry2016}. Initial investigations along these lines,
however, proved challenging, and we therefore settled on the simple
form described above.

If we set $D = 0$, then Eq.~\ref{eqn:local_grand_potential} reduces to
a standard approximate form encountered in
textbooks \cite{HansenMcDonaldBook}. If we then choose $m$ to attain
the surface tension of the free liquid-vapor interface, we have found
that the resulting density profiles compare poorly to results from
molecular simulations. While $\omega_{\rm coex}$ describes properties
of the bulk fluid, in order to simultaneously obtain $\gamma$ and
reasonable density profiles, we have taken the pragmatic approach of
introducing the second term in Eq.~\ref{eqn:local_grand_potential}. We
then parameterize $D$ and $m$ by the following procedure:

\begin{enumerate}[label=(\roman*)]
    \item We start with $D=0$ and find a value for $m$ such that the resulting
    functional gives the correct liquid--vapor surface tension $\gamma$.
    \item If the resulting density profile for the free liquid--vapor
    interface is too sharp compared to molecular simulation, this
    means that the contribution to $\varOmega_{\mrm{s}}$ from the
    local term is too large relative to that from the gradient
    term. We therefore reduce the relative contribution from
    $\omega_{\mrm{coex}}(\rho_{\mrm{s}})$ by guessing a negative value
    of $D$ to decrease the barrier height in the region
    $\rho_{\mrm{v}}<\rho_{\mrm{s}}<\rho_{\mrm{l}}$. If the resulting
    density profile is too wide, we instead guess a positive value of
    $D$.
    \item With this new value of $D$, we find a new value for
    $m$ that gives the correct surface tension $\gamma$.  \item We
    return to step (ii) if the resulting density profile still
    deviates significantly from the simulation data.
\end{enumerate}

We note that, since the interfacial width is more sensitive to $D$
than $m$, we do not need to adjust these parameters too much. As
a remark, were we to know the equation of state for the water models
that we investigate, then we would be able to determine $\omega_{\rm
coex}$ exactly. We would then focus our efforts entirely on
parameterizing $m(\rho_{\rm s})$ on properties of the free
liquid-vapor interface; in the absence of such data, the approach we
describe above provides a practical means to parameterizing
$\omega_{\rm coex}$ for our purposes.

In the main paper, we showed results using three different water
models: SPC/E \cite{Berendsen1987},
RPBE-D3 \cite{Hammer1999,Grimme2010} and mW \cite{Molinero2009}
(further details in Section \ref{sec:simulation_details}).  To
parameterize the functional for each water model, we use data from
simulations of the free liquid--vapor interface and the uniform fluid
at the liquid coexistence density at $300\,\mrm{K}$ for each water
model.  For the mW model, we also investigated $T = 426\,\mrm{K}$.
For SPC/E water, the liquid--vapor coexistence properties at
$300\,\mrm{K}$ have been determined by Vega and Miguel from direct
coexistence molecular dynamics (MD) simulations \cite{Vega2007}.  For
water described with the electronic density functional RPBE-D3, the
liquid--vapor coexistence curve (binodal) has been determined by
Schienbein and Marx \cite{Schienbein2018} through \emph{ab initio}
Gibbs ensemble Monte Carlo simulations. However, the surface tension
was not computed. Wohlfalhrt \emph{et al.} trained a neural
network potential \cite{Wohlfahrt2020} using energies and forces
computed at the RPBE-D3 level and were able to reproduce the
coexistence curve from Schienbein and Marx, in addition to determining
the surface tensions at different temperatures using direct
coexistence MD simulations.  We therefore use the liquid--vapor
coexistence properties at $300\,\mrm{K}$ from Wohlfahrt \emph{et
al.} \cite{Wohlfahrt2020} For the mW model, we use liquid--vapor
coexistence properties determined by Coe \emph{et al.} from grand
canonical Monte Carlo (GCMC) simulations \cite{Coe2022b}.  Since the
critical temperature of mW ($T_{\mrm{c}}=917\,\mrm{K}$) is much higher
than that of real water ($T_{\mrm{c}}=647\,\mrm{K}$), we investigate
$T=426\,\mrm{K}$ for mW as it gives the same scaled temperature as
ambient water $T/T_{\mrm{c}}\approx0.46$ (argued in
Ref.~\onlinecite{Coe2022b}).  We have also considered $T=300\,\mrm{K}$
where the surface tension for mW is in better agreement with the other
water models.  In instances where the interfacial density profile is
not given in the above references, we obtained it with direct
coexistence MD simulations of our own (in such cases, we checked for
consistency, e.g., by computing the surface tension). The bulk
simulations at constant density $\rho_{\rm u }=\rho_{\rm l}$ to
extract the bulk compressibility $\chi_{\rm u}$ are described in
Section \ref{sec:simulation_details}. We summarized the parameters
used in Table \ref{tab:square_gradient}.
\begin{table}[H]
    \centering
    \begin{tabular}{l  c c c c}
    \hline
    \hline
    Water model & SPC/E & RPBE-D3 & mW  & mW\\
    $T$ [K]     &   300  & 300    &  426  & 300 \\
    \hline
    $\rho_{\rm l}$ [$\mrm{nm^{-3}}$] & 33.234\ftb{a} & 30.06\ftb{b} & 32.203\ftb{c} & 33.405\ftb{c} \\
    $\rho_{\rm v}$ [$\mrm{nm^{-3}}$] & $4.747\times10^{-4}$\ftb{a} & $1.04\times10^{-3}$\ftb{b}  & $1.652\times10^{-3}$\ftb{c}  & $1.337\times10^{-5}$\ftb{c}\\
    $\gamma$ [$\mrm{mJ\, m^{-2}}$] & 63.6\ftb{a} & $68\pm7$\ftb{b} & $55.6\pm2$\ftb{c} &  65.8395\ftb{c} \\
    $k_{\mrm{B}}T\rho_{\mrm{u}}^{-1}\chi_{\mrm{u}}=S(0)$ &  0.0630 &    0.0598    &   0.0456 & 0.0265 \\
    \hline
    $C$ [$\mrm{J\, mol^{-2}\, nm^9}$] & 1.079 & 1.537 & 2.296 & 2.523 \\
    $D$ [$\mrm{J\, mol^{-2}\, nm^{15}}$] & 30 & 90 & -15 & -15\\
    $m$ [$\mrm{kJ\, mol^{-2}\, cm^3 \,\text{\AA}^2}$] & 1255 & 1500 & 1100 &  1148 \\
    \hline
    \hline
    \end{tabular}
   \caption{\textbf{Parameters for the slowly varying functional.}
   For the SPC/E, RPBE-D3 and mW water models, square gradient theory
   is parameterized based on liquid--vapor coexistence at temperature
   $T=300\,\mrm{K}$ (and $T=426\,\mrm{K}$ for mW). The first part
   of the table gives thermodynamic properties of water. The
   second part gives parameters for the square--gradient functional. \\
   \footnotesize
   \ftb{a}Determined by Vega and Miguel in
   Ref.~\onlinecite{Vega2007}. \\ 
   \ftb{b}Determined by Wohlfahrt \emph{et 
   al.} in Ref.~\onlinecite{Wohlfahrt2020}. \\
   \ftb{c}Determined by Coe in Ref.~\onlinecite{Coe2022b}. We have
   used data for $T=420\,\mrm{K}$ for $T=426\,\mrm{K}$ since it is the
   closest state point tabulated, with indicated uncertainties accounting for any
   discrepancy. \\}
\label{tab:square_gradient}
\end{table}

\subsection{Free liquid--vapor interface}

\begin{figure}[H]
    \centering \includegraphics[width=0.85\linewidth]{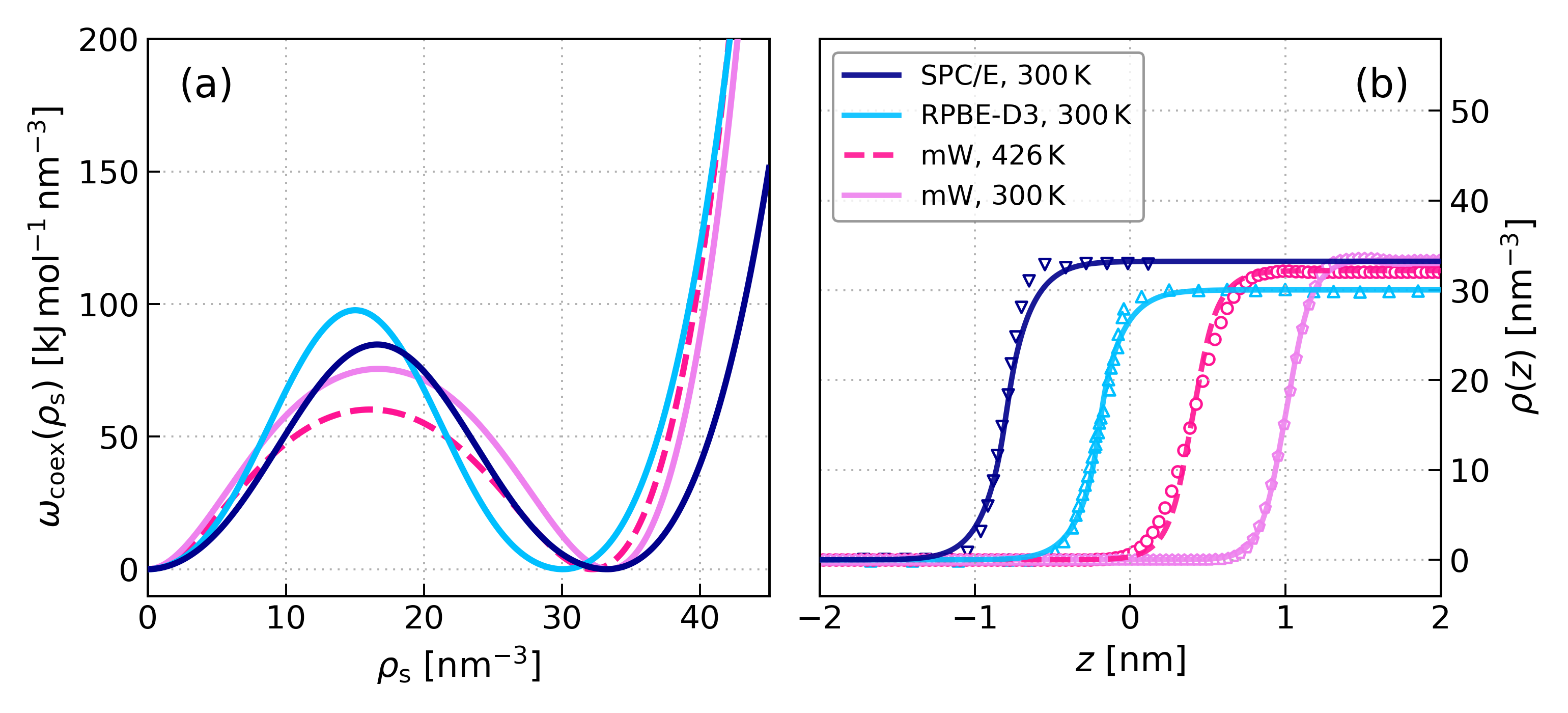}
    \caption{\textbf{Liquid--vapor interface from square gradient
    theory.} The local grand potential density at coexistence
    $\omega_{\mrm{coex}}(\rho_{\rm s})$ given in
    Eq.~\ref{eqn:square_gradient}, and corresponding liquid--vapor
    interface profiles $\rho(z)$ given in
    Eq.~\ref{eqn:local_grand_potential} are shown in (a) and (b),
    respectively. The square gradient theory is parameterized for the
    water models and state points indicated in the legend in (b). The
    function $\omega_{\mrm{coex}}(\rho_{\rm s})$ has two minima at
    $\rho_{\mrm{v}}$ and $\rho_{\mrm{l}}$ corresponding to the vapor
    and liquid phases, respectively.  The density profiles for SPC/E
    and RPBE-D3 are taken from Ref.~\onlinecite{Wohlfahrt2020} and
    Ref.~\onlinecite{Huang2002}, respectively. For mW, the
    density profiles are from our own direct coexistence
    simulations. We note that the density of the
    liquid phase for RPBE-D3 deviates the most from the experimental
    liquid density ($33.3\,\mrm{nm^{-3}}$), which is a shortcoming of
    this electronic density functional well-documented in the
    literature \protect\cite{Morawietz2016, Schienbein2018,
    Schmidt2009}. } \label{si:square_gradient}
\end{figure}

The resulting local grand potential density and liquid--vapor interface profiles from square gradient theory parameterized for each water model are shown in Fig.~\ref{si:square_gradient}.

\subsection{Bulk compressibility}

\begin{figure}[H]
    \centering \includegraphics[width=0.9\linewidth]{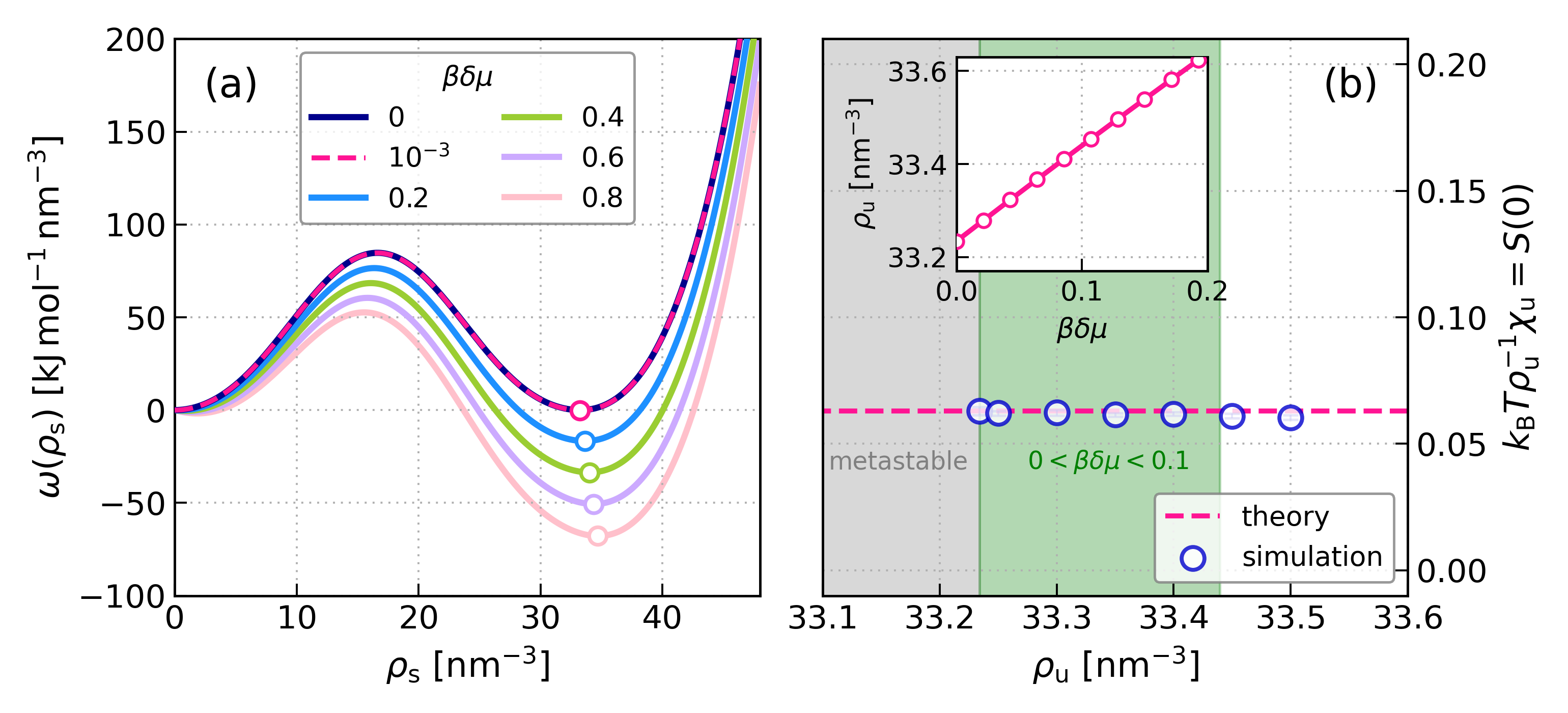}
    \caption{\textbf{Liquid bulk compressibility from square gradient
    theory}. From the simple square gradient form given in
    Eq.~\ref{eqn:square_gradient}, an increase in
    $\delta\mu=\mu-\mu_{\mrm{coex}}$ stabilizes the liquid phase
    further and shifts the liquid minimum to a higher density
    in a linear fashion, as shown in (a) for SPC/E water at
    $300\,\mrm{K}$. The circles indicate the global minimum of each
    curve, locating the stable liquid density at $\rho_{\mrm{u}}$.
    Note that for $\beta\delta\mu\approx10^{-3}$ the deviation from
    coexistence is relatively small. The linearity between $\rho_{\rm
    u}$ and $\delta\mu$ is depicted in the inset of (b); the gradient
    of this line gives the bulk compressibility $\chi_{\mrm{u}}$, as
    determined by the square gradient functional.  As seen in
    (b), as we move away from coexistence, this estimate for
    $\chi_{\mrm{u}}$ deviates more from values obtained from 
    simulations (detailed in Section \ref{sec:simulation_details}),
    which are indicated by the circles (error bars smaller than the
    symbols). In the region that is shaded gray ($\rho_{\mrm{u}}
    < \rho_{\mrm{l}}$), the liquid phase is metastable. The domain of
    applicability of the theory ($0<\beta\delta\mu<0.1$) is shaded
    green.}  \label{si:density_mu}
\end{figure}

The ensemble to naturally work in for cDFT is the grand canonical
ensemble.  Water under ambient conditions lies close to its
liquid--vapor coexistence \cite{Cerdeirina2011} with a chemical
potential deviation from coexistence $\beta\delta\mu\approx10^{-3}$.
For a bulk fluid at chemical potential $\mu$, the
corresponding uniform density $\rho_{\mrm{u}}$ is determined by the
global minimum of the local free energy density $\omega(\rho_{\rm
s})$. For the form that we adopted, as given in
Eq.~\ref{eqn:square_gradient}, the term linear in $\rho_{\rm s}$ adds
a bias to the double-well form of $\omega_{\mrm{coex}}(\rho_{\rm s})$,
making the liquid phase more stable at higher $\rho_{\rm s}$. This
behavior is demonstrated in Fig.~\ref{si:density_mu}(a) for the square
gradient theory parameterized on SPC/E water; note that we are using
large values of $\beta\delta\mu$ to exaggerate the behavior. In
Fig.~\ref{si:density_mu}(b) we plot the bulk compressibility of the
bulk liquid determined from simulations, and from square gradient
theory. Owing to the approximate form of 
$\omega_{\rm coex}(\rho_{\mrm{s}})$,
agreement with the simulation results is limited to $0
< \beta\delta\mu \lesssim 0.1$. In Section~\ref{sec:critical_drying},
we also show that this range of $\delta\mu$ is where the contact
theorem (a statistical mechanical sum rule) for a fluid at a planar
hard wall is also obeyed.

In the following, we determine the uniform liquid density
$\rho_{\mrm{u}}$ from the square gradient form corresponding to a
chemical potential deviation of $\beta\delta\mu\approx10^{-3}$ for each water model, as
summarized in Table \ref{tab:compressibility}. 

\begin{table}[H]
    \centering
    \begin{tabular}{l  c c c c}
    \hline
    \hline
    Water model & SPC/E & RPBE-D3 & mW  & mW\\
    $T$ [K]     &   300  & 300    &  426  & 300 \\
    \hline
    $\rho_{\rm u}$ [$\mrm{nm^{-3}}$]  & 33.236 &  30.079  & 32.204 & 33.406 \\
    \hline
    \hline
    \end{tabular}
\caption
{
\textbf{Bulk liquid density from square gradient theory.}
The table gives the stable liquid density $\rho_{\mrm{u}}$ that
corresponds $\beta\delta\mu\approx10^{-3}$ determined from the minimum
of $\omega(\rho_{\mrm{s}})$ for each water model and state point considered.}
\label{tab:compressibility}
\end{table}

\subsection{Adapting the reference slowly-varying density in the crossover regime}

\begin{figure}[H]
    \centering \includegraphics[width=0.9\linewidth]{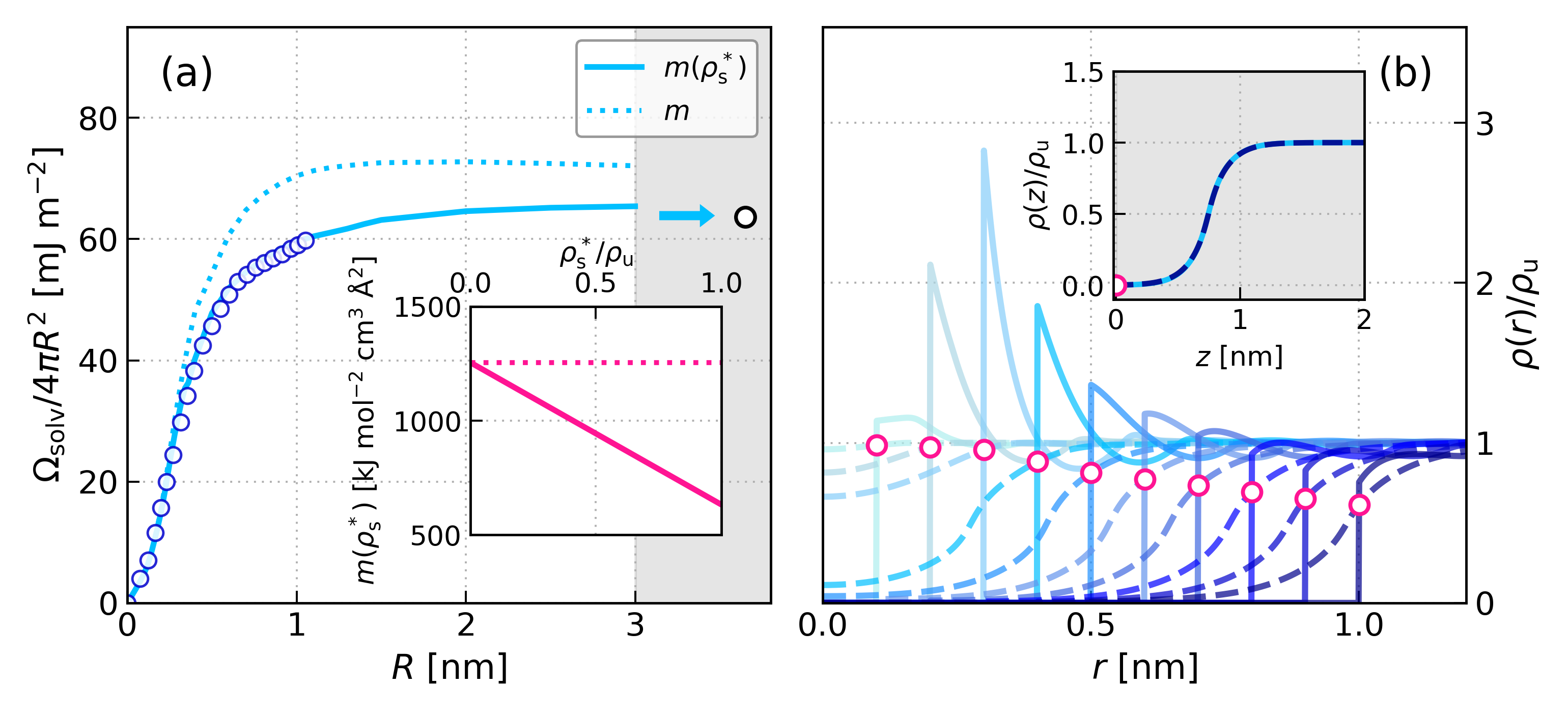
    } \caption{\textbf{Adapting the reference slowly-varying density
    in the crossover regime.} In (a), $\varOmega_{\mrm{solv}}/4\pi
    R^2$ for solvation of hard sphere solutes in SPC/E water at 300\,K
    and $\beta\delta\mu=10^{-3}$ predicted from the theory using a
    constant $m$ (dotted line) and using $m(\rho^{*}_\mrm{s})$ (solid
    line) where $\rho^{*}_\mrm{s}$ is a measure of how depleted the
    density profile is. The linear form of $m(\rho^{*}_\mrm{s})$
    employed is shown in the inset. For both approaches,
    $\varOmega_{\rm solv}/4\pi R^2 \sim \gamma$, indicated by the open
    circle. In (b), $\rho(r)$ and $\rho_{\mrm{s}}(r)$ for various
    solute sizes are shown in solid and dashed lines respectively. The
    case of the solvent in contact with a hard wall is shown in the
    inset. The values for $\rho^{*}_\mrm{s}$ are marked with open
    circles. }\label{si:adapt_m}
\end{figure}

By construction, the theory captures the solvation behavior in 
the limits of small solutes and macroscopic planar surfaces
most accurately, with a gradual crossover between these limits.
The exact behavior in the crossover regime, however, is inevitably
more challenging to capture. To explore this aspect, we
consider the solvation of hard spheres of
increasing radius $R$ in SPC/E water at 300\,K where simulation data
up to $R=1\,\mrm{nm}$ are available \cite{Huang2001}, as shown in
Fig.~\ref{si:adapt_m}.

Similar to previous studies with LCW theory 
\cite{Lum1999, Huang2000, Huang2000b, Huang2002}, 
the functional using the square gradient theory as parameterized in
the previous subsection with a constant value for $m$ predicts that
$\varOmega_{\mrm{solv}}/4\pi R^2$ first increases up to a value higher
than the liquid--vapor surface tension $\gamma$ at
$R\approx0.7\,\mrm{nm}$, before slowly decreasing to approach $\gamma$
from above as $R\to \infty$. (More precisely,
$\varOmega_{\mrm{solv}}/4\pi R^2 \sim \gamma + \gamma_{\rm wv}$, where
$\gamma_{\rm wv}$ is the surface tension between the vapor phase and
the wall. Owing the small vapor pressure of water, it is common to
neglect this contribution.)
Simulations, however, predict that $\varOmega_{\mrm{solv}}/4\pi R^2$
does not exceed the surface tension up to $R=1\,\mrm{nm}$.  This
observation suggests that in the crossover regime, the best
slowly-varying reference density slightly deviates from what is given
by $\tilde{\varOmega}_{\psi}[\rho_{\mrm{s}}]$ parameterized with constant
$m$. In principle, the coefficient of the square-gradient term could
depend on density, i.e., $m(\rho_{\rm s})$. [Note: this would
introduce a term proportional to $m^\prime(\rho_{\rm
s})|\nabla\rho_s|^2$ upon minimization.] Such a parameterization is
not straightforward. For practical purposes, we instead treat $m$ as a
function of a characteristic feature of $\rho_{\rm
s}^\ast$. Specifically, $\rho^{*}_{\mrm{s}}$ is the minimum value of
the density at positions around the solute where $\rho(r)$ intercepts
$\rho_{\mrm{s}}(r)$.  Since $\rho^*_{\mrm{s}}$ gives an indication of
how depleted the density profile is, adapting a constant $m$ to
$m(\rho^{*}_{\mrm{s}})$ allows us to quantitatively describe the
simulation data.  Here, we employ a linear dependence for
$m(\rho^{*}_{\mrm{s}})$,
\begin{equation}
    m(\rho^{*}_{\mrm{s}}) = m + t \rho^{*}_{\mrm{s}}.
\end{equation}
This enforces the constant $m$ value is recovered when $\rho_{\rm s}^\ast \approx \rho_{\rm v}$. For SPC/E water, we have used $t=-20323\,\mrm{kJ\, mol^{-2}\, cm^3 \,\text{\AA}^5}$. Note that the solvation
behaviors for spheres with $R<0.4\,\mrm{nm}$ and at a planar wall are
not sensitive to this change. For hard spheres,
$\rho^{*}_{\mrm{s}}$ amounts to the contact density; the definition we
describe facilitates its application to soft-core solutes. 

We emphasize that even with a constant $m$, the results for both the
solvation free energy and density are still
semi-quantitative. Therefore, in cases where we lack simulation data
in the crossover regime, we use a constant value of $m$ parameterized
on the free liquid-vapor interface.

\newpage

\section{Simulation details for correlation functions}  \label{sec:simulation_details}

In this section, we describe the procedure followed to obtain the
spherically symmetric direct correlation function of a bulk fluid from
simulation.

\subsection{Simulation details}

For each interatomic potential, MD simulations of bulk liquid water at
ambient conditions were performed.  All simulations were carried out
with the \texttt{LAMMPS} simulation package \cite{Thompson2022}.
Cubic simulation boxes with periodic boundary conditions were used.
For the structure factor calculation, the systems contained $13824$
water molecules for SPC/E and RPBE-D3, $15000$ water molecules in the
case of the mW model.
Dynamics were propagated using the velocity Verlet algorithm with a
time-step of $\SI{1}{\femto\second}$.  The temperature was maintained
at $T=\SI{300}{K}$ or $T=\SI{426}{K}$ using the CSVR
thermostat \cite{Bussi2007}.
The system was equilibrated for $200\,\mrm{ps}$ and production runs were
performed for at least $2\,\mrm{ns}$ in the NVT ensemble.


\subsection{Interatomic potentials}
Three different interatomic potentials for water with different levels
of complexity are considered:
\begin{enumerate}[label=(\roman*)]
    \item SPC/E: This simple point charge water
    model \cite{Berendsen1987} contains a Lennard-Jones centre on the
    oxygen atom and partial charges on the hydrogen and oxygen
    atoms. The geometries of water molecules were constrained using
    the \texttt{RATTLE} algorithm \cite{Andersen1983}. All
    Lennard-Jones interactions were truncated and shifted
    at 1\,nm. Long-ranged electrostatic interactions were
    evaluated using particle--particle particle--mesh Ewald
    summation \cite{hockney1988} (using a 1\,nm cut off in real
    space) such that the RMSE in the forces was a factor of $10^5$
    smaller than the force between two unit charges separated by a
    distance of 0.1\,nm \cite{Kolafa1992}.
    \item mW: The monatomic water potential \cite{Molinero2009} is a
    coarse-grained model of water in which there are no explicit
    hydrogens. It takes the form of a Stillinger--Weber
    potential \cite{Stillinger1985} and the tetrahedrality is achieved
    by three-body terms. For this single-site water model, the fluid
    density is only described by a singlet particle density so no
    angular component is neglected in our theory.
    \item RPBE-D3: An \textit{ab initio} approach to model water can
    be achieved through quantum mechanical calculations.  For the
    underlying electronic structure, the Revised
    Perdew--Burke--Ernzerhof (RPBE) generalized gradient approximation
    for the exchange--correlation functional \cite{Hammer1999}
    supplemented by Grimme's D3 dispersion
    correction \cite{Grimme2010} is used. Since thorough sampling of
    the liquid state's phase space at this level of theory is
    prohibitively expensive, we perform simulations using a surrogate
    neural network potential \cite{Behler2007} trained on forces and
    energies from \textit{ab initio} simulation trajectories as given
    in Ref.~\onlinecite{Wohlfahrt2020},
    using \texttt{LAMMPS} \cite{Thompson2022} with an interface to
    the \texttt{n2p2} package\cite{Singraber2019}.
\end{enumerate}

\subsection{Workflow for extracting the direct correlation function}

We denote the 3D spatial Fourier transform of function $f(\mbf{r})$ as
\begin{equation}
\hat{f}(\mbf{k}) = \int\!\!\mrm{d}\mbf{r}\,f(\mbf{r})\exp(-i\mbf{k}\cdot\mbf{r}),
\end{equation}
and the inverse Fourier transform of $\hat{f}(\mbf{k})$ as
\begin{equation}
f(\mbf{r}) = \frac{1}{(2\pi)^3}\int\!\!\mrm{d}\mbf{k}\,\hat{f}(\mbf{k})\exp(i\mbf{k}\cdot\mbf{r}).
\end{equation}
For a radially symmetric function $f(r)$, the corresponding expressions become
\begin{equation}
 \hat{f}(k) = 4\pi \int^\infty_0\!\!\mrm{d}r\,r^2\frac{\sin(kr)}{kr}f(r),
\end{equation}
and
\begin{equation}
f(r) = \frac{1}{(2\pi)^3} \int^\infty_0\!\!\mrm{d}k\,\frac{4\pi k}{r}\sin(kr)\hat{f}(k).
\end{equation}

\begin{figure}[b]
    \centering
    \includegraphics[width=0.9\linewidth]{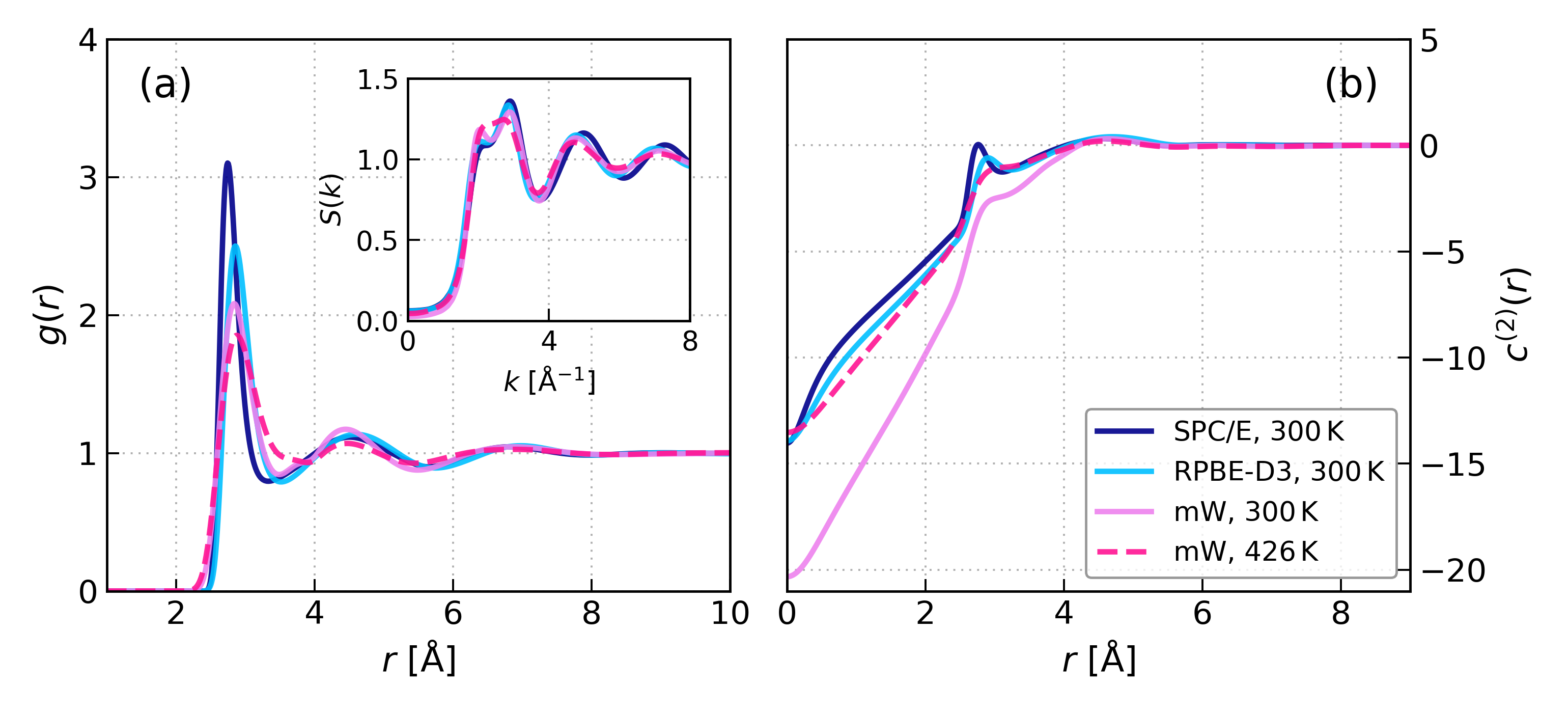}
    \caption{\textbf{Bulk structure of different water models from simulation.} The radial distribution functions $g(r)$ are shown in (a), with the structure factor $S(k)$ given in the inset. The direct correlation function determined from simulation to be used as input to cDFT is shown in (b).
    The water models and temperatures considered are denoted in the legend in (b).}
    \label{si:dcf_different_water}
\end{figure}

For a bulk system containing $N$ water molecules at density $\rho_\mrm{u}$, we extracted the radial distribution function $g(r)$ 
\begin{equation}
    g(r) = \frac{1}{N^2}\sum^N_{i,j}\left\langle \delta(r - |\mbf{r}_i - \mbf{r}_j|)\right\rangle,
\end{equation}
by generating a histogram of intermolecular distances and appropriate normalization. Note that we considered the positions of the oxygen as the positions of the water molecule in the case of SPC/E and RPBE-D3 water. 
The resulting radial distribution functions for the different water models are shown in Fig.~\ref{si:dcf_different_water}(a).
The high-wavevector part of the structure factor ($k>1\,\text{\AA}^{-1}$) are computed from \cite{HansenMcDonaldBook}
\begin{equation}
  \label{eqn:S-FT}
    S(k) = 1 + 4\pi\rho_\mrm{u}\int^\infty_0\!\!\mrm{d}r\, \frac{\sin(kr)}{k}\left[g(r)-1\right].
\end{equation}

 In the low-wavevector limit ($k<1\,\text{\AA}^{-1}$), we also calculate the structure factor directly from the simulation trajectories using\cite{Sedlmeier2011}
 \begin{equation}
 \label{eqn:S-D}
 \begin{split}
     S(k) & = \frac{1}{N}\sum^N_{i,j}\langle \exp[-i\mbf{k}\cdot(\mbf{r}_i-\mbf{r}_j)]\rangle \\
     & = \frac{1}{N} \left\langle \left[\sum^N_{i}\sin(\mbf{k}\cdot\mbf{r}_i)\right]^2 \right\rangle + \frac{1}{N} \left\langle \left[\sum^N_{i}\cos(\mbf{k}\cdot\mbf{r}_i)\right]^2 \right\rangle.
\end{split}
 \end{equation}
 The smallest wavevector accessible using this method from our simulations is $k\approx0.08\,\text{\AA}^{-1}$.
 Additionally, we also determine the isothermal compressibility, 
 \begin{equation}
     \kappa_T=-\frac{1}{V}\left(\frac{\partial V}{\partial P}\right)_T,
 \end{equation}
 where $V$ is the volume and $P$ is the pressure of the fluid, to calculate the $k\rightarrow0$ limit of the structure factor, given by the relation \cite{HansenMcDonaldBook}
 \begin{equation}
  \label{eqn:S-0}
  S(0)  = \rho_\mrm{u} k_{\mrm{B}} T \kappa_T.
 \end{equation}
We determine the isothermal compressibility by finite difference \cite{Motakabbir1990}
\begin{equation}
    \kappa_T = \frac{1}{\rho}\left(\frac{\partial\rho}{\partial P}\right)_T \approx \frac{\ln(\rho_1 / \rho_0)}{P_1 - P_0}.
\end{equation}
To evaluate this expression, the system is simulated in the NVT ensemble at densities $\rho_{0,1}=\rho_\mrm{u}\pm1 \,\mrm{nm^{-3}}$ and the resulting pressures $P_{0,1}$ are sampled.

The full structure factor is obtained by combining data obtained from Eqs.~\ref{eqn:S-FT}, \ref{eqn:S-D} and \ref{eqn:S-0}. Data at low $k$ were interpolated using Scipy's \texttt{UnivariateSpline} routine \cite{SciPy2020} and appropriate Savitzky--Golay smoothing filter \cite{Savitzky1964} to be as fine as $\delta k = 0.01\,\text{\AA}^{-1}$.
The resulting structure factors for the different water models are shown in the inset of Fig.~\ref{si:dcf_different_water}(a).

We then obtained the direct correlation function using the Ornstein--Zernike relation \cite{Ornstein1914, HansenMcDonaldBook} 
\begin{equation}
    1-\rho_{\mrm{u}} \hat{c}^{(2)}(k)= \frac{1}{1+\rho_{\mrm{u}}\hat{h}(k)} = \frac{1}{S(k)}, 
\end{equation}
where $\hat{c}(k)$ and $\hat{h}(k)$ are the Fourier transforms of the direct correlation function $c^{\mrm{(2)}}(r)$ and the total correlation function $h(r)=g(r)-1$. The resulting direct correlation functions in real space for the different water models are shown in Fig.~\ref{si:dcf_different_water}(b).

\subsection{Extracting solvation free energies of hard spheres}

When the solutes of interest are small hard spheres with radius
$R\lesssim0.4\,\mrm{nm}$, the solvation free energies can be determined from
simulations of the bulk liquid as
\begin{equation}
    \varOmega_{\mrm{solv}} = -k_{\mrm{B}}T\ln p (0;v),
\end{equation}
where $p(0;v)$ is the probability of forming a spherical cavity of volume
$v=4\pi R^3 / 3$ containing $N_v=0$ solvent
molecules \cite{Hummer1996}. For all water models and state points
considered, we computed $\varOmega_{\mrm{solv}}$ for hard spheres of
radii up to $\sim 0.4\,\mrm{nm}$.

\newpage

\section{Coarse-graining procedure} \label{sec:coarse_graining}

For the potential used in the pseudofunctional, we have prescribed
\begin{equation}
\label{eq:unbalancing_potential}
\psi(\mbf{r}) = -  k_{\mrm{B}}T\!\int\!\!\mrm{d}\mbf{r}'\,c^{(2)}_{\mrm{r},1}(\mbf{r},\mbf{r}')\,[\rho(\mbf{r}')-\rho_{\mrm{s}}(\mbf{r}')],
\end{equation}
where $c^{(2)}_{\mrm{r},1}$ denotes the
slowly-varying part of the two-body direct correlation function of the auxiliary density.
This is formally similar to the ``unbalancing potential'' introduced
by Weeks and co-workers \cite{Lum1999, Weeks2002}.
Eq.~\ref{eq:unbalancing_potential} is an integral of the
rapidly-varying component of the density field $\delta_{\rm s}\rho$
over a slowly-varying function $c^{(2)}_{\mrm{r,1}}$. This can be
interpreted as a coarse-graining of the density field.
More concretely, we define a coarse-graining procedure
\begin{equation}
 \label{eqn:CG}
  \overline{\rho}(\mbf{r}) =
  \frac{\int\!\mrm{d}\mbf{r}^\prime\,c^{(2)}_{{\rm r},1}(\mbf{r},\mbf{r}^\prime)\rho(\mbf{r}^\prime)}
  {\int\!\mrm{d}\mbf{r}^\prime\,c^{(2)}_{{\rm r},1}(\mbf{r},\mbf{r}^\prime)}.
\end{equation}
The self-consistent equation for  $\rho_{\rm  s}$ from minimising the pseudofunctional then becomes
\begin{equation}
\label{eqn:scf-slow}
  \omega^\prime(\rho_{\rm s}(\mbf{r})) =
  m\nabla^2\rho_{\rm s}(\mbf{r}) + 2a(\mbf{r})[\overline{\rho}(\mbf{r})-\overline{\rho}_{\rm s}(\mbf{r})],
\end{equation}
with
\begin{equation}
 2a(\mbf{r}) = k_{\rm B}T\int\!\mrm{d}\mbf{r}^\prime\, c^{(2)}_{{\rm r},1}(\mbf{r},\mbf{r}^\prime).
\end{equation}
We can approximate the reference two-body direct correlation function
$c^{(2)}_{\mrm{r}}$ with its translationally invariant counterpart of
the uniform fluid $c^{(2)}_{\mrm{u}}$, such that
\begin{equation}
  \overline{\rho}(\mbf{r}) \approx
  \frac{\int\!\mrm{d}\mbf{r}^\prime\,c^{(2)}_{{\rm u},1}(|\mbf{r}-\mbf{r}^\prime|)\rho(\mbf{r}^\prime)}
  {\int\!\mrm{d}\mbf{r}^\prime\,c^{(2)}_{{\rm u},1}(|\mbf{r}-\mbf{r}^\prime|)},
\end{equation}
and
\begin{equation}
 2a(\mbf{r}) \approx 2a= k_{\rm B}T\int\!\mrm{d}\mbf{r}\, c^{(2)}_{{\rm u},1}(r).
\end{equation}
As an interesting aside, for simple fluids like a Lennard--Jones
fluid, one might consider using the random phase approximation
(RPA), \cite{HansenMcDonaldBook}
\begin{equation}
    c^{(2)}_{{\rm u}}(|\mbf{r}-\mbf{r}^\prime|) \approx c^{(2)}_{{\rm u},0}(|\mbf{r}-\mbf{r}^\prime|) -\beta u_1(|\mbf{r}-\mbf{r}^\prime|),
\end{equation}
where $c^{(2)}_{{\rm u},0}$ denotes the rapidly-varying component and $u_1$ denotes the long-ranged attractive part of the
interparticle potential (through Weeks--Chandler--Anderson
splitting \cite{Weeks1971}). In this case, our coarse-graining reduces
to
\begin{equation}
  \overline{\rho}(\mbf{r}) \approx
  \frac{\int\!\mrm{d}\mbf{r}^\prime\,u_{1}(|\mbf{r}-\mbf{r}^\prime|)\rho(\mbf{r}^\prime)}
  {\int\!\mrm{d}\mbf{r}\,u_{1}(r)},
\end{equation}
with
\begin{equation}
  2a= \int\!\mrm{d}\mbf{r}\, u_{1}(r).
\end{equation}
This is identical to the coarse-graining procedure specified by Weeks
in the context of local molecular field theory \cite{Weeks2002}.

For complex fluids such as water, making a reasonable range separation for the direct correlation function is not straightforward as for
simple liquids. Therefore, for practical purposes, we use a Gaussian
weight of width $\lambda$ to coarse-grain the density,
\begin{equation}
  \overline{\rho}(\mbf{r}) \approx \frac{1}{(2\pi\lambda^2)^{3/2}}
\int\!\mrm{d}\mbf{r}^\prime\,\exp\left(-\frac{|\mbf{r}-\mbf{r}^\prime|^2}{2\lambda^2}\right)\rho(\mbf{r}^\prime).
   \label{eqn:CG-g}
\end{equation}
This leaves $a$ and $\lambda$ as parameters, for
which we have used $a = 200$\,kJ\,cm$^3$\,mol$^{-2}$ and $\lambda=
0.08\,\text{nm}$ for SPC/E water. Below, we will provide an analysis, for
SPC/E water at $300$\,K, that explains how we arrived at these values. 
Similar values are used for the other water models, which are given in Section~\ref{sec:comparemdft}.

\subsection{Hard sphere splitting}

For complex fluids, it is generally difficult to separate out the part
of the correlation function that is slowly-varying in space.  To make
progress, we therefore assume that the short-ranged correlations in
the uniform fluid can be approximated by that of a uniform hard-sphere
fluid at the same density,
\begin{equation}
c^{(2)}_{\mrm{u}}(r) \approx c_{\mrm{hs}}^{(2)}(r) + c_{\mrm{u},1}^{(2)}(r),
\label{eq:mDF_split}
\end{equation}
where $c^{(2)}_{\mrm{hs}}$ is the two-body direct correlation function
of the hard sphere fluid.  The coarse-graining procedure becomes
\begin{equation}
\overline{\rho}^{({\mrm{hs}})}(\mbf{r}) = \frac{\!\int\!\mrm{d}\mbf{r}'\,[c^{(2)}_{\mrm{u}}(|\mbf{r}-\mbf{r}'|) - c_{\mrm{hs}}^{(2)}(|\mbf{r}-\mbf{r}'|)]\,\rho(\mbf{r}')}{\!\int\!\mrm{d}\mbf{r}'\,[c^{(2)}_{\mrm{u}}(|\mbf{r}-\mbf{r}'|) - c_{\mrm{hs}}^{(2)}(|\mbf{r}-\mbf{r}'|)]},
\end{equation}
and
\begin{equation}
2 a^{({\mrm{hs}})} =  k_{\mrm{B}}T\!\int\!\mrm{d}\mbf{r}\,[c^{(2)}_{\mrm{u}}(r) - c_{\mrm{hs}}^{(2)}(r)].
\end{equation}
%
This is effectively treating water as a simple van der Waals fluid.



Here, we have used $c^{(2)}_{\mrm{u}}$ obtained from simulation of
bulk SPC/E water and the Percus--Yevick expression \cite{Percus1958}
for $c_{\mrm{hs}}^{(2)}$, which is given by\cite{HansenMcDonaldBook}
\begin{equation}
    c^{(2)}_{\mrm{hs}}(r) = 
    \begin{cases}
    -\sigma_1 -  \frac{6\eta\sigma_2}{d}r - \frac{\eta \sigma_1}{2 d^3}  r^3 & \text{if } r \leq d \\
    0 & \text{if } r > d,  \\
    \end{cases}  
\end{equation}
where $d$ is the hard sphere diameter, $\eta=\frac{\pi d^3\rho}{6}$ is
the packing fraction with $\rho=\rho_{\mrm{u}}=33.2\,\mrm{nm^{-3}}$
and
\begin{equation}
    \sigma_1= \frac{(1+2\eta)^2}{(1-\eta)^4}, \quad \sigma_2= -\frac{(2+\eta)^2}{4(1-\eta)^4}.
\end{equation}
It is reasonable to choose $d$ to be at least the molecular diameter,
and acceptable values could range from $0.275$--$0.295\,\text{nm}$,
corresponding to the range of the first peak in $g(r)$. We then find
$a^{({\mrm{hs}})} \approx 40$--$500$\,kJ\,cm$^3$\,mol$^{-2}$.

\begin{figure}[H]
    \centering \includegraphics[width=0.9\linewidth]{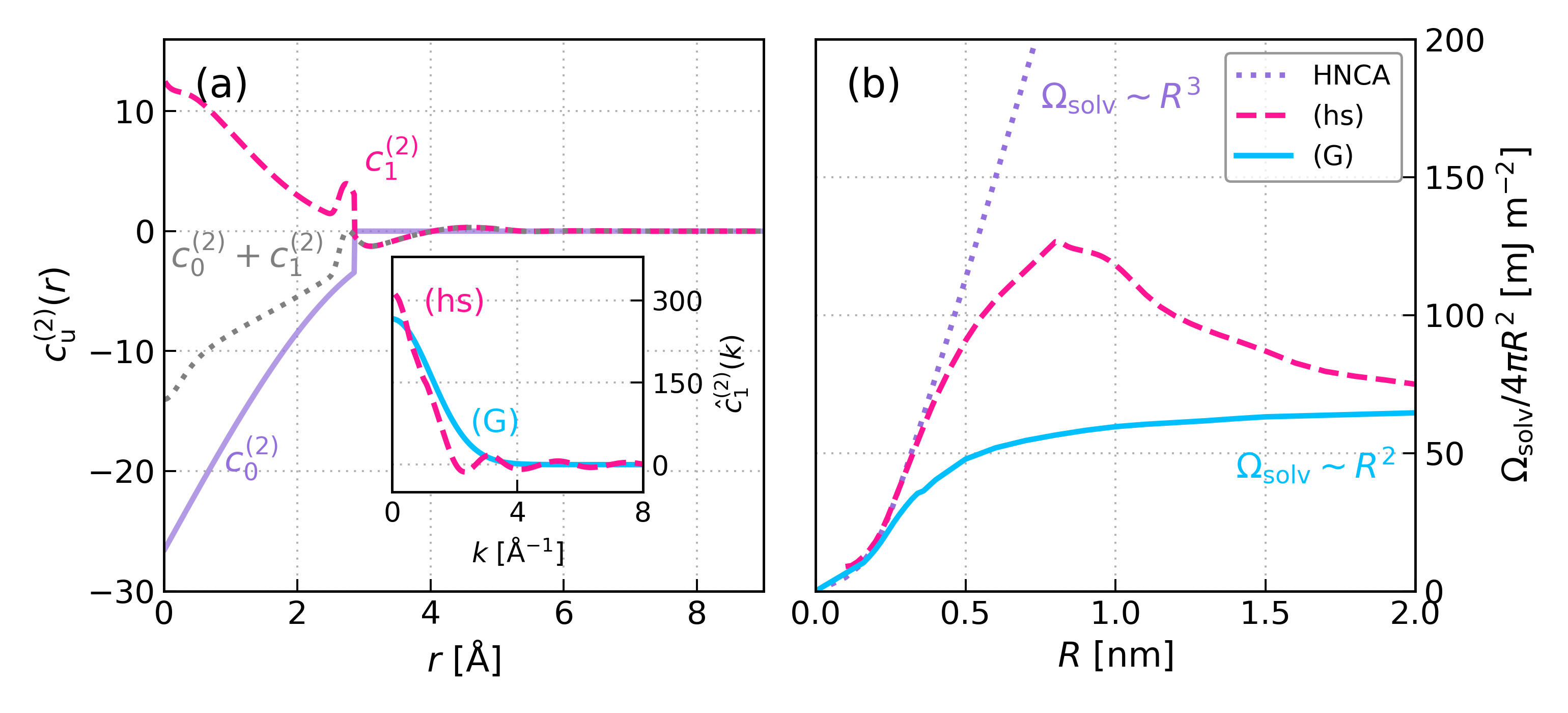} \caption{\textbf{Hard
    sphere splitting of the direct correlation function} (a) The bulk
    direct correlation function of SPC/E water
    ($\rho_{\mrm{u}}=33.2\,\mrm{nm^{-3}}$, $300\,\mrm{K}$) is split
    into a hard-sphere component $c^{(2)}_{\mrm{u},0}$ (Percus--Yevick
    approximation\protect\cite{Percus1958,HansenMcDonaldBook} with
    hard sphere diameter $d=0.286\,\text{nm}$, density
    $\rho=\rho_{\mrm{u}}$) and the remaining component
    $c^{(2)}_{\mrm{u},1}$. While $c^{(2)}_{\mrm{u},1}$ by definition
    should be a function that is slowly-varying over the molecular
    length scale, the hard sphere splitting still leaves
    rapidly-varying oscillations in $c^{(2)}_{\mrm{u},1}$. Instead, by
    using Eq.~\ref{eqn:dcf_gaussian} ($a =
    200$\,kJ\,cm$^3$\,mol$^{-2}$, $\lambda= 0.08$\,nm), these
    rapidly-varying oscillations are smoothed out, as shown in the
    inset of (a) in reciprocal space. In (b), we show
    $\varOmega_{\mrm{solv}}/4\pi R^2$ with the HNCA (dotted purple),
    our theory with the hard-sphere splitting (dashed pink) and our
    theory using a Gaussian weight (solid
    blue). } \label{si:dcf_coarse_grain}
\end{figure}

The hard sphere splitting of the direct correlation of SPC/E is shown
in Fig.~\ref{si:dcf_coarse_grain}(a) using $d=0.286$\,nm.
While such a splitting allows one to separate out some of the
short-ranged correlations, we see that $c^{(2)}_{\mrm{u}}(r) -
c_{\mrm{hs}}^{(2)}(r)$ is not completely slowly-varying. This reflects
the fact that water is not a simple fluid, and that the hard-sphere
fluid is not a suitable reference system to describe its behavior
quantitatively.  We can circumvent the problem of finding an
appropriate splitting for the direct correlation function by carrying
out the coarse graining using a Gaussian weight, as given in
Eq.~\ref{eqn:CG-g}.  This is equivalent to approximating the component
of the direct correlation function responsible for slowly-varying
density $c^{(2)}_{\mrm{u},1}$ with the form
\begin{equation}
    c^{(2)}_{\mrm{u},1}(|\mbf{r}-\mbf{r}^\prime|) \approx  \frac{2\beta a}{(2\pi\lambda^2)^{3/2}}\exp\left(-\frac{|\mbf{r}-\mbf{r}^\prime|^2}{2\lambda^2}\right).
    \label{eqn:dcf_gaussian}
\end{equation}
With an appropriate choice for $a$ and $\lambda$, as will be
rationalized in the following subsection, this form of
$c^{(2)}_{\mrm{u},1}$ resembles the slowly-varying behavior of
$c^{(2)}_{\mrm{u}}(r) - c_{\mrm{hs}}^{(2)}(r)$, as shown in the inset
of Fig.~\ref{si:dcf_coarse_grain}(a).

We determined the free energies of solvating hard sphere solutes using
this hard sphere splitting. The results are shown in
Fig.~\ref{si:dcf_coarse_grain}(b), together with the HNCA and our
theory using a Gaussian coarse graining.  While the hard sphere
splitting is able to remedy the HNCA's failure to capture the
hydrophobic crossover, $\varOmega_{\mrm{solv}}$ for solutes with
$R>0.4\,\mrm{nm}$ is still overestimated in comparison to using a
Gaussian coarse graining (and results from molecular simulations, as
described in the main article).

\subsection{Coarse-graining parameters}

\begin{figure}[H]
    \centering \includegraphics[width=0.9\linewidth]{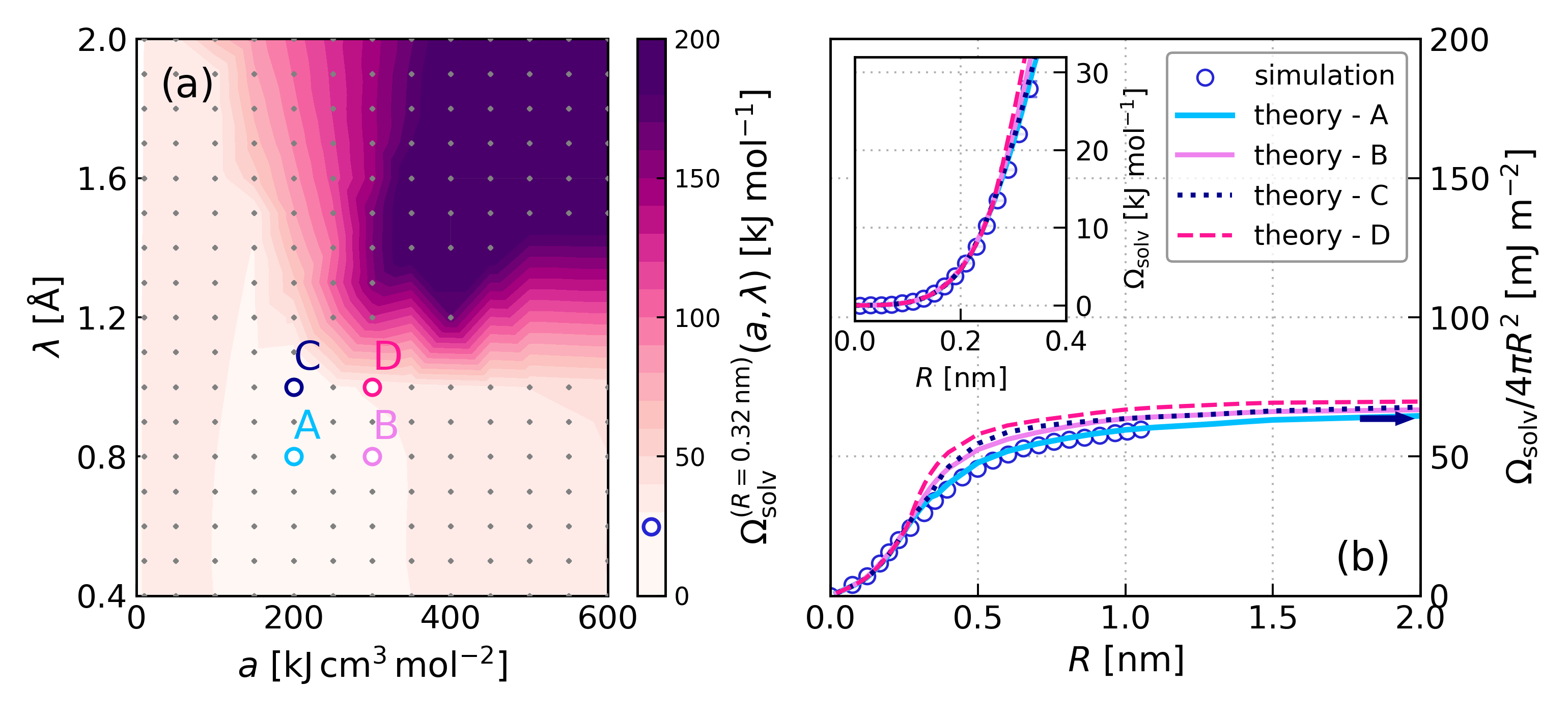} \caption{\textbf{Determining
    the coarse-graining parameters.}  The solvation free energy of a
    hard sphere with $R=0.32\,\mrm{nm}$ associated with different sets
    of parameters $(a,\lambda)$ for the coarse-graining procedure is
    shown in (a). The four sets of parameters selected all give
    $\varOmega_{\mrm{solv}}$ in good agreement with the value obtained
    from simulation, which is indicated by a marker on the colorbar.
    The parameters for each set are: ($a =
    200$\,kJ\,cm$^3$\,mol$^{-2}$, $\lambda= 0.8\,\text{\AA}$) for A;
    ($a = 300$\,kJ\,cm$^3$\,mol$^{-2}$, $\lambda= 0.8\,\text{\AA}$)
    for B; ($a = 200$\,kJ\,cm$^3$\,mol$^{-2}$, $\lambda=
    1.0\,\text{\AA}$) for C; and ($a = 300$\,kJ\,cm$^3$\,mol$^{-2}$,
    $\lambda= 1.0\,\text{\AA}$) for D.  In (b), set A is shown to give
    the best agreement for $\varOmega_{\mrm{solv}}/4\pi R^2$ both for
    $R\le 0.4$\,nm (shown in the inset, simulation data are our own)
    and across the crossover regime at $R\approx 1$\,nm (simulation
    data from Ref.~\onlinecite{Huang2001}).  The arrow indices
    $\gamma=63.6\,\mrm{mJ\,m^{-1}}$ for SPC/E
    water \cite{Vega2007}. In all cases, we have used the same
    parameterization for $m(\rho_{s}^\ast)$ as described in
    Section~\ref{sec:slowlyvarying}.} \label{si:sensitivity}
\end{figure}

To determine the coarse-graining parameters, we first use
$c^{(2)}_{\mrm{u}}(r) - c_{\mrm{hs}}^{(2)}(r)$ from the hard sphere
splitting described above as a guide. For plausible values of $d$
based on the position of water's oxygen--oxygen radial distribution
function ($0.275\lesssim d/\text{nm} \lesssim 0.295$), we estimate a
broad range for $a$ ($a^{({\mrm{hs}})} \approx
40$--$500$\,kJ\,cm$^3$\,mol$^{-2}$).  As we need to perform a
simulation of the bulk fluid to parameterize $c^{(2)}_{\rm u}$, we can
narrow down this estimated range at negligible further computational
cost by comparing $\varOmega_{\mrm{solv}}(a,\lambda)$ obtained from
theory to the result from molecular simulations for a solute on the
verge of the crossover regime. In Fig.~\ref{si:sensitivity}(a) we show
$\varOmega_{\mrm{solv}}(a,\lambda)$ for $R=0.32\,\mrm{nm}$,
identifying a narrow range of the coarse-graining parameters
$a\approx100$--$300$\,kJ\,cm$^3$\,mol$^{-2}$ and $\lambda\approx
0.06$--$0.11\,\text{nm}$. Similar to $m(\rho^*_{\rm s})$ discussed in
Section~\ref{sec:slowlyvarying}, the range of acceptable $a$ and
$\lambda$ gives an extra flexibility in describing the crossover
regime, as shown in Fig.~\ref{si:sensitivity}(b). For SPC/E water, we
settled on $a = 200$\,kJ\,cm$^3$\,mol$^{-2}$ and $\lambda= 0.08$\,nm.


\newpage

\section{Numerical procedure for LCW-style cDFT minimisation}

\subsection{Self-consistent cycle}

The pair of self-consistent equations that we need to solve are:

\begin{subequations}
\begin{align}
 &\omega^\prime(\rho_{\rm s}(\mbf{r})) =
  m\nabla^2\rho_{\rm s}(\mbf{r}) + 2a[\overline{\rho}(\mbf{r})-\overline{\rho}_{\rm s}(\mbf{r})], \label{eqn:a} \\
 &\rho(\mbf{r}) =
 \rho_{\rm s}(\mbf{r})
 \exp\left[-\beta\phi(\mbf{r}) +
 \int\!\mrm{d}\mbf{r}^\prime\,c^{(2)}_{\rm s}(\mbf{r},\mbf{r}^\prime)\delta_{\rm s}\rho(\mbf{r}^\prime)\right]. \label{eqn:b}
\end{align}
\end{subequations}
The self-consistent cycle consists of the following steps:
\begin{enumerate}[leftmargin=1.5cm,rightmargin=1.5cm,label=(\roman*)]
    \item Make an initial guess for the slowly-varying density field, e.g., $\rho_{\mrm{s}}(\mbf{r})=\rho_\mrm{u}$ and the fluid density, i.e., $\rho(\mbf{r})=\rho_\mrm{u}\exp\left[-\beta\phi(\mbf{r})\right]$.
    \item Fix $\rho_{\mrm{s}}(\mbf{r})$ and solve for a new fluid density $\rho(\mbf{r})$ from Eq.~\ref{eqn:b} using Picard iteration:  
            \begin{enumerate}[label=(\alph*)]
                \item Calculate the right hand side of Eq.~\ref{eqn:b} using the current $\rho^{(i)}(\mbf{r})$; call this $\tilde{\rho}^{(i)}(\mbf{r})$. 
                \item Calculate the new full density $\rho^{(i+1)}(\mbf{r})=(1-\alpha)\rho^{(i)}(\mbf{r}) + \alpha \tilde{\rho}^{(i)}(\mbf{r})$. 
                \item Set the new density $\rho^{(i+1)}(\mbf{r})$ as the current density $\rho^{(i)}(\mbf{r})$. Go to step (a) until the solution for $\rho(\mbf{r})$ is converged.
            \end{enumerate}
    \item Fix $\rho(\mbf{r})$ and solve for a new slowly-varying density
    $\rho_{\mrm{s}}(\mbf{r})$ from Eq.~\ref{eqn:a} using Picard iteration. We will recast Eq.~\ref{eqn:a} in an iterative form in the next subsection.
            \begin{enumerate}[label=(\alph*)]
                \item Calculate the right hand side of the iterative equation
                using the current $\rho_{\mrm{s}}^{(i)}(\mbf{r})$; call this $\tilde{\rho}_{\mrm{s}}^{(i)}(\mbf{r})$. 
                \item Calculate the new slowly-varying density $\rho_{\mrm{s}}^{(i+1)}(\mbf{r})=(1-\alpha_{\mrm{s}})\rho^{(i)}(\mbf{r}) + \alpha_{\mrm{s}} \tilde{\rho}_{\mrm{s}}^{(i)}(\mbf{r})$. 
                \item Set the new density $\rho_{\mrm{s}}^{(i+1)}(\mbf{r})$ as the current density $\rho_{\mrm{s}}^{(i)}(\mbf{r})$. Go to step (a) until the solution for $\rho_{\mrm{s}}(\mbf{r})$ is converged. 
            \end{enumerate}
    \item Repeat step (ii) and (iii) until both solutions for $\rho_{\mrm{s}}(\mbf{r})$ and $\rho(\mbf{r})$ are converged. 
\end{enumerate}
The values of the mixing parameters $\alpha$ and $\alpha_{\mrm{s}}$ used depend on the system in question and typically $\alpha\in[0.01,0.05]$ and $\alpha_{\mrm{s}}\in[0.05,0.2]$. Solutions are considered converged when difference between the current and new density profiles is less than the tolerance ($\mrm{rtol}\,=10^{-5}$, $\mrm{atol}\,=10^{-8}$) using Numpy's \texttt{allclose} routine \cite{harris2020array}.

\subsection{Solving for the slowly-varying density field}

To recast Eq.~\ref{eqn:a} in an iterative form, we can rewrite the term involving $\nabla^2\rho_{\mrm{s}}$ in terms of the coarse-grained density field $\overline{\rho}_{\mrm{s}}$\cite{Lum1999}. By a second order Taylor expansion of 
$\rho(\mbf{r}^\prime)$ about $\rho(\mbf{r})$ in Eq.~\ref{eqn:CG-g}, we can write
\begin{equation}
\begin{split}
    \bar{\rho}(\mbf{r})&  \approx \int\!\mrm{d}\mbf{r}'\,  \frac{1}{(2\pi\lambda^2)^{3/2}}\exp\left(-\frac{|\mbf{r}-\mbf{r}'|^2}{2\lambda^2}\right) \left[\rho(\mbf{r}) + \frac{1}{2}\nabla^2_{\mbf{r}'}\rho(\mbf{r}')\vert_{\mbf{r}}(\mbf{r}-\mbf{r}')^2 + ...\right]\\
    & = \rho(\mbf{r}) + \frac{1}{2}\lambda^2 \nabla^2\rho(\mbf{r}).
\end{split}
\label{eqn:Taylor-expand-CG}
\end{equation}
Therefore, we can re-express the Laplacian of the slowly-varying density as
\begin{equation}
    \nabla^2\rho_{\mrm{s}}(\mbf{r}) = \frac{2}{\lambda^2}[\overline{\rho}_{\mrm{s}}(\mbf{r}) - \rho_{\mrm{s}}(\mbf{r})].
    \label{eqn:laplacian}
\end{equation}
By substitution of Eq.~\ref{eqn:laplacian} into Eq.~\ref{eqn:a}, we reach an iterative form of Eq.~\ref{eqn:a}
\begin{equation}
  \rho_{\mrm{s}}(\mbf{r}) = \overline{\rho}_{\mrm{s}}(\mbf{r}) -  \frac{\lambda^2}{2m}\omega'(\rho_{\mrm{s}}) + \frac{a\lambda^2}{m} \left[ \overline{\rho}(\mbf{r}) - \overline{\rho}_{\mrm{s}}(\mbf{r})\right],
  \label{eqn:recast-scf}
\end{equation}
which is solved numerically by Picard iteration. We also note that for the coarse-graining procedure, the
Gaussian convolution is done in Fourier space, i.e
\begin{equation}
    \hat{\bar{\rho}}(\mbf{k}) =  \hat{\rho}(\mbf{k}) \exp\left(-\frac{1}{2}|\mbf{k}|^2 \lambda^2\right).
\end{equation}

\subsection{Solving for the rapidly-varying density field}

By using the approximation 
\begin{equation}
 c^{(2)}_\mrm{s}(\mbf{r},\mbf{r}^\prime) \approx c^{(2)}_\mrm{u}(|\mbf{r}-\mbf{r}^\prime|)\rho_{\mrm{s}}(\mbf{r}) \rho_{\mrm{s}}(\mbf{r}^\prime) \rho_{\mrm{u}}^{-2},
\end{equation}
for the direct correlation function of the slowly-varying density,  Eq.~\ref{eqn:b} becomes
\begin{equation}
 \rho(\mbf{r}) = \rho_{\mrm{s}}(\mbf{r}) \exp\left[ -\beta\phi(\mbf{r})  + \frac{\rho_{\mrm{s}}(\mbf{r})}{\rho_\mrm{u}^2}\int\!\!\mrm{d}\mbf{r}'\,c^{(2)}_{\mrm{u}}(|\mbf{r}-\mbf{r}'|)    \rho_{\mrm{s}}(\mbf{r}')[\rho(\mbf{r}') -\rho_{\mrm{s}}(\mbf{r}') ]\right],
\end{equation}
where the integral
\begin{equation}
f_1(\mbf{r}) = \int\!\!\mrm{d}\mbf{r}'\,c^{(2)}_{\mrm{u}}(|\mbf{r}-\mbf{r}'|)    \rho_{\mrm{s}}(\mbf{r}')[\rho(\mbf{r}') -\rho_{\mrm{s}}(\mbf{r}') ],
\end{equation}
can be performed in Fourier space.
By defining
\begin{equation}
    f_2(\mbf{r}) = \rho_{\mrm{s}}(\mbf{r})[\rho(\mbf{r}) -\rho_{\mrm{s}}(\mbf{r}) ],
\end{equation}
the convolution theorem allows us to compute $\hat{f}_1$ as a product in Fourier space
\begin{equation}
    \hat{f}_1(\mbf{k}) = \hat{c}^{(2)}_{\mrm{u}}(|\mbf{k}|)  \hat{f}_2(\mbf{k}).
\end{equation}

\subsection{External potentials}

We have considered spherical solutes with three different external potential forms:
\begin{enumerate}[label=(\roman*)]
    \item Hard sphere solutes: For an ideal hydrophobic solute, the external potential is
    \begin{equation}
  \phi(r) =
    \begin{cases}
      \infty&\quad  \quad \quad \text{for $r \leq R$}\\
     0 &\quad  \quad \quad \text{for $ r> R$},
    \end{cases}      
\end{equation}
    where $R$ is the hard sphere radius.
    \item Lennard-Jones solutes: For solvation of a single Lennard-Jones particle with a fixed well-depth of $\epsilon_{\mrm{s}}=0.5\,\mrm{kJ\,mol^{-1}}$ of increasing size $\sigma_{\mrm{s}}$ in SPC/E water (with Lennard--Jones parameter as $\epsilon_{\mrm{w}}=0.65\,\mrm{kJ\,mol^{-1}}$, $\sigma_{\mrm{w}}=0.3166\,\mrm{nm}$), as considered in Ref.~\onlinecite{Fujita2017}, the solute--solvent interaction is given by
    \begin{equation}
    \phi(r) = 4\epsilon_{\mrm{sw}}\left[ \left( \frac{R}{r} \right)^{12} - \left( \frac{R}{r} \right)^{6} \right],
    \end{equation}
    where the effective strength is $\epsilon_{\mrm{sw}}=\sqrt{\epsilon_{\mrm{s}}\epsilon_{\mrm{w}}}$ and the effective radius is  $R=\frac{1}{2}(\sigma_{\mrm{s}}+\sigma_{\mrm{w}})$.
    \item For solvation of solutes with attractive tails following Coe \emph{et al.} \cite{Coe2022,Coe2023}, the external potential is
    \begin{equation}
    \label{eqn:HSwAtt}
  \phi(r) =
    \begin{cases}
      \infty&\quad  \quad \quad \text{for $r \leq R$}\\
     \epsilon_{\rm sf} \bigg[ 
     \dfrac{2\sigma_{\mrm{s}}^9}{15}\left(\dfrac{1}{r_{-}^9}-\dfrac{1}{r_{+}^9}\right)
     + \dfrac{3\sigma_{\mrm{s}}^9}{20(r+r_{\rm min})}\left(\dfrac{1}{r_{+}^8}-\dfrac{1}{r_{-}^8}\right) \\
     \quad\quad +\, \sigma_{\mrm{s}}^3\left(\dfrac{1}{r_{+}^3}-\dfrac{1}{r_{-}^3}\right)
     + \dfrac{3\sigma_{\mrm{s}}^3}{2(r+r_{\rm min})}\left(\dfrac{1}{r_{-}^2}-\dfrac{1}{r_{+}^2}\right)
     \bigg]&\quad  \quad \quad \text{for $ r> R$},
    \end{cases}    
\end{equation}
where $\epsilon_{\rm sf}$ is the effective solute--fluid attraction strength, $R$ is the solute hard sphere radius, $r_{\rm min}$ is the location of the attractive tail and $r_{\pm}=r+r_{\rm min}\pm R$.
We considered solutes with $R$ and $\epsilon_{\rm sf}$ as a multiple of the size and attractive strength of a water ``molecule'' in the mW model ($\sigma_{\mrm{mw}}=0.23925\,\mrm{nm}$, $\epsilon_{\mrm{mw}}=25.895\,\mrm{kJ\,mol^{-1}}$).
\end{enumerate}

\newpage

\section{Comparison to mDFT } \label{sec:comparemdft}

\subsection{Theory}

In mDFT \cite{Zhao2011mdft,Jeanmairet2013,Jeanmairet2013jcp,Jeanmairet2016,Borgis2020}, 
the grand potential functional is given by
\begin{equation}
\label{eqn:omega_rho-mdft}
\begin{split}
\varOmega_{\phi}[\rho]
= &   \varOmega_{0}[\rho_\mrm{u}] +  \!\int\!\!\mrm{d}\mbf{r}\,\phi(\mbf{r})\rho(\mbf{r})+  k_{\mrm{B}}T\!\int\!\!\mrm{d}\mbf{r}\,\left[\rho(\mbf{r})\ln\left(\frac{\rho(\mbf{r})}{\rho_{\mrm{u}}}\right) - \delta_{\mrm{u}}\rho(\mbf{r}) \right] \\
& -\frac{k_{\mrm{B}}T}{2}\!\int\!\!\mrm{d}\mbf{r}\!\!\int\!\!\mrm{d}\mbf{r}'\,c^{(2)}_{\mrm{u}}(|\mbf{r}-\mbf{r}'|)\,\delta_{\mrm{u}}\rho(\mbf{r})\,\delta_{\mrm{u}}\rho(\mbf{r}')  + \mcl{F}_{\rm bridge}[\rho],
\end{split}
\end{equation}
where $\varOmega_{0}[\rho_\mrm{u}]$ is the grand potential of the reference uniform density.
The ``bridge'' functional, $\mcl{F}_{\rm bridge}$, accounts for contributions to $\mcl{F}^{(\rm ex)}_{\rm intr}$ beyond quadratic order in $\delta_{\rm u}\rho(\mbf{r})$. Neglecting $\mcl{F}_{\rm bridge}$ amounts to the hypernetted-chain approximation (HNCA) of integral equation theories \cite{HansenMcDonaldBook}.
The resulting equilibrium solvent
density is 
\begin{equation}
\label{eqn:scf-mDFT}
\rho(\mbf{r}) =
 \rho_{\rm u}
 \exp\left[-\beta\phi(\mbf{r}) +
\int\!\!\mrm{d}\mbf{r}^\prime\,c^{(2)}_{\rm u}(|\mbf{r}-\mbf{r}^\prime|)\,\delta_{\rm u}\rho(\mbf{r}^\prime) + \frac{\delta \mcl{F}_{\rm bridge}}{\delta \rho(\mbf{r})}\right].
\end{equation}

Recently, Borgis \textit{et al.} proposed a bridge functional based on the weighted density approximation (WDA) \cite{Tarazona1985,Curtin1985} parameterized with thermodynamic properties of the bulk solvent \cite{Borgis2021}. 
The functional was demonstrated to accurately predict solvation free energies for a database of solutes with a variety of shape and size.
We shall refer to the theory as the HNC+WDA.
In this ansatz, the bridge functional takes a cubic form \cite{Borgis2021}
\begin{equation}
\mcl{F}_{\mrm{bridge}}[\rho] = k_\mrm{B} T \rho_{\mrm{u}} \int\!\mrm{d}\mbf{r}\,A_{\mrm{b}}\left(\frac{\overline{\rho}(\mbf{r})}{\rho_{\rm u}}-1\right)^3,
\label{eqn:bridgefunctional}
\end{equation}
where  $\bar{\rho}(\mbf{r})$ is a weighted density given by  a convolution of the solvent density with a Gaussian function
\begin{equation}
    \overline{\rho}(\mbf{r})= \int\!\mrm{d}\mbf{r}'\, \frac{1}{(2\pi\lambda_{\mrm{b}}^2)^{3/2}}\exp\left(-\frac{|\mbf{r}-\mbf{r}'|^2}{2\lambda_{\mrm{b}}^2}\right)\rho(\mbf{r}'),
\end{equation}
where $\lambda_\mrm{b}$ is the coarse-graining length.

The two parameters in this functional, $A_{\mrm{b}}$ and $\lambda_\mrm{b}$, need to be determined. The parameter $A_{\mrm{b}}$ is fixed by imposing the liquid--vapor coexistence condition, ensuring that 
\begin{equation}
\varOmega_0[\rho_{\mrm{l}}]=\varOmega_0[\rho_{\mrm{v}}].
\label{eqn:coexistence-condition}
\end{equation} 
If we assume that $\rho_{\mrm{l}}\approx\rho_\mrm{u}$ and $\rho_{\mrm{v}}\approx0$, then one finds
\begin{equation}
    A_{\mrm{b}} = 1 - \frac{1}{2}\rho_{\mrm{u}}\hat{c}^{(2)}_{\mrm{u}}(0),
\end{equation}
where $\hat{c}^{(2)}_{\mrm{u}}(0)=\int\mrm{d}\mbf{r}c_{\mrm{u}}^{(2)}(r)$.
Using the standard compressibility relation \cite{HansenMcDonaldBook}, $A_{\mrm{b}}$ can be directly related to the bulk compressibility $\chi_\mrm{u}$ (or the $k\rightarrow0$ limit of the structure factor) 
\begin{equation}
    A_{\mrm{b}} = \frac{1}{2} \left(1+\frac{1}{k_{\mrm{B}}T\rho_{\mrm{u}}^{-1}\chi_{\mrm{u}}}\right) = \frac{1}{2} \left(1+\frac{1}{S(0)}\right),
    \label{eqn:fix-a}
\end{equation}
leaving $\lambda_{\mrm{b}}$ as the only parameter to be determined.

One approach to determine $\lambda_{\rm b}$ is to require that
$\varOmega_{\mrm{solv}}/4\pi R^2 \sim \gamma$ for hard sphere solutes
with large $R$. This approach, however, does not guarantee the
behaviour for small solutes. Alternatively, one can choose
$\lambda_{\mrm{b}}$ such that $\varOmega_{\mrm{solv}}$ for small hard
sphere solutes ($R<0.4\,\mrm{nm}$) agrees well with estimates from
molecular simulation (which can be computed from bulk simulations, see
Section ~\ref{sec:simulation_details}). With this second approach,
however, it is not guaranteed that $\varOmega_{\mrm{solv}}/4\pi
R^2 \sim \gamma$ for large solutes.

\subsection{Numerical results}

We now proceed to show our results using the HNC+WDA functional 
for the different water models we have investigated, at the state points considered in Section \ref{sec:slowlyvarying}. Here, we have parameterized $\lambda_{\mrm{b}}$ using simulation data for $\varOmega_{\mrm{solv}}$ of hard spheres with $R<0.4\,\mrm{nm}$ with the corresponding water models and state points.
For comparison, we also show the results using the LCW-style cDFT.
The parameters used for each approach are given in Tables~\ref{tab:mdftparameter} and \ref{tab:lcwcdftparameter}.

\begin{table}[H]
    \centering
    \begin{tabular}{ l  c c c c}
    \hline
    \hline
    Water model & SPC/E & RPBE-D3 & mW  & mW\\
    $T$ [K]     &   300  & 300    &  426  & 300 \\
    \hline
    $\rho_{\rm u}$ [$\mrm{nm^{-3}}$] & 33.236 &  30.079  & 32.204 & 33.406 \\
    $A_{\mrm{b}}$ & 8.5640 & 8.7297 & 11.3269 & 18.9831\\
    $\lambda_{\mrm{b}}$ [$\text{\AA}$] & 1.0 & 1.0 & 0.95 & 1.0 \\
    \hline
    \hline
    \end{tabular}
\caption
{\textbf{Parameters of the HNC+WDA functional from mDFT.} For each
water model and temperature considered, the density $\rho_{\rm u}$ has
been determined to have a small chemical potential deviation from
liquid--vapor coexistence from square gradient theory. The parameter
$A_{\mrm{b}}$ is determined by the bulk compressibility, as given in
Eq.~\ref{eqn:fix-a} by imposing the liquid--vapour coexistence
condition. The stated values of $\lambda_{\rm b}$ give $\varOmega_{\rm solv}$ 
in best agreement with results from molecular simulations for $R<0.4$\,nm.}
\label{tab:mdftparameter}
\end{table}

\begin{table}[H]
    \centering
    \begin{tabular}{ l  c c c c}
    \hline
    \hline
    Water model & SPC/E & RPBE-D3 & mW  & mW\\
    $T$ [K]     &   300  & 300    &  426  & 300 \\
    \hline
    $a$ [kJ\,cm$^3$\,mol$^{-2}$] & 200 & 300 & 300 & 300\\
    $\lambda$ [nm] & 0.08 & 0.08 & 0.11 & 0.08 \\
    $m$ [$\mrm{kJ\, mol^{-2}\, cm^3 \,\text{\AA}^2}$] & 1255 & 1550 & 1100 & 1148 \\
    $t$ [$\mrm{kJ\, mol^{-2}\, cm^3 \,\text{\AA}^5}$] & -20323 & -21115 & -21115 & -18370 \\
    \hline
    \hline
    \end{tabular}
\caption
{\textbf{Parameters of LCW-style cDFT for different 
water models.} This table gives the parameters used to 
give the results presented in Fig.~\ref{si:compareMDFT}. }
\label{tab:lcwcdftparameter}
\end{table}

\begin{figure}[H]
    \centering
    \includegraphics[width=\linewidth]{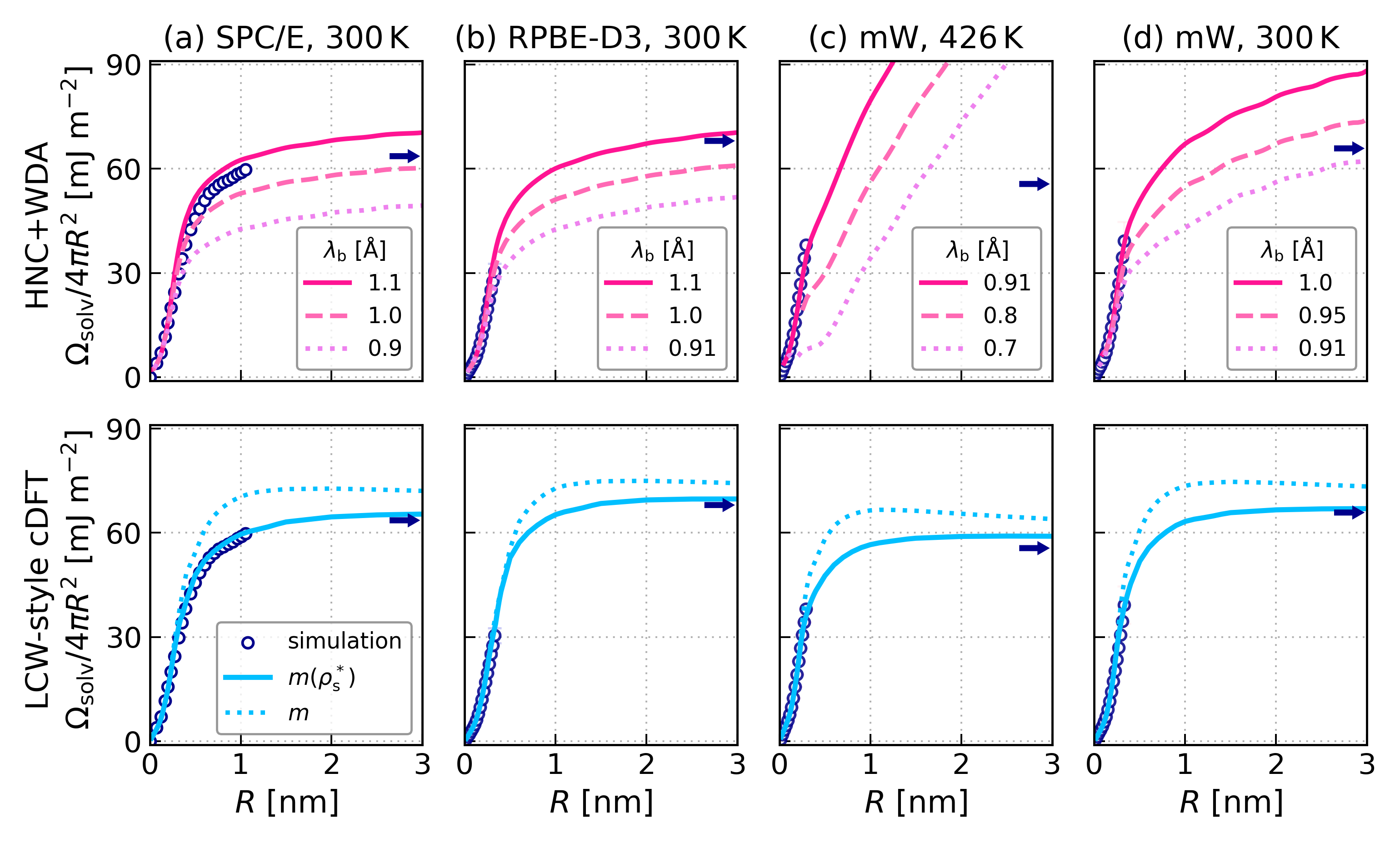}
    \caption{\textbf{Assessing the robustness of mDFT and our
    LCW-style cDFT to different interatomic potentials for water.} We
    show $\varOmega_{\mrm{solv}}/4\pi R^2$ for hard sphere solutes of
    increasing radius $R$, predicted from the HNC+WDA approximation 
    in the top panels (using various values of the
    coarse-graining parameter $\lambda_{\mrm{b}}$ indicated in the legend)
    and our LCW-style cDFT in the lower panels [dotted lines
    indicate our theory parameterized with constant $m$, while solid
    lines indicate our theory with $m(\rho_{\rm s}^\ast)$]. Results
    are shown for (a) the SPC/E water model at 300\,K, (b) the RPBE-D3
    electronic functional at 300\,K, (c) the mW water model at 426\,K,
    and (d) the mW water model at 300\,K. In each panel, we indicate
    $\gamma$ with an arrow, and simulation data with
    circles. }  \label{si:compareMDFT}
\end{figure}

The results from the HNC+WDA functional for solvation of hard spheres are shown in Fig.~\ref{si:compareMDFT}.
We also show the results obtained from the cDFT approach that we have
presented. By construction, the HNC+WDA functional agrees very well
with simulation data for $R<0.4\,\mrm{nm}$. As the solute size
increases, the HNC+WDA functional begins to disagree with our
LCW-style theory. In the case of SPC/E and RPBE-D3 water as seen in
Figs.~\ref{si:compareMDFT}(a) and (b), with
$\lambda_{\mrm{b}}\approx1\,\text{\AA}$, while the HNC+WDA functional
can capture ``hydrophobic crossover'' similar to our approach,
$\varOmega_{\rm solv}/4\pi R^2$ is not guaranteed to converge to
$\gamma$ at the $R\rightarrow\infty$ limit.  The results for the mW
model of water, shown in Figs.~\ref{si:compareMDFT}(c) and (d), are
dramatic; HNC+WDA fails, even qualitatively, to describe the expected
behavior at large $R$. This highlights the difficulty in capturing
faithfully the behavior at each length scale when only a single
functional is constructed. In contrast, the theory from this work,
which solves for two functionals self-consistently, gives results that
are physically reasonable at both length scales.

An obvious point of difference between our approach and mDFT is that
the latter can account for orientational degrees of freedom. However,
the fact differences are most pronounced for the mW model (which lacks
orientational degrees of freedom) suggest this is not the root of the
discrepancy. Moreover, Ref.~\onlinecite{Borgis2020} has shown that for
hydrophobic solvation, results are not significantly impacted by
neglecting orientational degrees of freedom.

\begin{figure}[H]
    \centering
    \includegraphics[width=\linewidth]{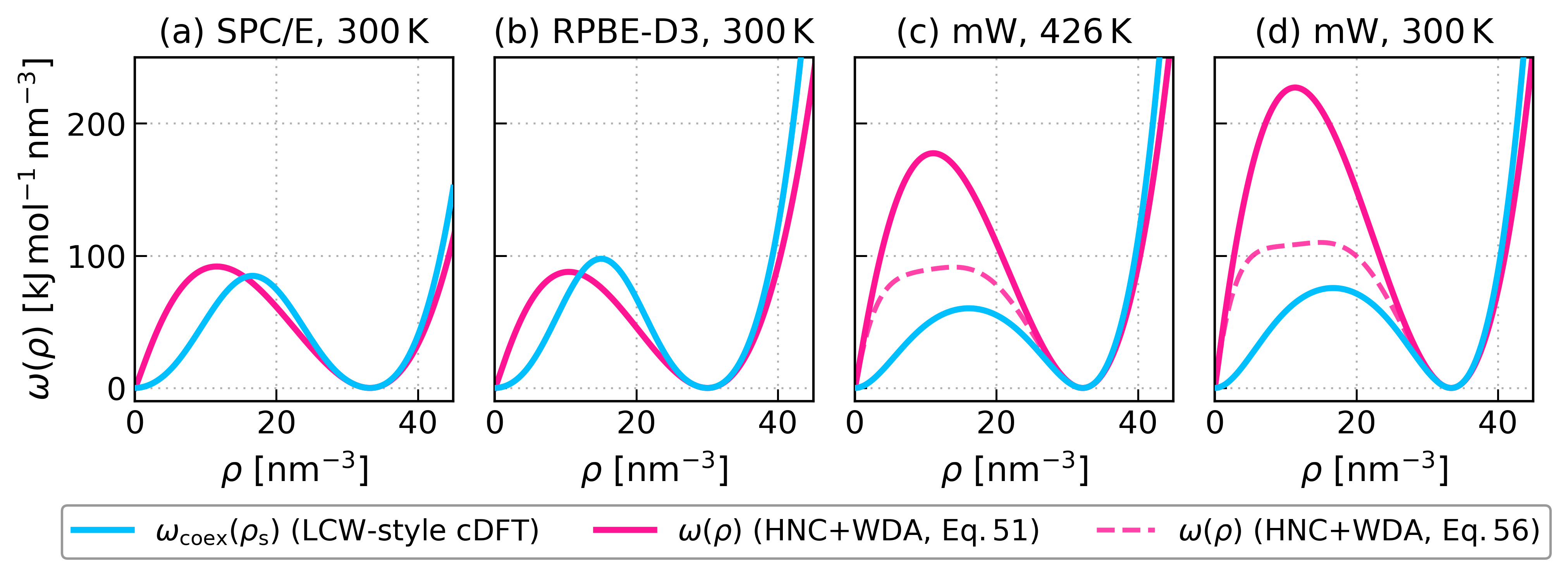}
    \caption{\textbf{Incorporating liquid--vapor coexistence information in the HNC+WDA approach and LCW-style cDFT.}  The local grand potential density of each approach
    is shown for (a) the SPC/E water model at 300\,K, (b) the RPBE-D3
    electronic functional at 300\,K, (c) the mW water model at 426\,K,
    and (d) the mW water model at 300\,K. For the LCW-style cDFT, $\omega_{\rm coex}(\rho_{\rm s})$ is parameterized based on coexistence information, which is used in solving for the slowly-varying reference density in Eq.~\ref{eqn:a}. For the HNC+WDA approach, using a cubic form for the bridge functional as in Eq.~\ref{eqn:bridgefunctional} leads to an overestimation of the barrier height in mW. While this problem can be remedied with an extra parameter in Eq.~\ref{eqn:bridgefunctional2} [the dashed lines correspond to $B_{\rm b}=-35$ in (c) and $B_{\rm b}=-65$ in (d)], we are unable to solve Eq.~\ref{eqn:scf-mDFT} self-consistently.
    }  \label{si:compareMDFTgrandpotential}
\end{figure}

To better understand the differences between results for the different
water models when using HNC+WDA, it is instructive to look at the
local grand potential, as shown in
Fig.~\ref{si:compareMDFTgrandpotential}. It is clear that for mW, when
using the simple cubic bridge functional
(Eq.~\ref{eqn:bridgefunctional}), the barrier between the liquid and
vapor densities is much larger compared to SPC/E and RPBE-D3. One
approach to remedy the results for mW might be to use a more flexible
form for the bridge functional by including higher-order terms. In
fact, in the original HNC+WDA implementation~\cite{Borgis2020}, the
following form was used:
\begin{equation}
\mcl{F}_{\mrm{bridge}}[\rho] = k_\mrm{B} T \rho_{\mrm{u}} \int\!\mrm{d}\mbf{r}\,\left[A_{\mrm{b}}\left(\frac{\overline{\rho}(\mbf{r})}{\rho_{\rm u}}-1\right)^3 + B_{\mrm{b}}\left(\frac{\overline{\rho}(\mbf{r})}{\rho_{\rm u}}\right)^2\left(\frac{\overline{\rho}(\mbf{r})}{\rho_{\rm u}}-1\right)^4 h(\rho(\mbf{r})-\rho_{\mrm{u}})\right].
\label{eqn:bridgefunctional2}
\end{equation}
The extra parameter $B_{\mrm{b}}$ can be used to adjust the barrier
height in $\omega(\rho)$, as shown by the dashed lines in
Fig.~\ref{si:compareMDFTgrandpotential}. However, with such a choice
of $B_{\rm b}$, we have found it challenging to find self-consistent
solutions for the density.

\newpage

\section{Measure of hydrophobicity and critical drying} \label{sec:critical_drying}

\subsection{Local compressibility}

A measure of the hydrophobicity, or more generally solvophobicity, of
a solute or substrate in the grand canonical ensemble is the local
compressibility, as proposed by Evans and Stewart \cite{Evans2015}
\begin{equation}
    \chi(\mbf{r}) = \left( \frac{\partial\rho(\mbf{r})}{\partial\mu} \right)_T,
\end{equation}
i.e., as the derivative of the local density $\rho(\mbf{r})$ with
respect to the chemical potential $\mu$ at a fixed temperature $T$.
In this work, we also consider the contribution to $\chi(\mbf{r})$
from the slowly-varying density $\rho_{\mrm{s}}(\mbf{r})$, which is
given as
\begin{equation}
    \chi_{\mrm{s}}(\mbf{r}) = \left( \frac{\partial\rho_{\mrm{s}}(\mbf{r})}{\partial\mu} \right)_T.
\end{equation}
In practice, we compute $\chi(\mbf{r})$ through a finite difference
\begin{equation}
    \chi(\mbf{r}) \approx \frac{\rho(\mbf{r};\mu + \Delta\mu) - \rho(\mbf{r};\mu)}{\Delta\mu},
    \label{eqn:chi-finitedifference}
\end{equation}
and similarly for $\chi_{\mrm{s}}(\mbf{r})$
\begin{equation}
    \chi_{\mrm{s}}(\mbf{r}) \approx \frac{\rho_{\mrm{s}}(\mbf{r};\mu + \Delta\mu) - \rho_{\mrm{s}}(\mbf{r};\mu)}{\Delta\mu},
\end{equation}
where $\Delta\mu$ is a small finite change in chemical potential.
Since there are available GCMC results from Refs.~\onlinecite{Coe2022,
Coe2023}, we will focus on mW water at $426\,$K
($T/T_{\mrm{c}}=0.46$).

\subsection{Water in contact with a planar hard wall}

\begin{figure}[H]
    \centering
    \includegraphics[width=0.9\linewidth]{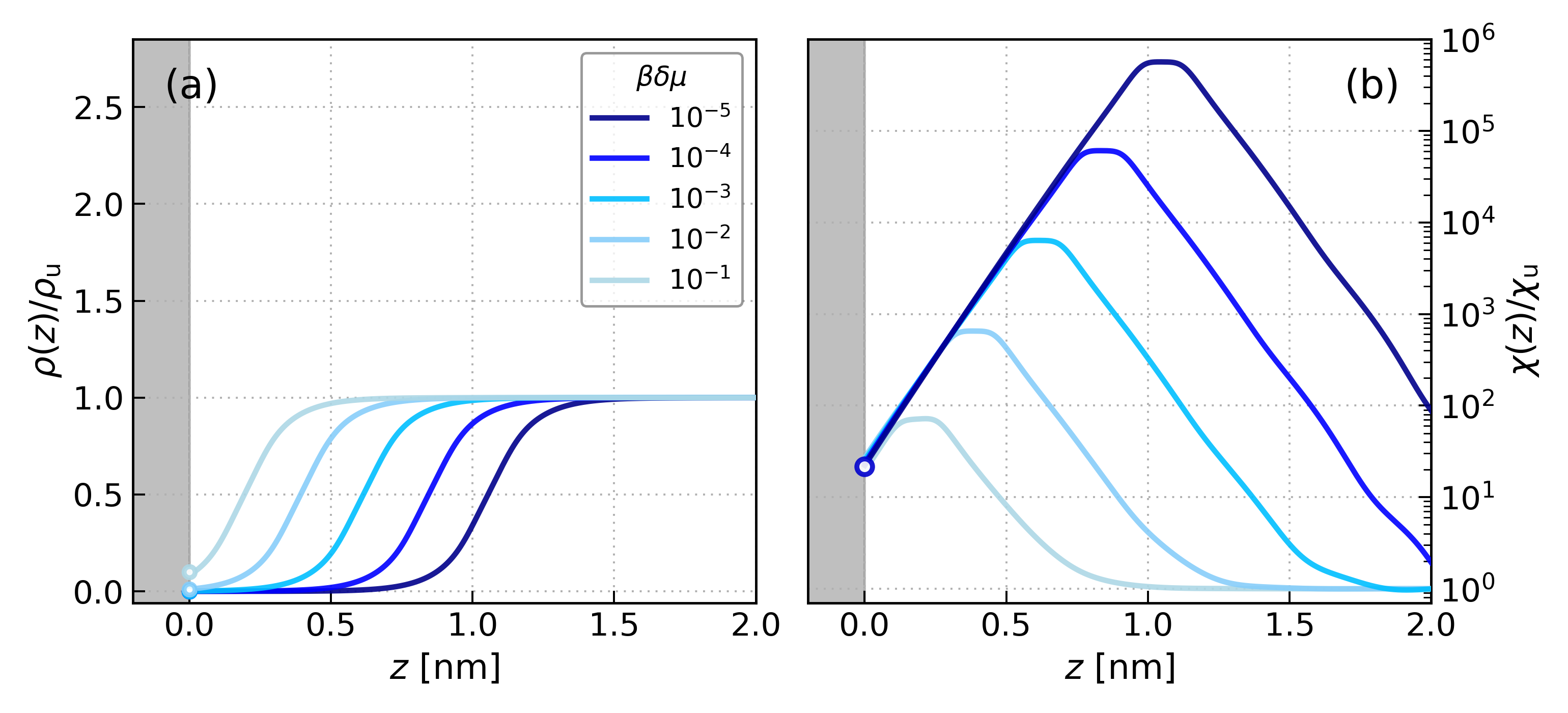}
    \caption{\textbf{Critical drying for water at a planar hard
    wall from LCW-style cDFT. } Density and local compressibility
    profiles for different chemical potentials from cDFT. The fluid
    modelled mW at $426\,\mrm{K}$ is in contact with a planar hard
    wall. The expected values at contact from the respective contact
    theorem for $\rho(0^{+})$ and $\chi(0^{+})$ are marked as open
    circles.}  \label{si:critical_drying}
\end{figure}


To show the critical drying behavior prominently, we determine the
density profile $\rho(z)$ and local compressibility profile $\chi(z)$
of the fluid in contact with a planar hard wall at various chemical
potentials approaching the coexistence point, as shown in
Fig.~\ref{si:critical_drying}.  The equilibrium $\rho(z)$ at each
$\beta\delta\mu$ is determined carefully by starting with different
initial guesses with the vapor layer intruding up to various points
away from the wall, and the solution with the lowest grand potential is
selected.  We determine $\chi(z)$ with a finite difference with
$\Delta\mu$ at least two times smaller than $\delta\mu$.

As $\beta\delta\mu\rightarrow0$, an intruding vapor layer between the
wall and the liquid develops that becomes thicker with $\delta\mu$. This behaviour is in agreement with what is expected as one
moves away from the critical drying point \cite{Coe2022,Evans2015,
Evans2015prl}.

For a fluid in contact with a hard wall, the contact theorem 
\cite{Evans2015} for the density profile is 
\begin{equation}
    \rho(0^{+})= \beta P,
\end{equation}
where $P$ is the bulk pressure. In our square gradient
functional, the equivalent relation the contact density should follow
is
\begin{equation}
  \rho(0^{+}) = \beta\delta \mu \rho_{\mrm{u}} - \beta \omega_{\mrm{coex}}(\rho_{\mrm{u}}),
\end{equation}
consistent with the common tangent construction.  For the local
compressibility, the contact theorem \cite{Evans2015} is
\begin{equation}
    \frac{\chi(0^+)}{\chi_{\mrm{u}}} = \frac{1}{S(0)},
\end{equation}
Our results for both $\rho(z)$ and $\chi(z)$ obey these relationship 
well for $\beta\delta\mu\leq10^{-1}$, establishing the limit for
which the theory is applicable away from coexistence.

We define the equilibrium vapor layer thickness $\ell_{\mrm{eq}}$ as
the distance from the wall to where the density first approaches
$\rho(z)/\rho_{\mrm{u}}=0.5$. From binding potential analysis as laid
out in Ref.~\onlinecite{Evans2017}, the divergence behaviors expected
approaching the drying point are
\begin{equation}
    \ell_{\rm eq}\sim - \ln \delta\mu,
\end{equation}
and
\begin{equation}
  \chi(\ell_{\rm eq}) \sim \delta\mu^{-1}.
\end{equation}
As in Ref.~\onlinecite{Evans2017}, we evaluate the local
compressibility at $z=\ell_{\mrm{eq}}$ using
\begin{equation}
    \chi(\ell_{\rm eq}) = \left(\frac{\partial \rho(z)}{\partial\mu}\right)_{T,\,z=\ell_{\rm eq}} = -\rho'(\ell_{\rm eq}) \left(\frac{\partial \ell_{\rm eq}}{\partial \delta\mu}\right)_T,
\end{equation}
where the prime denotes differentiation with respect to $z$.  For a
range of $10^{-5}\leq \beta \delta\mu \leq 7 \times 10^{-1}$, we
verified that the predictions for $\ell_{\mrm{eq}}$ and
$ \chi(\ell_{\rm eq})$ from the theory follow the scaling expected for
critical drying behavior excellently, as shown in the main paper.

\subsection{Water around hydrophobic spherical solutes}

\begin{figure}[H]
    \centering
    \includegraphics[width=0.95\linewidth]{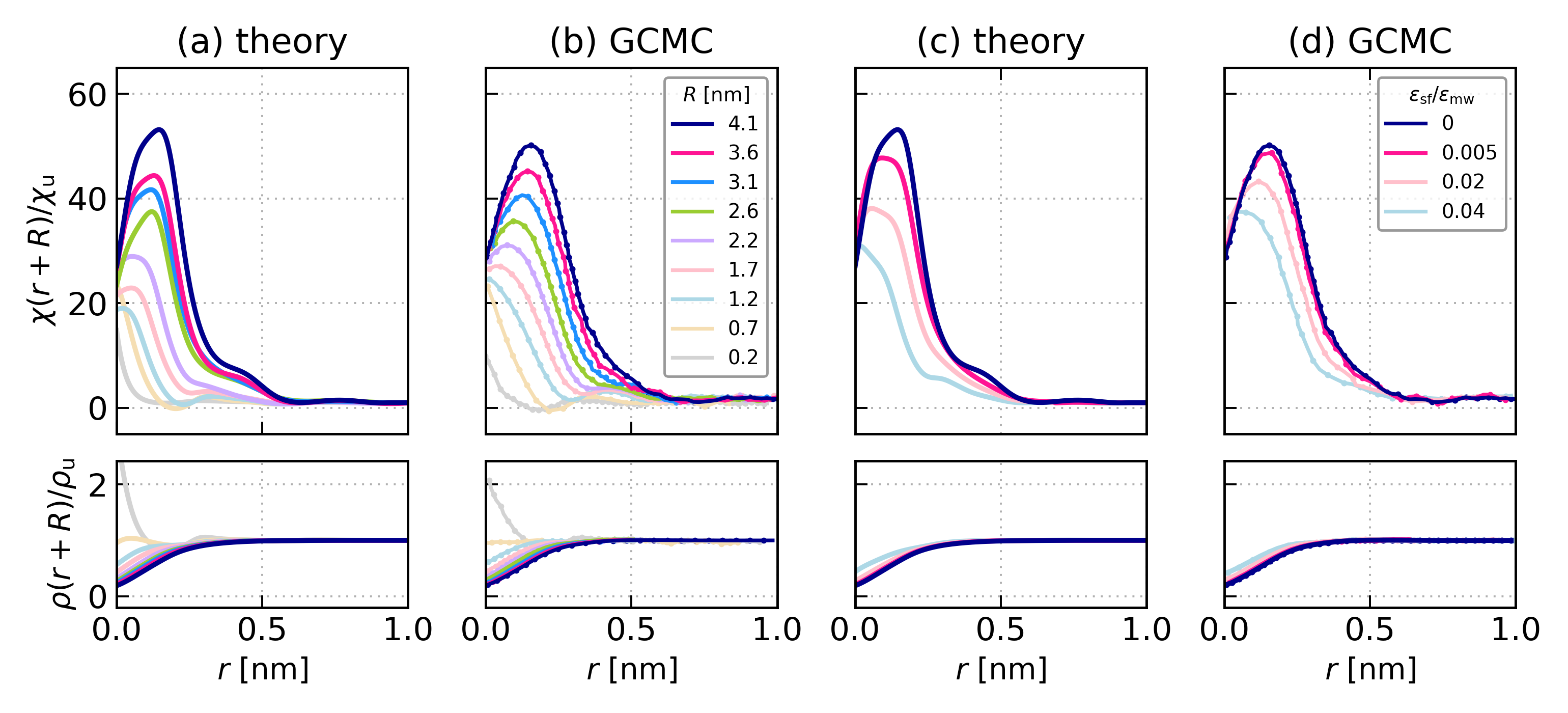}
    \caption{\textbf{Measuring hydrophobicity across length scales
    with cDFT.} The compressibility $\chi(r)$ (top panels) and density
    profiles $\rho(r)$ (bottom panels) are shown for hard sphere
    solutes of different radii $R$ [panel (a) shows results from cDFT,
    panel (b) shows results from GCMC simulations].  We also show
    results obtained with a solute with a hard repulsive core
    ($R=4.1\,\mrm{nm}$) but with different solute--solvent attractive
    strengths, indicated by $\epsilon_{\mrm{sf}}/\epsilon_{\mrm{mw}}$
    [panel (c) shows results from cDFT, panel (d) shows results from
    GCMC simulations]. The results from cDFT are in overall good
    agreement with GCMC data from
    Ref.~\onlinecite{Coe2023}.}  \label{si:localcompressibility}
\end{figure}

Having established that the LCW-style cDFT obey the scaling relations
for critical drying, we proceed to compute the local compressibility
$\chi(r)$ of the fluid around spherical hydrophobic solutes.  We
consider two cases. First we investigate how $\chi$ changes as we
increase the radius $R$ of a hard sphere solute.  Second, we consider
how $\chi$ changes as we vary the attractive interaction strength
between a large solute and water (see Eq.~\ref{eqn:HSwAtt}). The
resulting compressibility and density profiles from our cDFT approach
are shown in Fig.~\ref{si:localcompressibility}. We have also shown
the GCMC simulations results from Coe \emph{et al.} from
Ref.~\onlinecite{Coe2023}. Overall, our results are in good agreement
the molecular simulations, indicating that the LCW-style cDFT approach
contains the essential physics of critical drying.

\newpage

\section{Temperature dependence of hydrophobic solvation}

\subsection{Parameterization}

\begin{figure}[H]
    \centering
    \includegraphics[width=\linewidth]{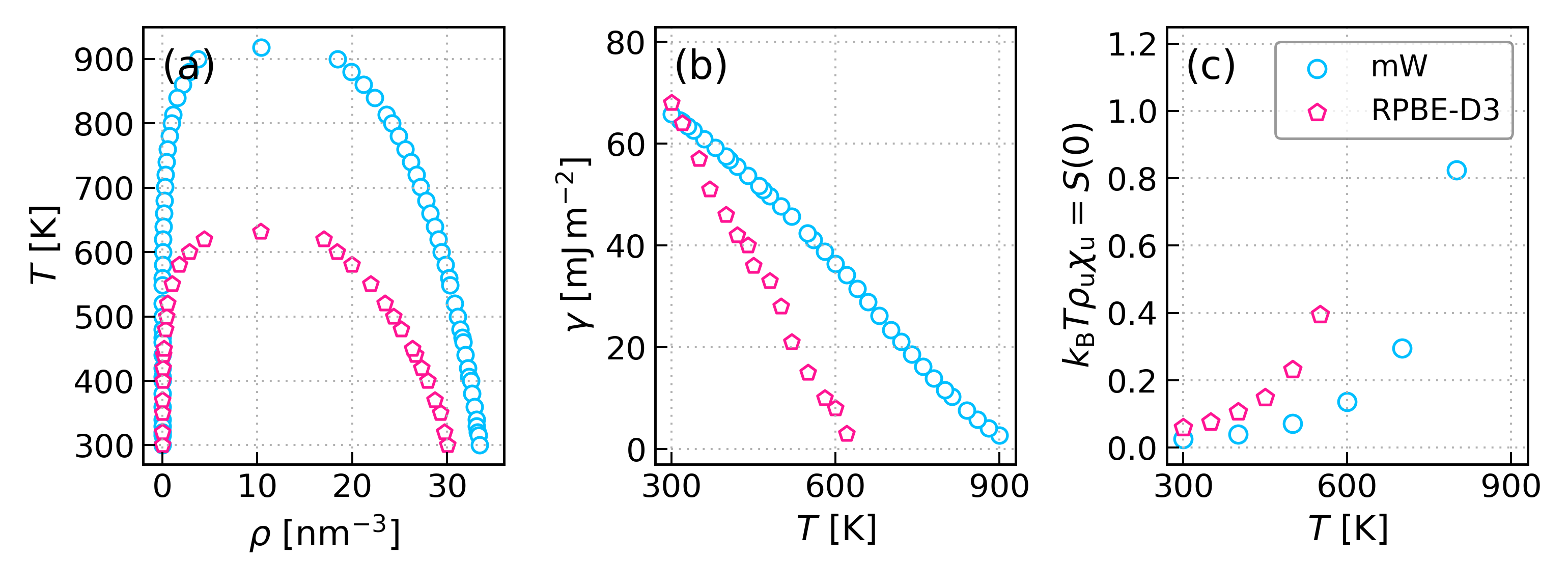}
    \caption{\textbf{Temperature dependence of inputs of theory
    from simulation.} For the binodal curves in (a) and liquid--vapor
    surface tensions in (b), simulation data for RPBE-D3 and mW are in
    taken from Refs.~\onlinecite{Wohlfahrt2020,Coe2022b},
    respectively.  For the bulk compressibilities in (c), we have
    obtained these values with our own simulations of the bulk fluid
    at the liquid coexistence density
    $\rho_{\mathrm{u}}=\rho_{\mathrm{l}}$.}  \label{si:temperature_dependence_data}
\end{figure}

In this section, we assess how well our theory captures the
temperature dependence of hydrophobic solvation, paying particular
attention to the ``entropic crossover" present in water but not in
simple liquids.  We will consider solvation in RPBE-D3 and mW at state
points neighbouring the triple point up to the critical point.  The
fluid is in the liquid phase at coexistence with density
$\rho_{\mrm{u}}=\rho_{\mrm{l}}$. We show the available data points
from simulations for these water models in
Fig.~\ref{si:temperature_dependence_data}.

For each water model at each temperature considered, we 
parameterized the slowly-varying functional as described in 
Section~\ref{sec:slowlyvarying} and summarized
the parameters for the theory in Table~\ref{tab:temp_parameter}.

\begin{table}[H]
    \centering
    \begin{tabular}{c | c c | c c c  c| c c c }
    \hline
    \hline
    & $T$ [K] & $T/T_{\mrm{c}}$ &$\rho_{\rm l}$ [nm$^{-3}$] & $\rho_{\rm v}$ [nm$^{-3}$] & $\gamma$ [$\mathrm{mJ\,m}^{-2}$] & $S(0)$ & $C$ & $D$ & $m$ \\
    \hline
    \multirow{6}{*}{\rotatebox[origin=c]{90}{RPBE-D3}}  
    & 300 & 0.47 & 30.0517 & 0.001003 & 68 & 0.0598 & 1.537 & 90   & 1500 \\
    & 350 & 0.55 & 29.2829 & 0.006686 & 57 & 0.0757 & 1.532 & 50   & 1750 \\
    & 400 & 0.63 & 27.9458 & 0.047133 & 46 & 0.1070 & 1.429 & 20   & 2100 \\
    & 450 & 0.71 & 26.3412 & 0.157111 & 36 & 0.1489 & 1.391 & -20  & 2600 \\
    & 500 & 0.79 & 24.3356 & 0.424535 & 28 & 0.2324 & 1.286 & -30  & 2800 \\
    & 550 & 0.87 & 21.9622 & 1.042952 & 15 & 0.3951 & 1.204 & -100 & 2400 \\
    \hline
    \multirow{6}{*}{\rotatebox[origin=c]{90}{mW}} 
    & 300 & 0.33 & 33.4053 & 0.000013 & 65.8395 & 0.0265 & 2.523 & -15 & 1148 \\
    & 400 & 0.44 & 32.5038 & 0.000895 & 57.5166 & 0.0401 & 2.414 & -25 & 1215 \\
    & 500 & 0.54 & 31.1015 & 0.011361 & 47.7113 & 0.0712 & 1.943 & -30 & 1480 \\
    & 600 & 0.65 & 29.3987 & 0.064951 & 36.4794 & 0.1362 & 1.448 & -30 & 1680 \\
    & 700 & 0.76 & 27.1951 & 0.265254 & 23.4635 & 0.2956 & 0.998 & -35 & 1850 \\
    & 800 & 0.87 & 24.1901 & 0.951576 & 11.6765 & 0.8246 & 0.618 & -50 & 1950\\
    \hline
    \hline
    \end{tabular}
    \caption{\textbf{Parameters of the functional at different
    temperatures}.  The liquid and vapor densities at coexistence,
    $\rho_{\mrm{l}}$ and $\rho_{\mrm{v}}$, and surface tensions
    $\gamma$ have been determined with simulations for RPBE-D3 in
    Ref.~\onlinecite{Wohlfahrt2020} and for mW in
    Ref.~\onlinecite{Coe2022b}. The critical temperature is at
    $T_{\mrm{c}}=632\,$K for RPBE-D3 and $T_{\mrm{c}}=917.6\,$K for
    mW. We have determined the compressibilities of the bulk liquid
    phase with our own simulations, given as the $k\rightarrow0$ limit
    of the structure factors $S(k)$.  The final three columns of the
    table give the parameters for the slowly-varying functional in our
    theory: $C$ given in $\mrm{J\, mol^{-2}\, nm^9}$, $D$ given in
    $\mrm{J\, mol^{-2}\, nm^{15}}$ and $m$ given in $\mrm{kJ\,
    mol^{-2}\, cm^3 \, \AA^2}$.}
\label{tab:temp_parameter}
\end{table}

\begin{figure}[H]
    \centering
    \includegraphics[width=0.95\linewidth]{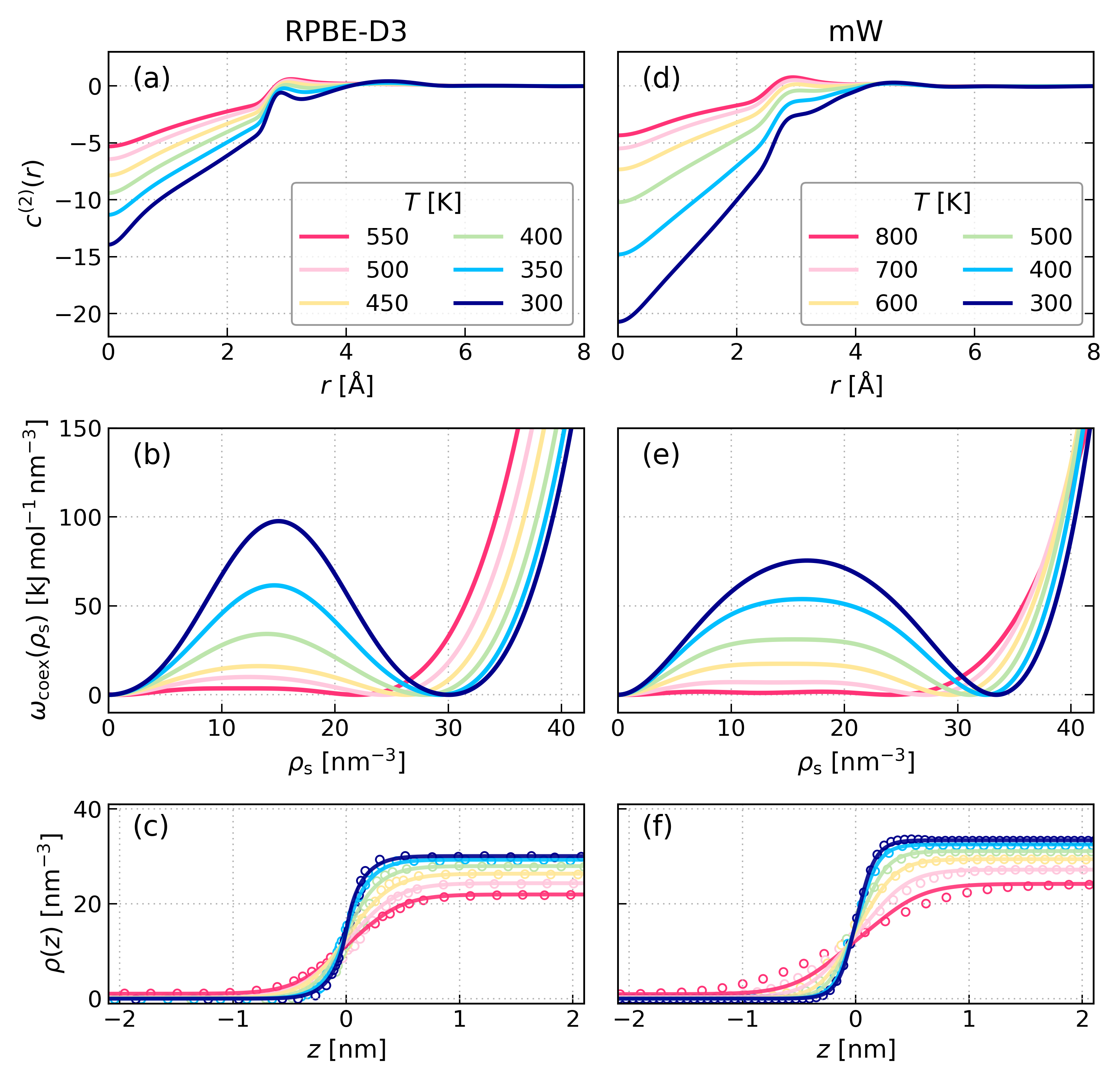}
    \caption{\tbf{Comparison of the bulk liquid local structure and
    liquid--vapor interface of RPBE-D3 and mW.}  The direct
    correlation functions of the bulk liquid at different temperatures
    are extracted from our own bulk simulations and are shown in (a)
    and (d).  The local grand potential densities
    $\omega_{\mrm{coex}}(\rho_\mrm{s})$ parameterized based on bulk
    and coexistence properties for each water model are shown in (b)
    and (e). The free liquid--vapor interfaces for each water model at
    different temperatures are shown in (c) and (d), with the solid
    lines being predictions from the theory and open circles data from
    direct coexistence simulations.  For RPBE-D3, the density profiles
    are taken from Ref.~\onlinecite{Wohlfahrt2020}. For mW, the
    density profiles are from our own direct coexistence
    simulations.}  \label{si:dcf_temp}
\end{figure}

The direct correlation functions of the uniform liquid phases are
obtained from bulk simulations via the Ornstein--Zernike relation as
described in Section~\ref{sec:simulation_details}.  We show the
temperature dependence of these functions as well as the square
gradient theory of the free liquid--vapor interface for each water
model in Fig.~\ref{si:dcf_temp}.

\subsection{Entropic crossover}

\begin{figure}[H]
    \centering \includegraphics[width=0.95\linewidth]{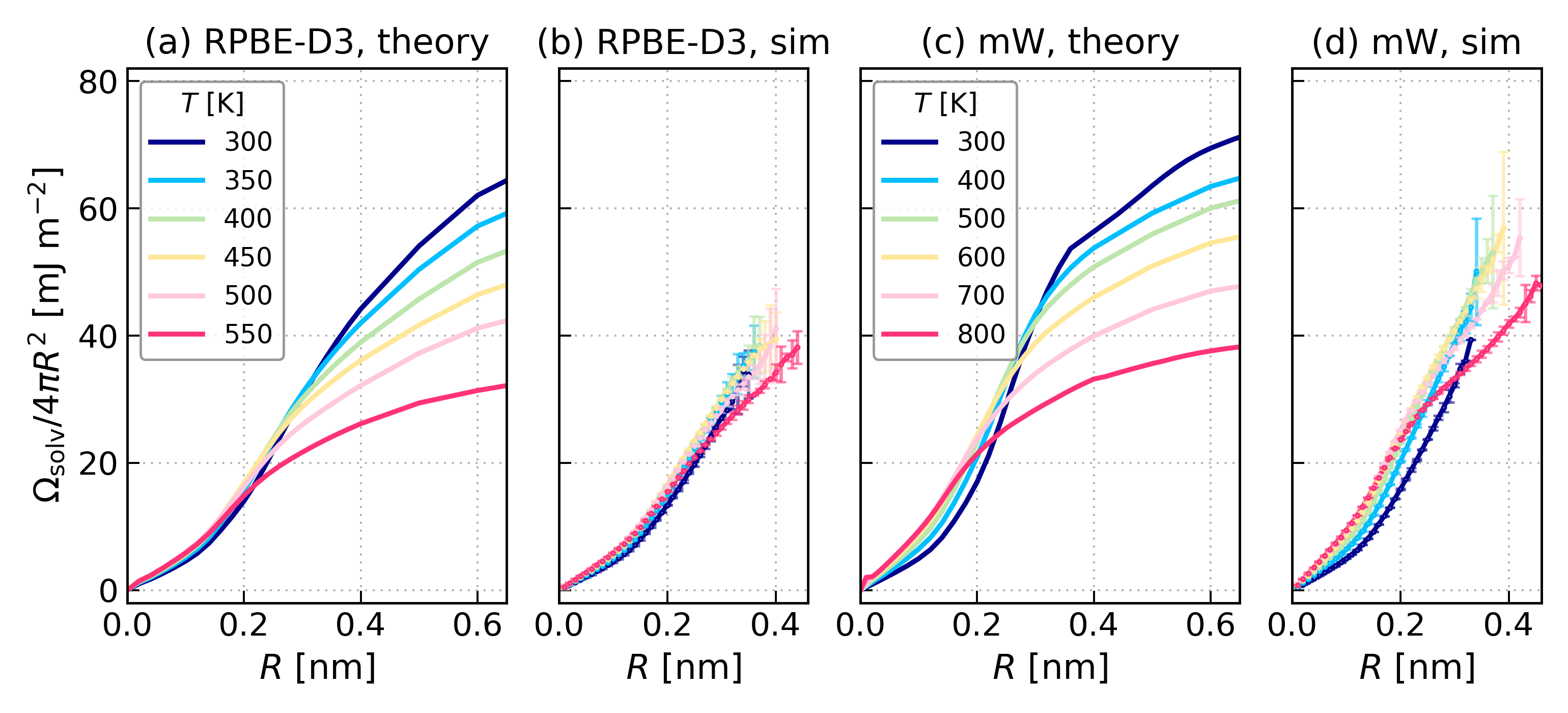}
    \caption{\tbf{Capturing the entropic crossover with cDFT.}  The
    solvation free energy at different temperatures are shown for hard
    sphere solutes of different radii $R$ in RPBE-D3 and mW
    water. Panels (a) and (c) show the results from cDFT while panels
    (b) and (c) show the results from simulations. The results from
    cDFT are in good agreement with simulation, capturing the entropic
    crossover in both water models.}  \label{si:entropic_crossover}
\end{figure}

With the functions and parameters determined, we then determine the
solvation free energy of hard spheres of increasing radii $R$ at
different temperatures from the theory, as shown in
Fig.~\ref{si:entropic_crossover}. For both RPBE-D3 and mW water, the
theory predicts an ``entropic crossover'', whereby
$\varOmega_{\mrm{solv}}$ increases with $T$ for small solutes with
$R \lesssim 0.3\,\mrm{nm}$ and decreases with $T$ for larger solutes
with $R \gtrsim 0.3\,\mrm{nm}$, in good agreement with data from
simulations.

\bibliography{si_references}